\documentclass{wileysev}

\usepackage{chapterbib}
\usepackage[linktoc=none]{hyperref}
\usepackage[usenames,dvipsnames]{xcolor}
\usepackage{w-bookps,amsmath,wasysym,pifont,graphicx}
\usepackage[textsize=small,textwidth=5.5cm,shadow,colorinlistoftodos,disable]{todonotes}
\usepackage{datetime,lipsum,xspace,tikz,soul}

\makeatletter\patchcmd{\@addmarginpar}{%
\ifodd\c@page}{\ifodd\c@page\@tempcnta\m@ne}{}{}
\makeatother\reversemarginpar

\DeclareGraphicsExtensions{.pdf,.eps,.png,.jpg}

\definecolor{blue}{RGB}{74,128,156}
\definecolor{green}{RGB}{155,187,90}
\definecolor{orange}{RGB}{232,153,42}
\definecolor{maroon}{RGB}{194,131,135}
\definecolor{gray}{RGB}{153,153,153}
\definecolor{light}{RGB}{235,235,235}
\definecolor{mid}{RGB}{169,169,169}
\definecolor{dark}{RGB}{121,121,121}

\newcommand{\adhoc}{\emph{ad-hoc}\xspace}

\newcommand{\tab}{~~}

\newcommand{\ovr}{\varrho}
\newcommand{\ovrg}[2]{\ovr_{#1}(#2)}
\newcommand{\obj}{\mathcal{F}}
\newcommand{\objf}[1]{\mathcal{F}(#1)}
\newcommand{\hml}{\mathcal{H}}
\newcommand{\haml}[1]{\mathcal{H}(#1)}
\newcommand{\hami}[2]{\mathcal{H}_{#1}(#2)}
\newcommand{\cns}{\mathcal{G}}
\newcommand{\cons}[2]{\mathcal{G}_{#1}(#2)}
\newcommand{\mdq}{\mathcal{Q}}
\newcommand{\modq}[1]{\mathcal{Q}(#1)}
\newcommand{\cmp}[1]{\mathcal{O}(#1)}

\newcommand{\prbi}[2]{\mathrm{P}_{#1}(#2)}
\newcommand{\neigh}[1]{\Gamma_{#1}}
\newcommand{\avg}[1]{\langle#1\rangle}

\newcommand{\argmax}[0]{\arg\!\max}
\newcommand{\given}[2]{#1\,\vert\,#2}
\newcommand{\holds}[2]{#1\,\colon\,#2}
\newcommand{\size}[1]{\vert#1\vert}
\newcommand{\set}[1]{\{#1\}}
\newcommand{\term}[1]{'#1'}

\newcommand{\chpref}[2]{\todon{Create link to Chapter #1 entitled ``#2''}{Chapter~#1\xspace}}
\newcommand{\secref}[1]{Section~\ref{sec:#1}\xspace}
\newcommand{\secsref}[2]{Sections~\ref{sec:#1}--\ref{sec:#2}\xspace}
\newcommand{\figref}[1]{Figure~\ref{fig:#1}\xspace}
\newcommand{\tblref}[1]{Table~\ref{tbl:#1}\xspace}
\renewcommand{\eqref}[1]{Equation~(\ref{eq:#1})\xspace}
\newcommand{\eqsref}[2]{Equations~(\ref{eq:#1}) and~(\ref{eq:#2})\xspace}
\newcommand{\algref}[1]{Algorithm~\ref{alg:#1}\xspace}

\newcommand{\cref}[1]{Ref.~\cite{#1}\xspace}
\newcommand{\crefs}[1]{Refs.~\cite{#1}\xspace}
\newcommand{\dm}{Doreian~and~Mrvar~\cite{DM96}\xspace}
\newcommand{\rb}{Reichardt~and~Bornholdt~\cite{RB06a}\xspace}
\newcommand{\rak}{Raghavan~et~al.~\cite{RAK07}\xspace}
\newcommand{\tk}{Tib{\'e}ly~and~Kert{\'e}sz~\cite{TK08}\xspace}
\newcommand{\lhlc}{Leung~et~al.~\cite{LHLC09}\xspace}
\newcommand{\bc}{Barber~and~Clark~\cite{BC09a}\xspace}
\newcommand{\lm}{Liu~and~Murata~\cite{LM09b}\xspace}
\newcommand{\lmm}{Liu~and~Murata~\cite{LM09c,LM09d}\xspace}
\newcommand{\sbd}{\v{S}ubelj~and~Bajec~\cite{SB11d}\xspace}
\newcommand{\sbdd}{\v{S}ubelj~and~Bajec~\cite{SB11d,SB10d}\xspace}

\newcommand{\sbbb}{\v{S}ubelj~and~Bajec~\cite{SB11b,SB12u}\xspace}
\newcommand{\sbg}{\v{S}ubelj~and~Bajec~\cite{SB12u,SB14g}\xspace}
\newcommand{\hlszd}{Han~et~al.~\cite{HLSZD16}\xspace}
\newcommand{\zch}{Zhang~et~al.~\cite{ZCH15}\xspace}
\newcommand{\lljt}{Li~et~al.~\cite{LLJT15}\xspace}
\newcommand{\cgg}{Cordasco~and~Gargano~\cite{CG10,CG11}\xspace}
\newcommand{\xs}{Xie~and~Szymanski~\cite{XS11}\xspace}
\newcommand{\xss}{Xie~and~Szymanski~\cite{XSL11,XS12}\xspace}
\newcommand{\gre}{Gregory~\cite{Gre10}\xspace}
\newcommand{\wlgwt}{Wu~et~al.~\cite{WLGWT12}\xspace}
\newcommand{\brsv}{Boldi~et~al.~\cite{BRSV11}\xspace}

\newcommand{\comm}[1]{\todo[color=gray!60,inline,nolist]{{\bf Comment: }#1}\xspace}
\newcommand{\todon}[2]{\todo[color=orange!75,linecolor=orange]{{\bf Note: }#1}{#2}\xspace}

\newcommand{\degs}[5]{$#3$~~\tikz[scale=2]{
\draw[line cap=round] (0, -0.025) -- (0, 0.025) (0, 0) -- (1, 0) (1, -0.025) -- (1, 0.025); 
\draw[ultra thick,color=dark,line cap=round] (#1, -0.05) -- (#1, 0.05) (#1, 0) -- (#2, 0) (#2, -0.05) -- (#2, 0.05);}~~$#4$}

\begin{document}

\listoftodos[List of todos]  \tableofcontents

%
%

\booktitle{Advances in Network Clustering and Blockmodeling}
\offprintinfo{Advances in Network Clustering and Blockmodeling}{P.\ Doreian, V.\ Batagelj \& A.\ Ferligoj}

\chapter{Label Propagation for Clustering}
\chapterauthors{Lovro \v{S}ubelj \chapteraffil{University of Ljubljana, 
Faculty of Computer and Information Science, Ljubljana, Slovenia}}

%
%

\todon{Adjust text to Wiley template!}{Label propagation is a heuristic method initially proposed for community detection in networks~\cite{RAK07,Gre10}, while the method can be adopted also for other types of network clustering and partitioning~\cite{BC09a,LM09c,SB14g,HLSZD16}. Among all the approaches and techniques described in this book, label propagation is neither the most accurate nor the most robust method. It is, however, without doubt one of the simplest and fastest clustering methods. Label propagation can be implemented with a few lines of programming code and applied to networks with hundreds of millions of nodes and edges on a standard computer, which is true only for a handful of other methods in the literature.}

In this chapter, we present the basic framework of label propagation, review different advances and extensions of the original method, and highlight its equivalences with other approaches. We show how label propagation can be used effectively for large-scale community detection, graph partitioning, identification of structurally equivalent nodes and other network structures. We conclude the chapter with a summary of the label propagation methods and suggestions for future research.

%
%

\section{\label{sec:prop}Label Propagation Method}

The label propagation method was introduced by \rak for detecting non-overlapping communities in large networks. There exist multiple interpretations of network communities~\cite{FH16,SDRL17} as described in~\chpref{4}{Community detection}. For instance, a community can be seen as a densely connected group, or cluster, of nodes that is only loosely connected to the rest of the network, which is also the perspective that we adopt here. 

\begin{figure}[t]\centerline{%
	\includegraphics[width=.225\textwidth]{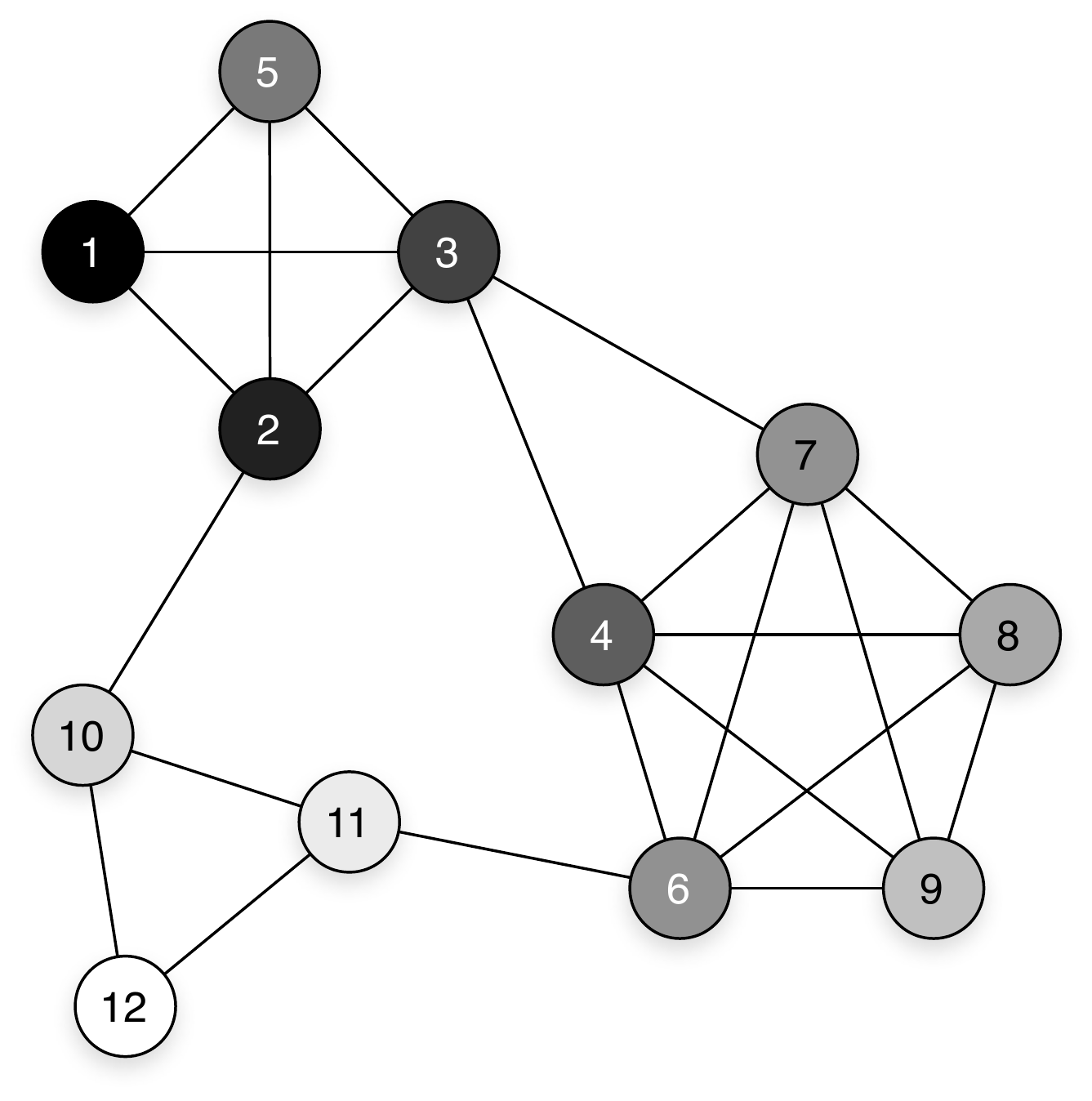}\hskip.033\textwidth
	\includegraphics[width=.225\textwidth]{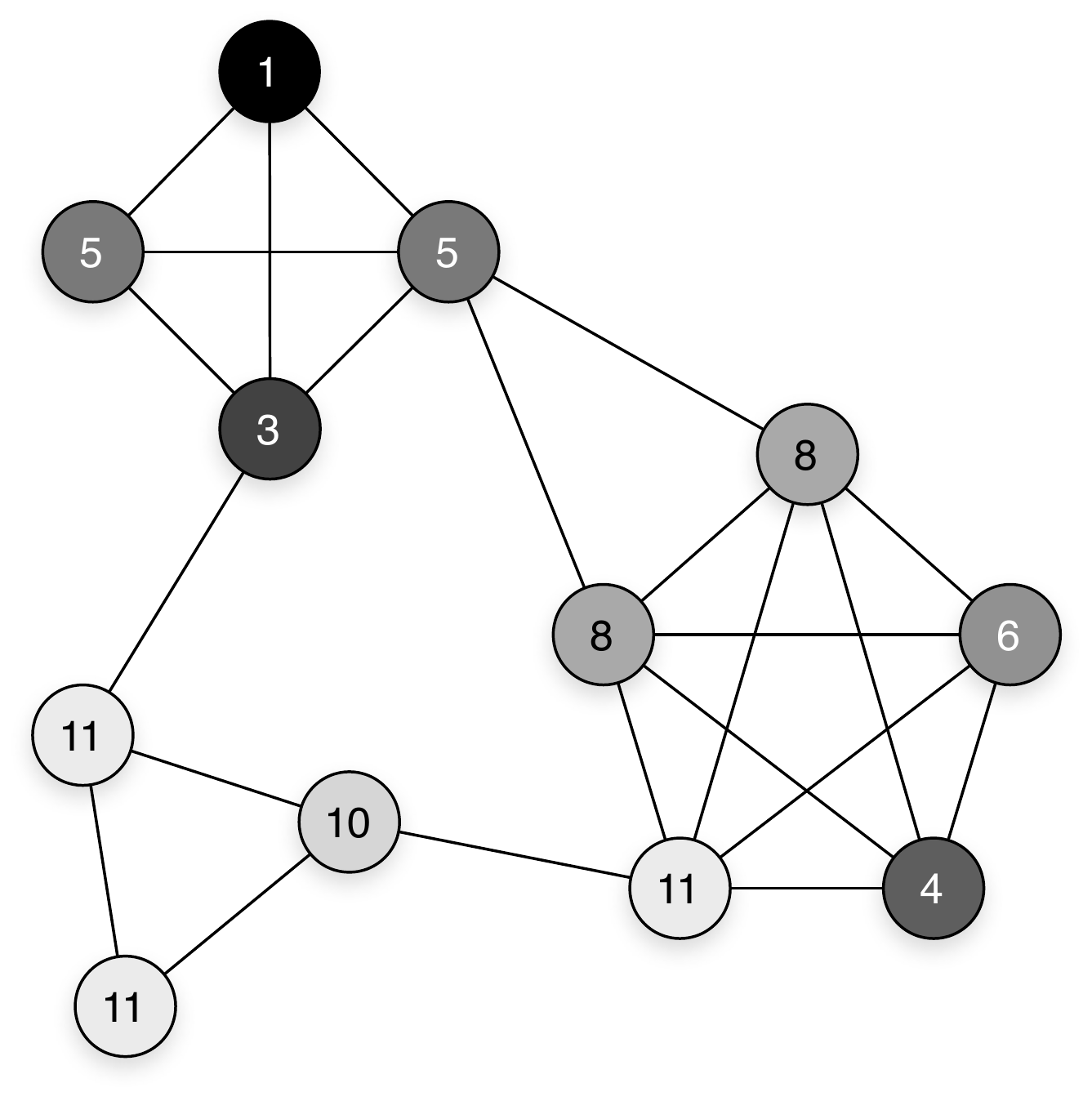}\hskip.033\textwidth
	\includegraphics[width=.225\textwidth]{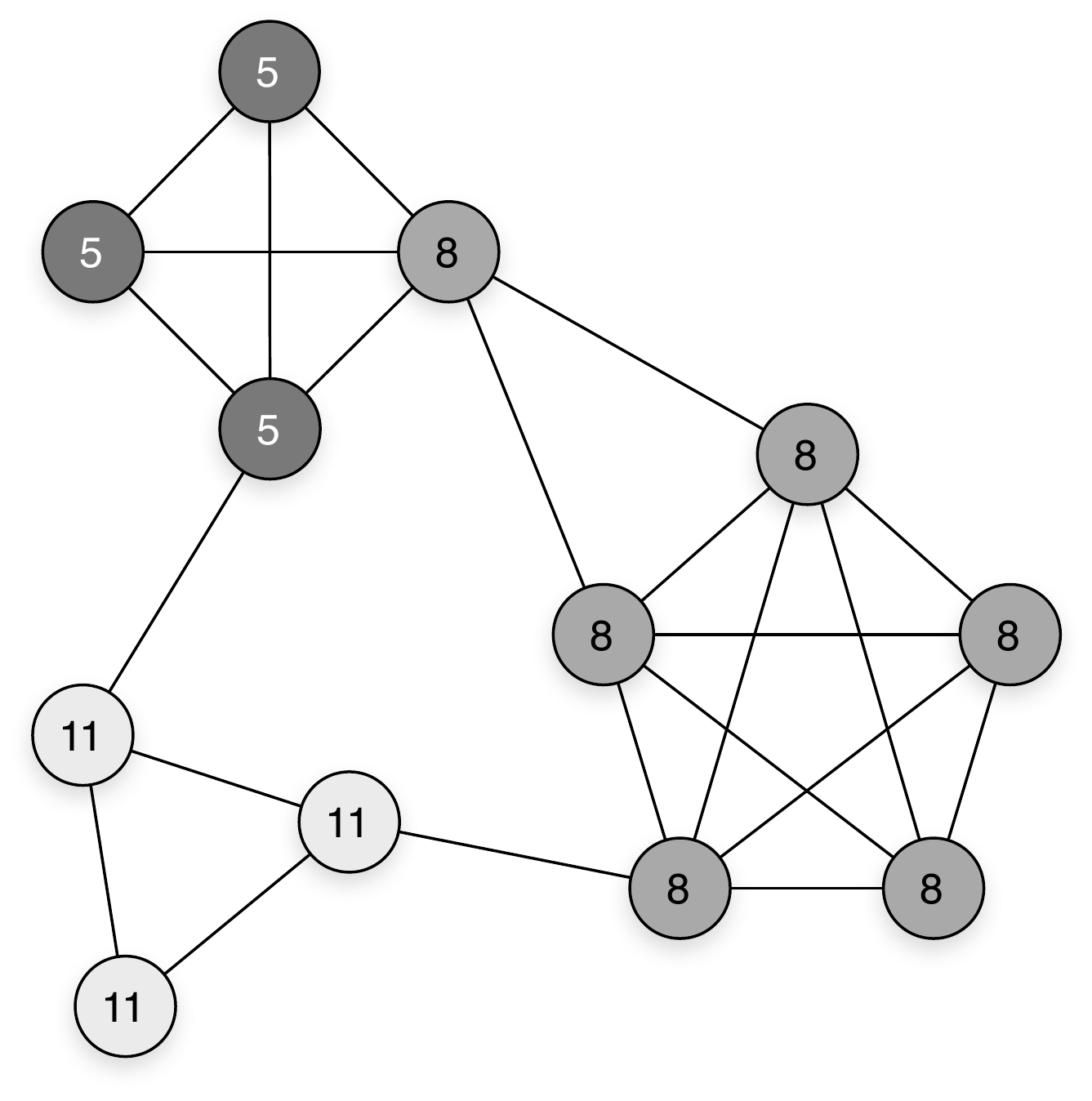}\hskip.033\textwidth
	\includegraphics[width=.225\textwidth]{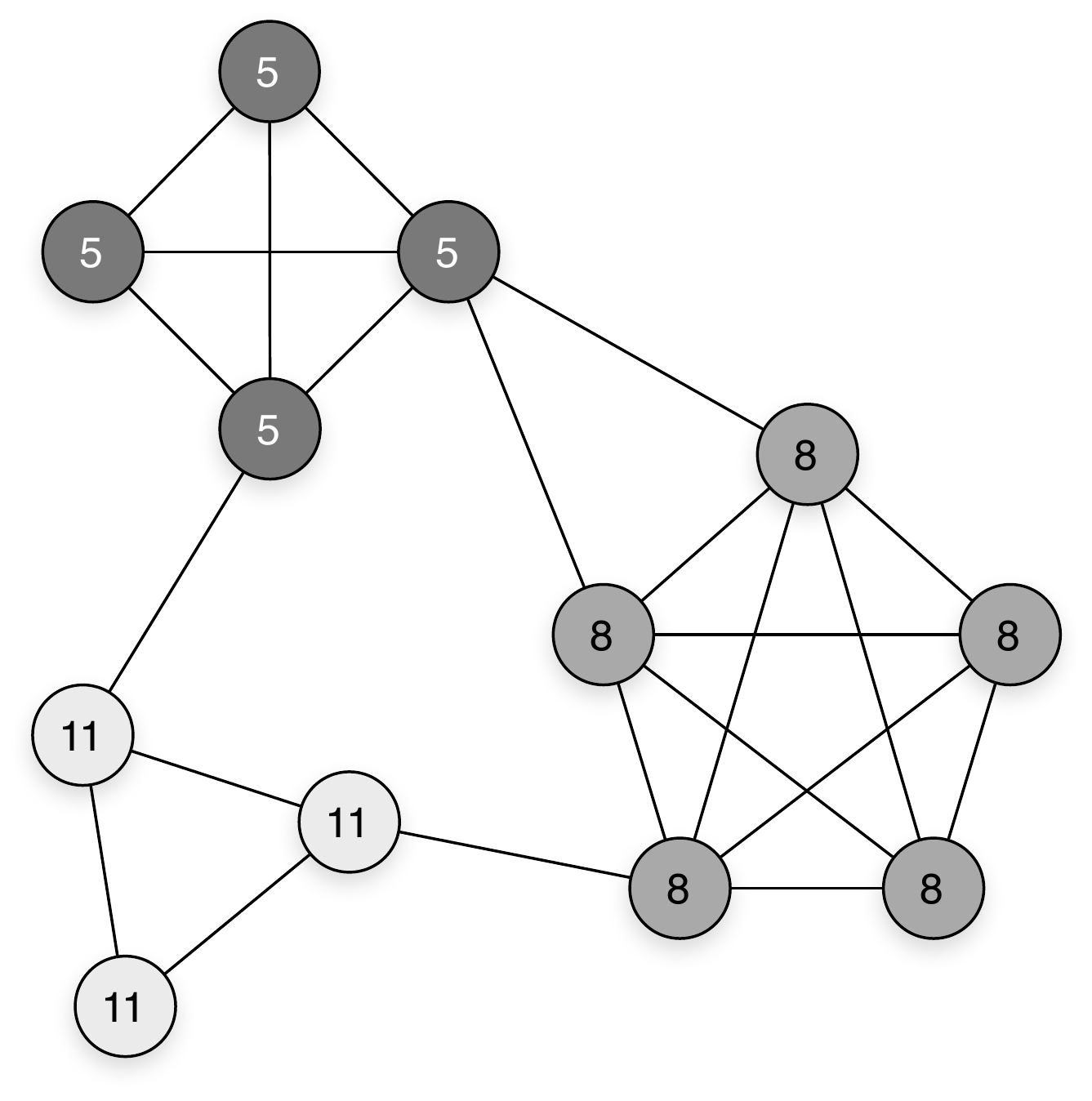}}
	\caption[The label propagation method applied to a small network.]{Label propagation in a small network with three communities. The labels and shades of the nodes represent community assignments at different iterations of the label propagation method.}
	\label{fig:prop}
\end{figure}

For the sake of simplicity, we describe the basic label propagation framework for the case of detecting communities in simple undirected networks. Consider a network with $n$ nodes and let $\neigh{i}$ denote the set of neighbors of node $i\in\set{1,\ldots,n}$. Furthermore, let $g_i$ be the group assignment or community label of node $i$ which we would like to infer. The label propagation method then proceeds as follows. Initially, the nodes are put into separate groups by assigning a unique label to each node as $g_i=i$. Then, the labels are propagated between the nodes until an equilibrium is reached. At every iteration of label propagation, each node $i$ adopts the label shared by most of its neighbors $\neigh{i}$. Hence,
\begin{equation}
	g_i = \argmax_g\size{\set{\given{j\in\neigh{i}}{g_j=g}}}.
	\label{eq:prop}
\end{equation}
Due to having numerous edges within communities, relative to the number of edges towards the rest of the network, nodes of a community form a consensus on some label after only a couple of iterations of label propagation. More precisely, in the first few iterations, the labels form small groups in dense regions of the network, which then expand until they reach the borders of communities. Thus, when the propagation converges meaning that~\eqref{prop} holds for all of the nodes and the labels no longer change, connected groups of nodes sharing the same label are classified as communities. \figref{prop} demonstrates the label propagation method on a small network, where it correctly identifies the three communities in just three iterations. In fact, due to the extremely fast structural inference of label propagation, the estimated number of iterations in a network with a billion edges is about one hundred~\cite{SB11d}.

Label propagation is not limited to simple networks having, at most, one edge between each pair of nodes. Let $A$ be the adjacency matrix of a network, where $A_{ij}$ is the number of edges between nodes $i$ and $j$, and $A_{ii}$ is the number of self-edges or loops on node $i$. The label propagation rule in~\eqref{prop} can be written as
\begin{equation}
	g_i = \argmax_g\sum_{j}A_{ij}\delta(g_j,g),
	\label{eq:lpa}
\end{equation}
where $\delta$ is the Kronecker delta operator that equals one when its arguments are the same and zero otherwise. Furthermore, in weighted or valued networks, the label propagation rule becomes
\begin{equation}
	g_i = \argmax_g\sum_{j}W_{ij}\delta(g_j,g), 
	\label{eq:lpw}
\end{equation}
where $W_{ij}$ is the sum of weights on the edges between nodes $i$ and $j$, and $W_{ii}$ is the sum of weights on the loops on node $i$. Label propagation can also be adopted for multipartite and other types of networks, which is presented in~\secref{nets}. However, there seems to be no obvious extension of label propagation to networks with directed arcs, since propagating the labels exclusively in the direction of arcs enables the exchange of labels only between mutually reachable nodes.


\subsection{\label{sec:ties}Resolution of Label Ties}

\begin{figure}[t]\centerline{%
	\includegraphics[width=.2\textwidth]{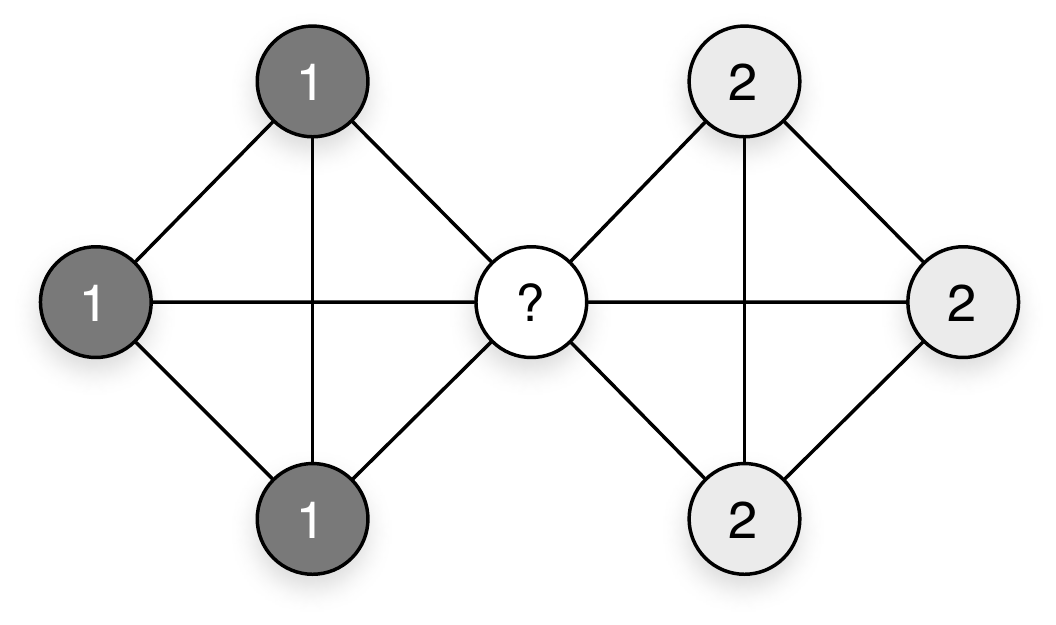}\hskip.05\textwidth
	\includegraphics[width=.2\textwidth]{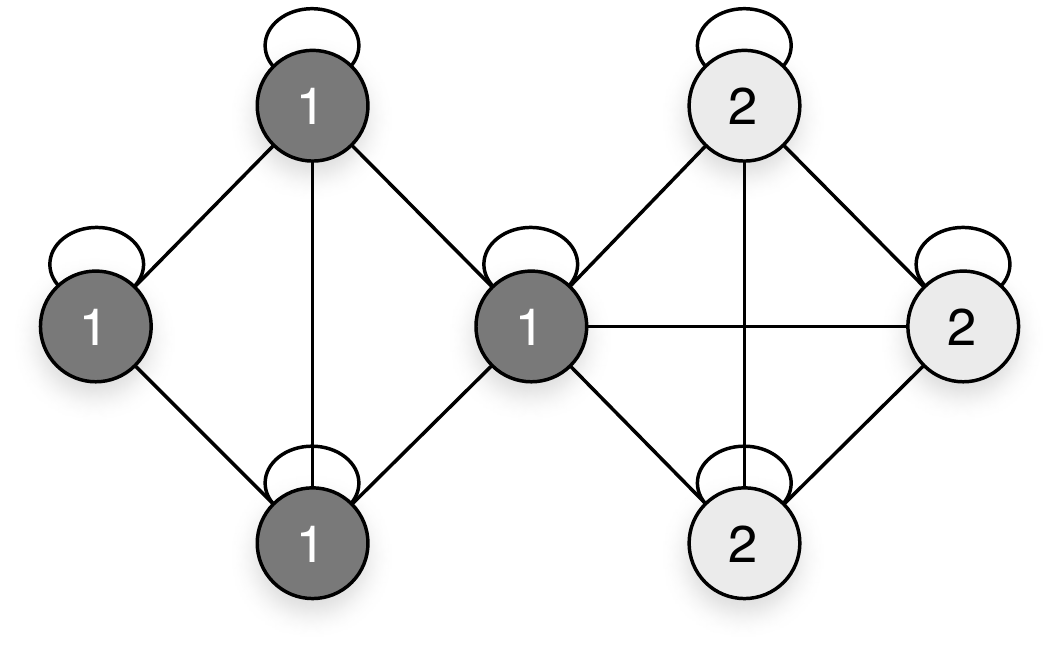}\hskip.05\textwidth
	\includegraphics[width=.2\textwidth]{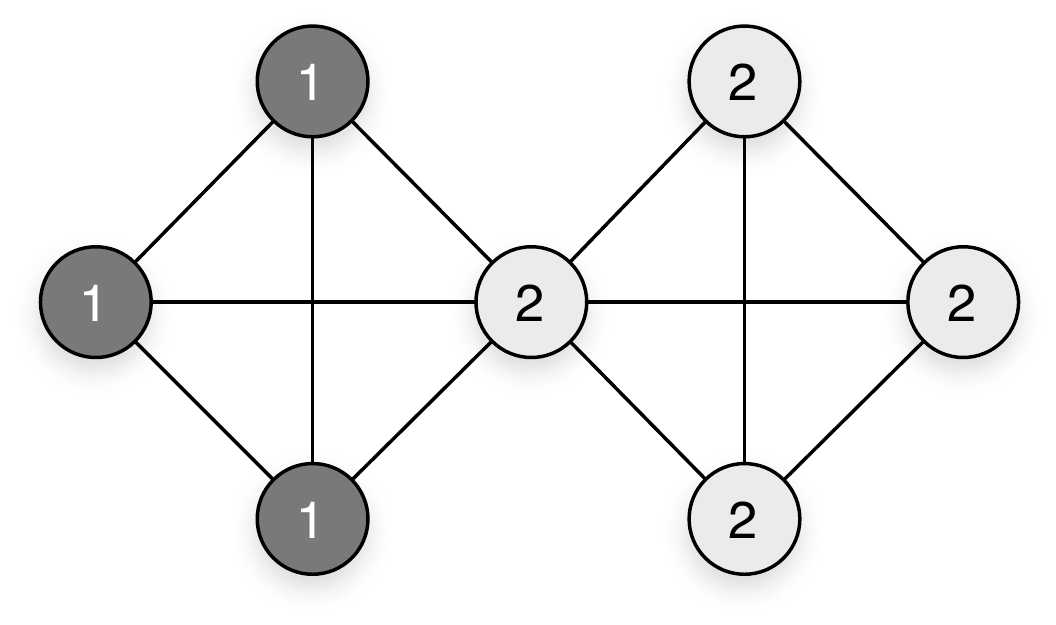}}
	\caption[Resolution of ties between the maximal labels of nodes.]{Resolution of ties between the maximal labels of the central nodes of the networks. The labels and shades of the nodes represent their current community assignments.}
	\label{fig:ties}
\end{figure}

At each step of label propagation, a node adopts the label shared by most of its neighbors denoted by the maximal label. There can be multiple maximal labels as shown in the left side of~\figref{ties}. In that case, the node chooses one maximal label uniformly at random~\cite{RAK07}. Note, however, that the propagation might never converge, especially when there are many nodes with multiple maximal labels in their neighborhoods. This is because their labels could constantly change and label convergence would never be reached. The problem is particularly apparent in networks of collaborations between the authors of scientific papers, where a single author often collaborates with others in different research communities.

The simplest solution is always to select the smallest or the largest maximal label according to some predefined ordering~\cite{CG11}, which has obvious drawbacks. \lhlc proposed a seemingly elegant solution to include also the concerned node's label itself into the maximal label consideration in~\eqref{lpa}. This is equivalent to adding a loop on each node in a network. Nevertheless, the label inclusion strategy might actually create ties when there is only one maximal label in a node's neighborhood, which happens in the case of the central node of the network in the middle of~\figref{ties}.

Most label propagation algorithms implement the label retention strategy introduced by \bc. When there are multiple maximal labels in a node's neighborhood, and one of these labels is the current label of the node, the node retains its label. Otherwise, a random maximal label is selected to be the new node label. The main difference to the label inclusion strategy is that the current label of a node is considered only when there actually exist multiple maximal labels in its neighborhood. For example, the network in the right side of~\figref{ties} is at equilibrium under the label retention strategy.

Random resolution of label ties represents the first of two sources of randomness in the label propagation method hindering its robustness and consequently also the stability of the identified communities. The second is the random order of label propagation. 


\subsection{\label{sec:order}Order of Label Propagation}

\begin{figure}[t]\centerline{%
	\includegraphics[width=.175\textwidth]{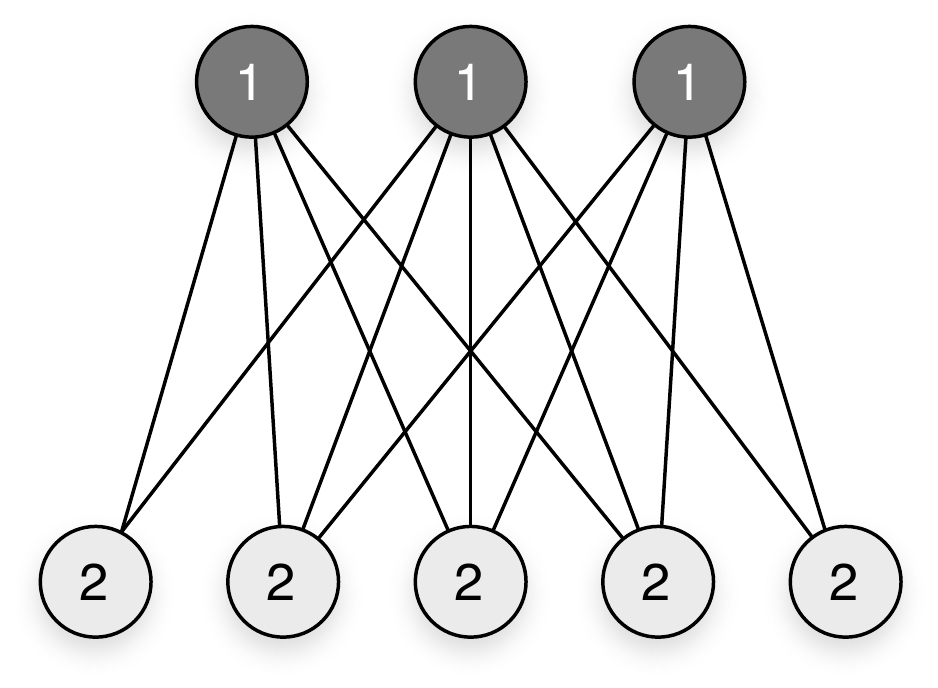}\hskip.025\textwidth
	\includegraphics[width=.175\textwidth]{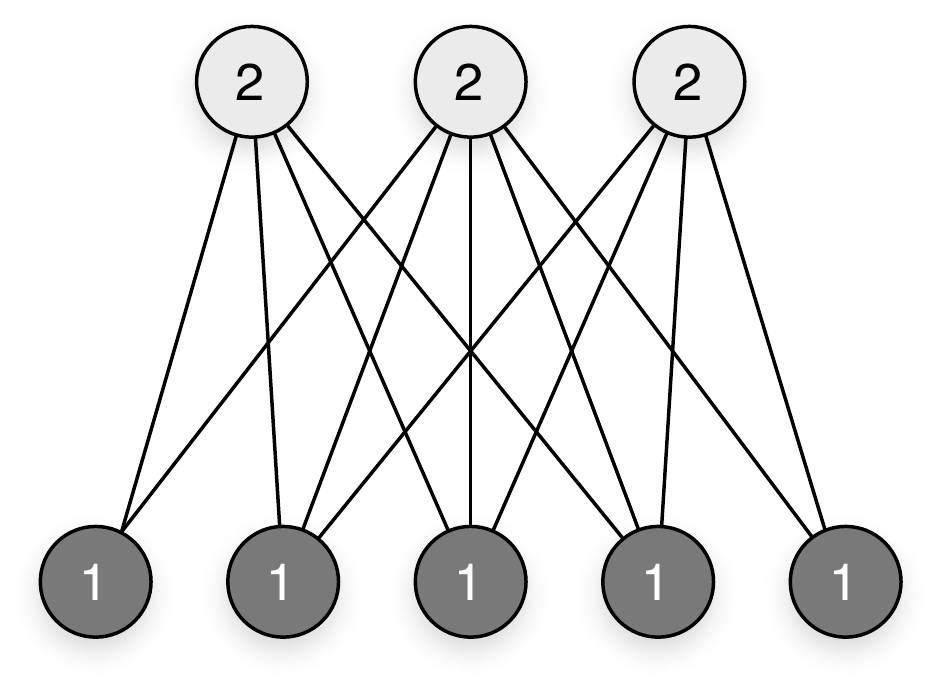}\hskip.075\textwidth
	\includegraphics[width=.175\textwidth]{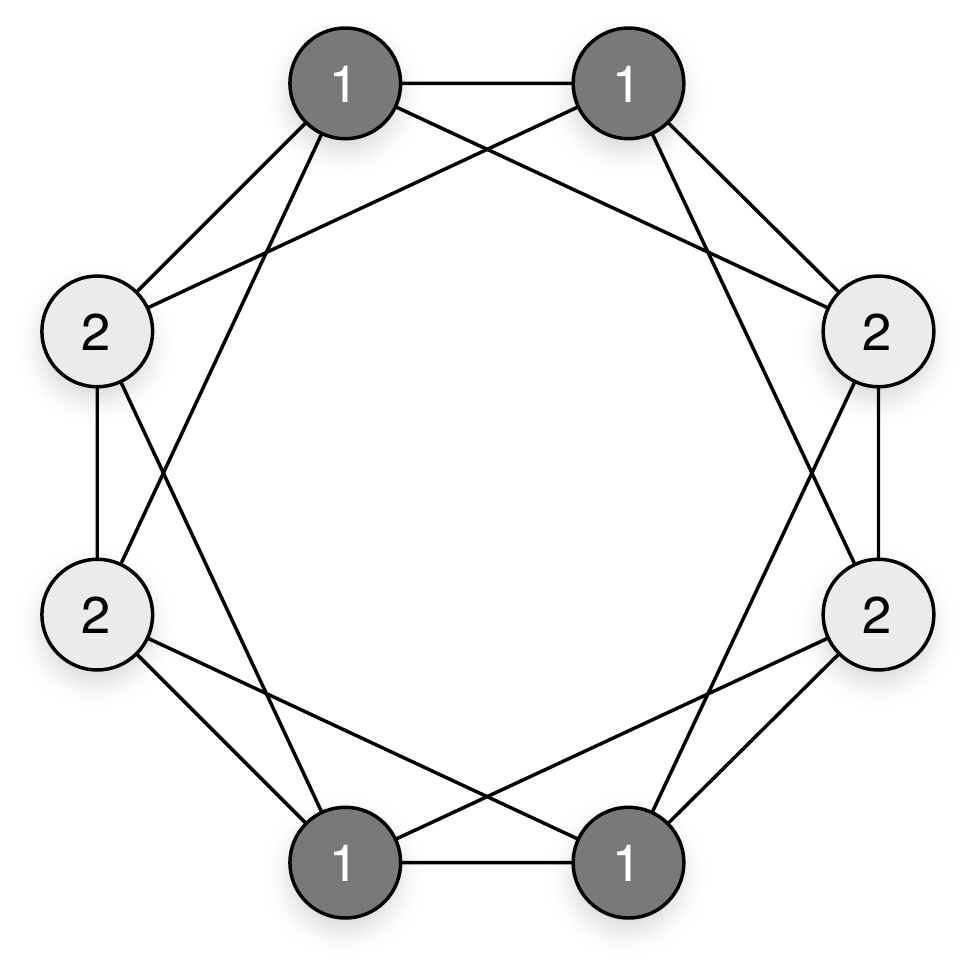}\hskip.025\textwidth
	\includegraphics[width=.175\textwidth]{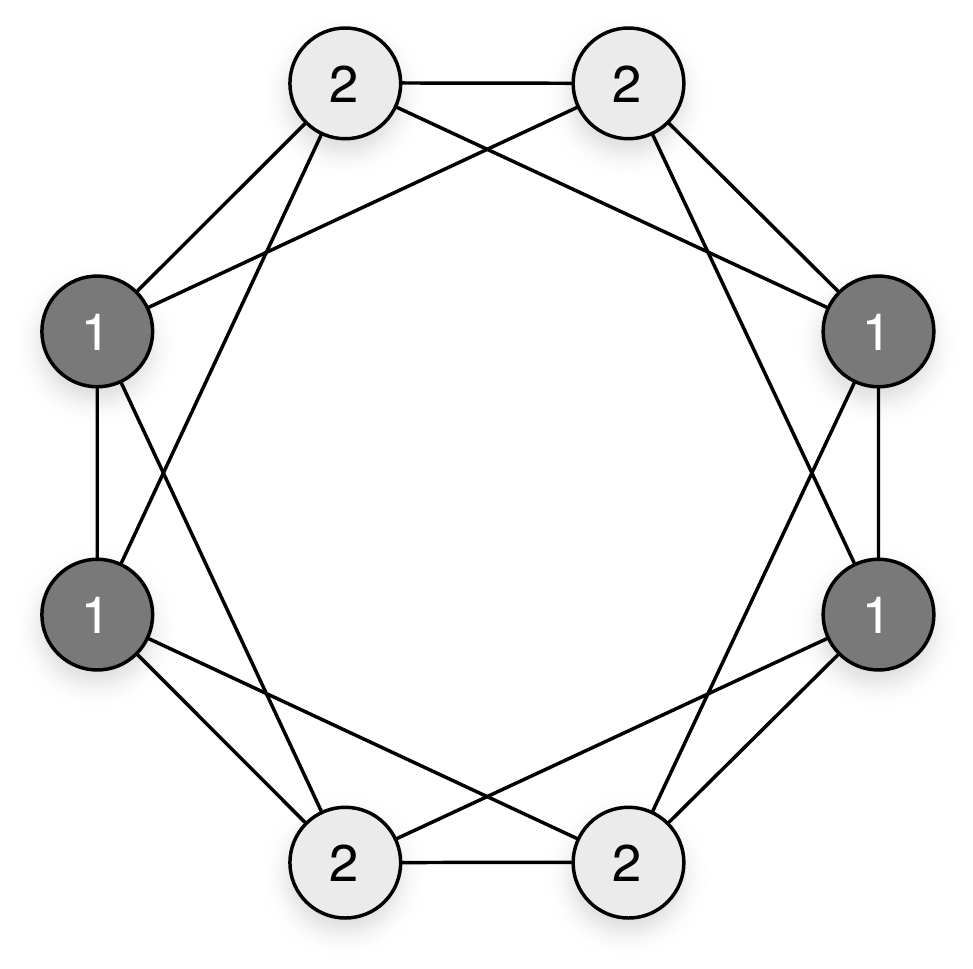}}
	\caption[Label oscillations in bipartite and non-bipartite networks.]{Label oscillations in bipartite and non-bipartite networks. The labels and shades of the nodes represent community assignments at two consecutive iterations of the label propagation~method.}
	\label{fig:oscil}
\end{figure}

The discussion above assumed that, at every iteration of label propagation, all nodes update their labels simultaneously. This is called synchronous propagation~\cite{RAK07}. The authors of the original method noticed that synchronous propagation can lead to oscillations of some labels in certain networks. Consider a bipartite or two-mode network with two types of nodes and edges only between the nodes of different type. Assume that, at some iteration of label propagation, the nodes of each type share the same label as in the example in the left side of~\figref{oscil}. Then, at the next iteration, the labels of the nodes would merely switch and start to oscillate between two equivalent label configurations. For instance, such behavior occurs in networks with star-like communities 
consisting of one or few central hub nodes that are connected to many peripheral nodes, while the peripheral nodes themselves are not directly connected. Note that label oscillations are not limited to bipartite or nearly bipartite networks~\cite{CG11} as seen in the example in the right side of~\figref{oscil}.

For this reason, most label propagation algorithms implement asynchronous propagation~\cite{RAK07}. At every iteration of label propagation, the labels of the nodes are no longer updated all together, but sequentially in some random order, which is different for each iteration. 
This is in contrast to synchronous propagation, which always considers the labels from the previous iteration. Due to random order of label updates, asynchronous propagation successfully breaks the cyclic oscillations of labels in~\figref{oscil}.

It must be stressed that asynchronous propagation with random tie resolution makes the label propagation method very unstable. In the case of the famous Zachary karate club network~\cite{Zac77}, the method identifies more than $500$ different community structures~\cite{TK08}, although the network consists of only $34$ nodes. Asynchronous propagation applied to large online social networks and web graphs can wrongly also produce a giant community occupying the majority of the nodes in a network~\cite{LHLC09}.


\subsection{\label{sec:conv}Label Equilibrium Criterium}

\rak defines the convergence of label propagation as the state of label equilibrium when \eqref{prop} is satisfied for every node in a network. Let $k_i$ denote the number of neighbors of node $i$ 
and let $k_i^g$ be the number of neighbors that share label $g$. The label propagation rule in~\eqref{prop} can be rewritten as
\begin{equation}
	g_i = \argmax_g k_i^g.
	\label{eq:lpk}
\end{equation}
The label equilibrium criterium thus requires that, for every node $i$, the following must hold
\begin{equation}
	\holds{\forall g}{k_i^{g_i} \geq k_i^{g}}.
	\label{eq:conv}
\end{equation}
In other words, all nodes must be labeled with the maximal labels in their neighborhoods. 

This criterion is similar, but not equivalent, to the definition of a strong community~\cite{RCCLP04}. Strong communities require that every node has strictly more neighbors in its own community than in all other communities together, whereas at the label equilibrium every node has at least as many neighbors in its own community than in any other community.

An alternative approach is to define the convergence of label propagation as the state when the labels no longer change~\cite{BC09a}. \eqref{conv} obviously holds for every node in a network and the label equilibrium is reached. Note, however, that this criterion must necessarily be combined with an appropriate label tie resolution strategy in order to ensure convergence when there are multiple maximal labels in the neighborhoods of nodes.


\subsection{\label{sec:comp}Algorithm and Complexity}

As mentioned in the introduction, the label propagation method can be implemented with a few lines of programming code. \algref{lpa} shows the pseudocode of the basic asynchronous propagation framework defining the convergence of label propagation as the state of no label change and implements the retention strategy for label tie resolution.

\begin{figure}[h]
	\begin{algorithm}\label{alg:lpa}
		{\bf label propagation} $\{$
		\tab for each node $i\in\set{1,\ldots,n}$ $\{$
		\tab\tab initialize node label $g_i$ with $i$;
		\tab $\}$   
		\tab until node labels change repeat $\{$   
		\tab\tab for each node $i\in\set{1,\ldots,n}$ in random order $\{$   
		\tab\tab\tab compute labels $\set{g}$ that maximize $k_i^g=\sum_{j}A_{ij}\delta(g_j,g)$;
		\tab\tab\tab if $g_i\notin \set{g}$ update $g_i$ with random label from $\set{g}$; 
		\tab\tab $\}$   
		\tab $\}$   
		\tab report connected components induced by node labels;
		$\}$
	\end{algorithm}
\end{figure}	

When the state of label equilibrium is reached, groups of nodes sharing the same label are classified as communities. These can, in general, be disconnected, which happens when a node propagates its label to two or more disconnected nodes, but is itself relabeled in the later iterations of label propagation. Since connectedness is a fundamental property of network communities~\cite{FH16}, groups of nodes with the same label are split into connected groups of nodes at the end of label propagation. Reported communities are thus connected components of the subnetworks induced by different node labels.

The label propagation method exhibits near-linear time complexity in the number of edges of a network denoted with $m$~\cite{RAK07,LHLC09}. At every iteration of label propagation, the label of node $i$ can be updated with a sweep through its neighborhood which has complexity $\cmp{k_i}$, where $k_i$ is the degree of node $i$. Since $\sum_i k_i=2m$, the complexity of an entire iteration of label propagation is $\cmp{m}$. A random order or permutation of nodes before each iteration of asynchronous propagation can be computed in $\cmp{n}$ time, while the division into connected groups of nodes at the end of label propagation can be implemented with a simple network traversal, which has complexity~$\cmp{n+m}$.

\begin{figure}[t]\centerline{%
	\includegraphics[height=.25\textwidth]{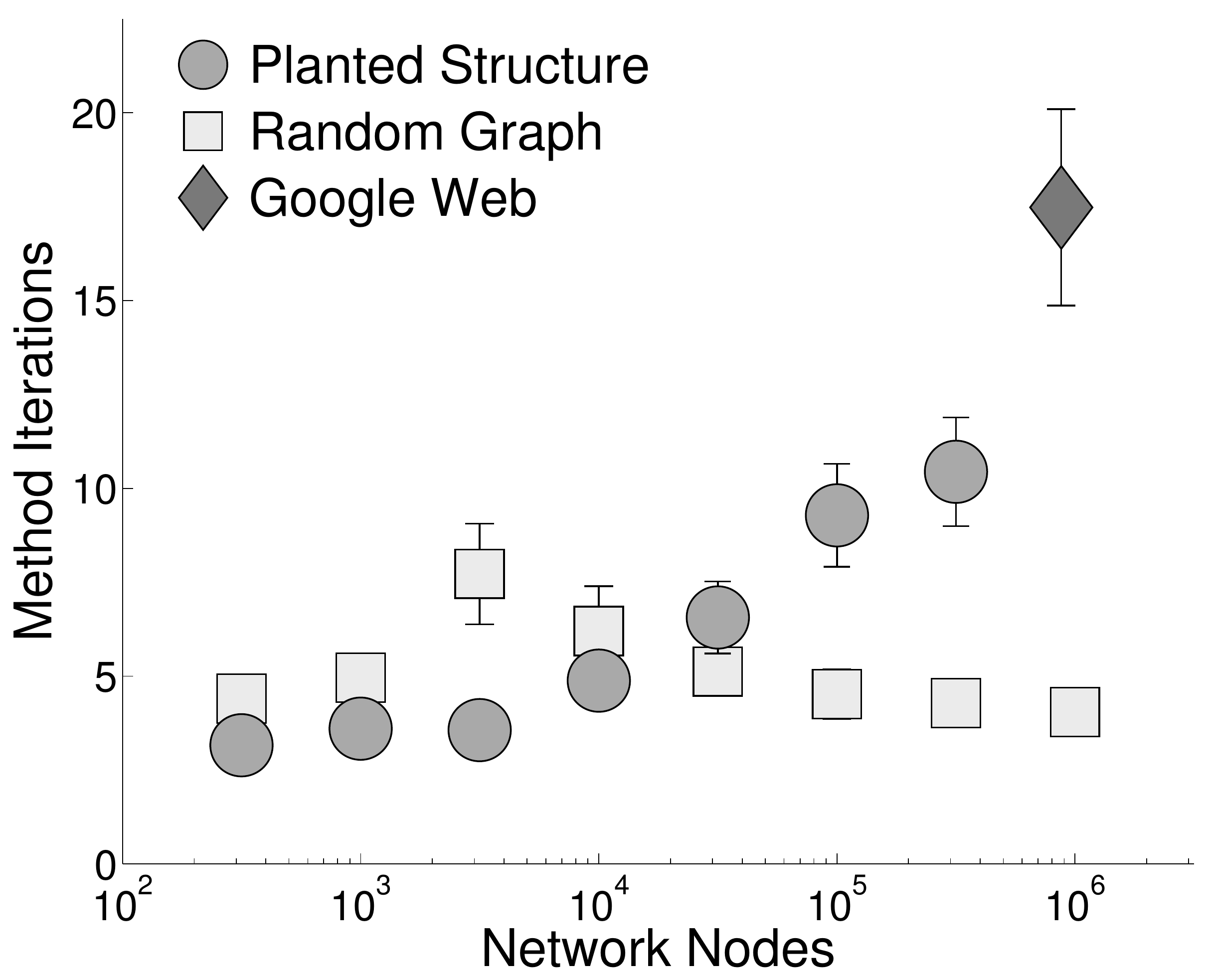}\hskip.033\textwidth
	\includegraphics[height=.25\textwidth]{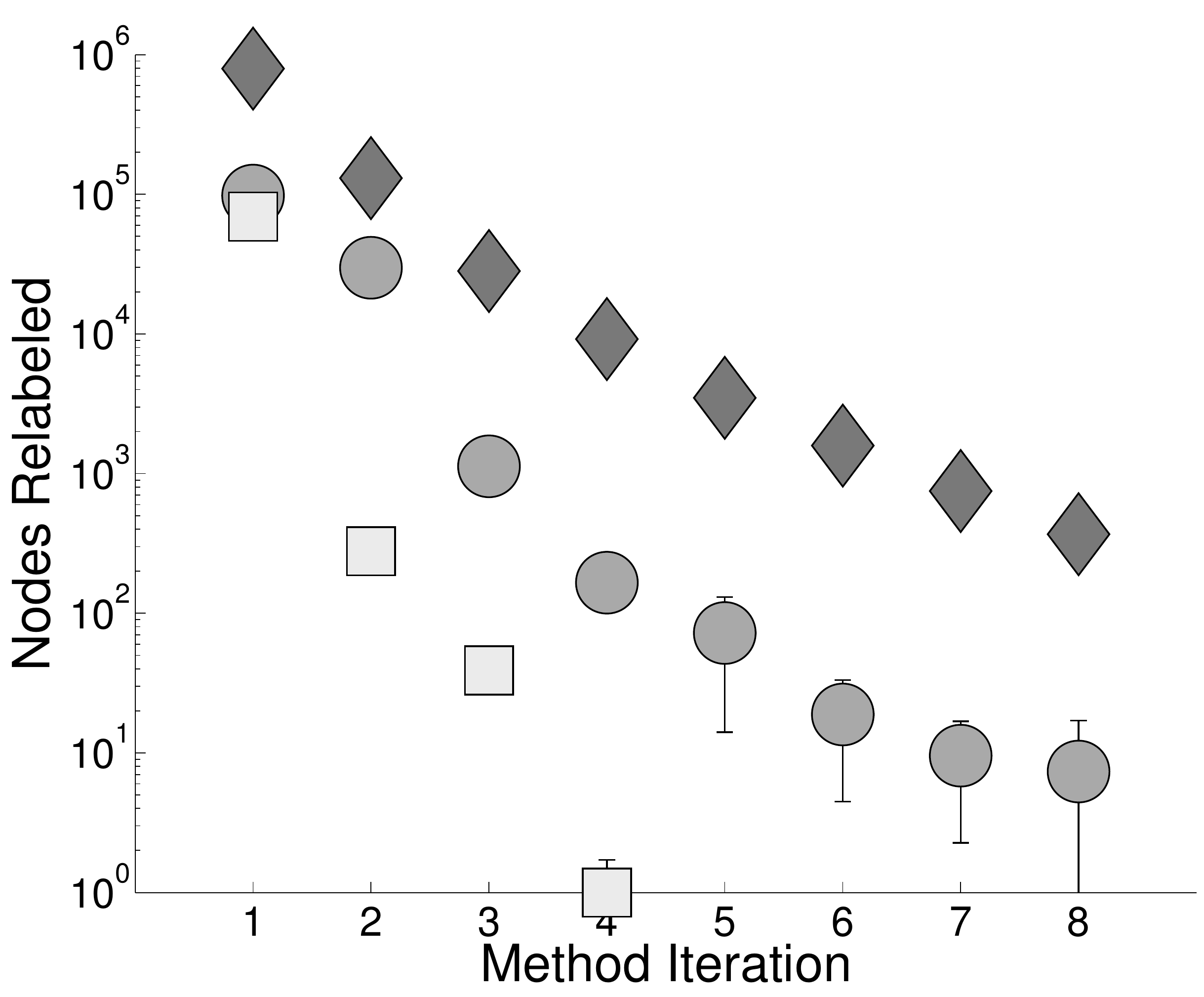}\hskip.033\textwidth
	\includegraphics[height=.25\textwidth]{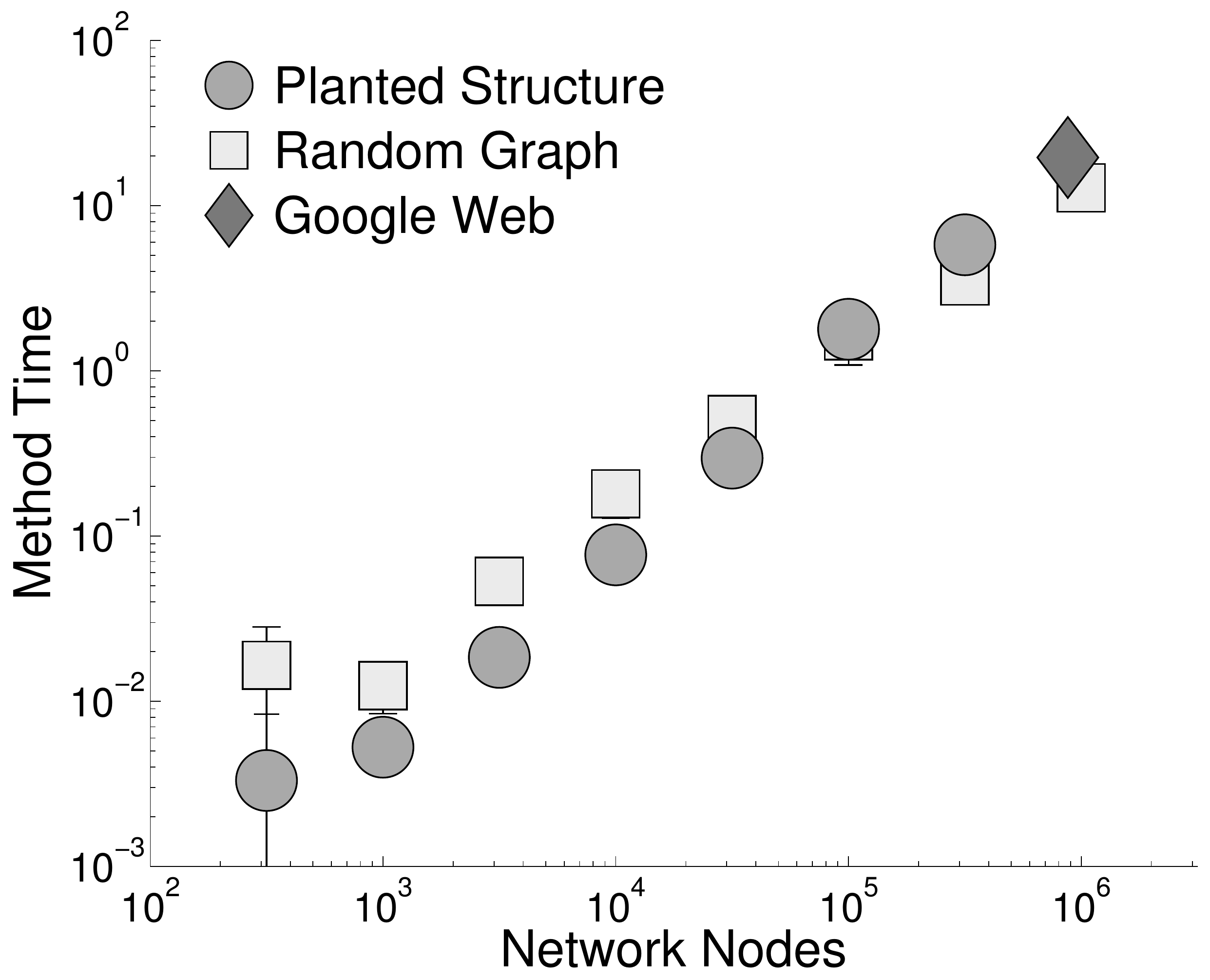}}
	\caption[Time complexity of the label propagation method.]{The number of iterations of label propagation, the number of relabeled nodes at first eight iterations and the running time in seconds. The markers are averages over $25$ runs of the label propagation method, while the error bars show standard deviation.}
	\label{fig:comp}
\end{figure}

The overall time complexity of label propagation is therefore $\cmp{cn+cm}$, where $c$ is the number of iterations before convergence. In the case of networks with a clear community structure, label propagation commonly converges in no more than ten iterations. Still, the number of iterations increases with the size of a network as can be seen in~\figref{comp}. \sbd estimated the number of iterations of asynchronous label propagation from a large number of empirical networks obtaining $c\approx 1.03m^{0.23}$. The time complexity of label propagation is thus approximately $\cmp{m^{1.2}}$, which makes the method applicable to networks with up to hundreds of millions of nodes and edges on a standard desktop computer as long as the network fits into its main memory.

The left side of~\figref{comp} shows the number of iterations of the label propagation framework in~\algref{lpa} in artificial networks with planted community structure~\cite{LFR08}, Erd\H{o}s-R\'{e}nyi random graphs~\cite{ER59} and a part of the Google web graph~\cite{LLDM09} available at KONECT\footnote{\url{http://konect.uni-koblenz.de}}. The web graph consists of $875,\!713$ nodes and $5,\!105,\!039$ edges, while the sizes of random graphs and artificial networks can be seen in~\figref{comp}. In random graphs having no structure, label propagation correctly classifies all nodes into a single group in about five iterations, regardless of the size of a graph. Yet, the number of iterations increases with the size in artificial networks with community structure, while the estimated number of iterations in a network with a billion edges is $113$~\cite{SB11d}.

Most nodes acquire their final label after the first few iterations of label propagation. The middle of~\figref{comp} shows the number of nodes that update their label at a particular iteration for the Google web graph, artificial networks having a planted community structure and random graphs with $10^5$ nodes. The number of relabeled nodes drops exponentially with the number of iterations (logarithmic scales are used). For example, the percentages of relabeled nodes of the web graph after the first five iterations are $90.7\%$, $14.9\%$, $3.2\%$, $1.1\%$ and $0.4\%$, respectively. Furthermore, the algorithm running time is only $19.5$ seconds as shown in the right side of~\figref{comp}.

%
%

\section{\label{sec:optm}Label Propagation as Optimization}

Here, we discuss the objective function of the label propagation method to shed light on label propagation as an optimization method.

At every iteration of label propagation, each node adopts the most common label in its neighborhood. Therefore, label propagation can be seen as a local optimization method seeking to maximize the number of neighbors with the same label or, equivalently, minimize the number of neighbors with different labels. From the perspective of node $i$, the label propagation rule in~\eqref{lpa} assigns its group label $g_i$ to maximize $\sum_{j}A_{ij}\delta(g_i,g_j)$, where $A$ is the adjacency matrix of the network. Hence, the objective function maximized by the basic label propagation method is
\begin{equation}
	\objf{\set{g}} = \sum_{ij}A_{ij}\delta(g_i,g_j),
	\label{eq:objf}
\end{equation}
where $\set{g}$ is the group labeling of network nodes~\cite{TK08,BC09a}. Notice that $\obj$ is non-negative and has the optimum of $2m$, where $m$ is the number of edges in a network.

\eqref{objf} has a trivial optimal solution of labeling all nodes in a network with the same label, corresponding to putting all nodes into one group. \eqref{lpa} then holds for every node and $\obj=2m$. However, starting with each node in its own group by assigning them unique labels when $\obj=0$, the label propagation process usually is trapped in a local optimum. For networks having a clear community structure, this corresponds to nodes of each community being labeled with the same label when $\obj=2m-2m'$, where $m'$ is the number of edges between communities. For example, the value of $\obj$ for the community structure revealed in the right side of~\figref{prop} is~$46-8=38$.

Network community structure is only a local optimum of the label propagation process, whereas the global optimal solution corresponds to a trivial, undesirable, labeling. Thus, directly optimizing the objective function of label propagation with some other optimization method trying to escape a local optimum might not yield a favorable outcome. Furthermore, a network can have also many local optima that imply considerably different community structures. As already mentioned in~\secref{order}, label propagation identifies more than $500$ different structures in the Zachary karate club network~\cite{Zac77} with $34$ nodes and more than $10^5$ in the \emph{Saccharomyces cerevisiae} protein interaction network~\cite{JMBO01} with $2,\!111$ nodes~\cite{TK08}. \rak suggested aggregating labelings from multiple runs of label propagation. However, this can fragment a network into very small communities~\cite{TK08}. A more suitable method for combining different labelings of label propagation is consensus clustering~\cite{LF12,ZJFP14,GCSKXLBNHGMK15}, but this comes with increased time complexity.

The above perspective on label propagation as an optimization method results from the following equivalence. \tk have shown that the label propagation in~\eqref{lpa} is equivalent to a ferromagnetic Potts model~\cite{Pot52,WuF82}. The $q$-state Potts model is a generalization of the Ising model as a system of interacting spins on a lattice, with each spin pointing to one of $q$ equally spaced directions. Consider the so-called standard $q$-state Potts model on a network placing a spin on each node~\cite{RB06a}. Let $\sigma_i$ denote the spin on node $i$ which can be in one of $q$ possible states, where $q$ is set equal to the number of nodes in a network $n$. The zero-temperature kinetics of the model are defined as follows. One starts with each spin in its own state as $\sigma_i=i$ and then iteratively aligns the spins to the states of their neighbors as in the label propagation process. The ground state is ferromagnetic with all spins in the same state, while the dynamics can also get trapped at a metastable state with more than one spin state. The Hamiltonian of the model can be written as
\begin{equation}
	\haml{\set{\sigma}} = -\sum_{ij}A_{ij}\delta(\sigma_i,\sigma_j),
	\label{eq:haml}
\end{equation}
where $\set{\sigma}$ are the states of spins on network nodes. By setting $\sigma_i=g_i$, minimizing the described Potts model Hamiltonian $\hml$ in~\eqref{haml} is equivalent to maximizing the objective function of the label propagation method $\obj$ in~\eqref{objf}.

As almost any other clustering method, the label propagation method is nondeterministic and can produce different outcomes on different runs. Therefore, throughout the chapter, we report the results obtained over multiple runs of the method.

%
%

\section{\label{sec:advs}Advances of Label Propagation}

\secref{prop} presented the basic label propagation method and discussed details of its implementation. \secref{optm} clarified the objective function of label propagation. In this section, we review different advances of the original method addressing some of the weaknesses identified in the previous sections. \secref{const} shows how to redefine the method's objective function by imposing constraints to use label propagation as a general optimization framework. \secref{prefs} demonstrates different heuristic approaches changing the method's objective function implicitly by adjusting the propagation strength of individual nodes. This promotes the propagation of labels from certain desirable nodes or, equivalently, suppresses the propagation from the remaining nodes. Finally, \secref{perf} discusses different empirically motivated techniques to improve the overall performance of the method.

Unless explicitly stated otherwise, the above advances are presented for the case of non-overlapping community detection in simple undirected networks. Nevertheless, \secref{nets} presents extensions of label propagation to other types of networks such as multipartite, multilayer and signed networks. Furthermore, in~\secref{grps}, we show how label propagation can be adopted to detect alternative types of groups such as overlapping or hierarchical communities and groups of nodes that are similarly connected to the rest of the network by structurally equivalent nodes as in~\chpref{6}{Partitioning valued network data}. Note that different approaches and techniques described in~\secsref{advs}{grps} can be combined. The advances of the basic label propagation method described in this section can be used directly with the extensions to other types of groups and networks described in the next sections.


\subsection{\label{sec:const}Label Propagation under Constraints}

As shown in~\secref{optm}, the objective function of label propagation has a trivial optimal solution of assigning all nodes to a single group. A standard approach for eliminating such undesirable solutions is to add constraints to the objective function of the method. Let $\hml$ be the objective function of label propagation expressed in the form of the ferromagnetic Potts model Hamiltonian as in~\eqref{haml}. The modified objective function minimized by label propagation under constraints is $\hml+\lambda\cns$, where $\cns$ represents a penalty term with imposed constraints with $\lambda$ being a regularization parameter weighing the penalty term $\cns$ against the original objective function $\hml$.

\bc proposed a penalty term $\cns_1$ borrowed from the graph partitioning literature requiring that nodes are divided into smaller groups of the same size.
\begin{equation}
	\cons{1}{\set{g}} = \sum_{g}n_g^2,
	\label{eq:cons1}
\end{equation}
where $n_g=\sum_i\delta(g_i,g)$ is the number of nodes in group $g$, $g_i$ is the group label of node $i$ and $n=\sum_gn_g$ is the number of nodes in a network. The penalty term $\cns_1$ has the minimum of $n$ when all nodes are in their own groups and the maximum of $n^2$ when all nodes are in a single group, which effectively guards against the undesirable trivial solution. The modified objective function $\hml_1=\hml+\lambda_1\cns_1$ can be written as 
\begin{equation}
	\hami{1}{\set{g}} = -\sum_{ij}(A_{ij}-\lambda_1)\delta(g_i,g_j),
	\label{eq:haml1}
\end{equation}
where $A$ is the adjacency matrix of a network. \eqref{haml1} is known as the constant Potts model~\cite{TVN11} and is equivalent to a specific version of the stochastic block model~\cite{ZCH15}, while the regularization parameter $\lambda_1$ can be interpreted as the threshold between the density of edges within and between different groups. The label propagation rule in~\eqsref{lpa}{lpk} for the modified objective function $\hml_1$ is
\begin{equation}\begin{split}
	g_i & = \argmax_g\sum_{j}(A_{ij}-\lambda_1)\delta(g_j,g) \\
	& = \argmax_g k_i^g-\lambda_1 n_g,
	\label{eq:lp1}
\end{split}\end{equation}
where $k_i^g=\sum_jA_{ij}\delta(g_j,g)$ is the number of neighbors of node $i$ in group $g$. \eqref{lp1} can be efficiently implemented with~\algref{lpa} by updating $n_g$.

An alternative penalty term $\cns_2$, which has been popular in the community detection literature, requires nodes being divided into groups having the same total degree~\cite{BC09a}. 
\begin{equation}
	\cons{2}{\set{g}} = \sum_{g}k_g^2,
	\label{eq:cons2}
\end{equation}
where $k_g=\sum_ik_i\delta(g_i,g)$ is the sum of degrees of nodes in group $g$ and $k_i$ is the degree of node $i$. The penalty term $\cns_2$ is again minimized when all nodes are in their own groups and maximized when all nodes are in a single group, avoiding the trivial solution. The modified objective function $\hml_2=\hml+\lambda_2\cns_2$ can be written as 
\begin{equation}
	\hami{2}{\set{g}} = -\sum_{ij}(A_{ij}-\lambda_2 k_ik_j)\delta(g_i,g_j),
	\label{eq:haml2}
\end{equation}
while the corresponding label propagation rule is
\begin{equation}\begin{split}
	g_i & = \argmax_g\sum_{j}(A_{ij}-\lambda_2 k_ik_j)\delta(g_j,g) \\
	& = \argmax_g k_i^g-\lambda_2 k_ik_g+\lambda_2 k_i^2\delta(g_i,g).
	\label{eq:lp2}
\end{split}\end{equation}
\eqref{lp2} can be efficiently implemented with~\algref{lpa} by updating $k_g$.

\eqref{haml2} is a special case of the Potts model investigated by~\rb and is a generalization of a popular quality function in community detection named modularity~\cite{NG04}. The modularity $\mdq$ measures the number of edges within network communities against the expected number of edges in a random graph with the same degree sequence~\cite{NSW01}. Formally,
\begin{equation}
	\modq{\set{g}} = \frac{1}{2m}\sum_{ij}\left(A_{ij}-\frac{k_ik_j}{2m}\right)\delta(g_i,g_j).
	\label{eq:modq}
\end{equation}
Notice that setting $\lambda_2=1/2m$ in~\eqref{haml2} yields $\hml_2=-2m\mdq$~\cite{BC09a}.

Label propagation under the constraints of~\eqref{lp2} can be employed for maximizing the modularity $\mdq$. Note, however, that the method might easily get trapped at a local optimum, not corresponding to very high $\mdq$. For example, the average $\mdq$ over $25$ runs for the Google web graph from~\figref{comp} is $0.763$. In contrast, the unconstrained label propagation gives a value of $0.801$. For this reason, label propagation under constraints is usually combined with a multistep greedy agglomerative algorithm~\cite{SC08}, one driving the method away from a local optimum. Using such an optimization framework, \lm revealed community structures with the highest values of $\mdq$ than ever reported for some commonly analyzed empirical networks. \hlszd recently adapted the same framework also for another popular quality function called map equation~\cite{RB08}.

The third variant of label propagation under constraints~\cite{BRSV11} is based on the absolute Potts model~\cite{RN10} with the modified objective function $\hml_3=\hml+\lambda_3\cns_3$ written as
\begin{equation}
	\hami{3}{\set{g}} = -\sum_{ij}(A_{ij}(\lambda_3+1)-\lambda_3)\delta(g_i,g_j).
	\label{eq:haml3}
\end{equation}
By setting $\lambda_1=\lambda_3/(\lambda_3+1)$ in~\eqref{haml1}, one derives $\hml_1=\hml_3/(\lambda_3+1)$ implying the method is in fact equivalent to the constant Potts model~\cite{TVN11}.


\subsection{\label{sec:prefs}Label Propagation with Preferences}

\lhlc have shown that adjusting the propagation strength of individual nodes can improve the performance of the label propagation method in certain networks. Let $p_i$ be the propagation strength associated with node $i$ called the node preference. Incorporating the node preferences $p_i$ into the basic label propagation rule in~\eqref{lpa} gives
\begin{equation}
	g_i = \argmax_g\sum_{j}p_jA_{ij}\delta(g_j,g),
	\label{eq:lpp}
\end{equation}
while the method objective function in~\eqref{haml} becomes
\begin{equation}
	\hami{p}{\set{g}} = -\sum_{ij}p_ip_jA_{ij}\delta(g_i,g_j).
	\label{eq:hamlp}
\end{equation}
In contrast to~\secref{const}, these node preferences impose constraints on the objective function only implicitly by either promoting or suppressing the propagation of labels from certain desirable nodes, as shown in the examples below. 

An obvious choice is to set the node preferences equal to the degrees of the nodes as $p_i=k_i$~\cite{LHLC09}. For instance, this improves the performance of community detection in networks with high degree nodes in the center of each community. \sbdd proposed estimating the most central nodes of each community or group during the label propagation process using a random walk diffusion. Consider a random walker utilized on a network limited to the nodes of group $g_i$ and let $p_i$ be the probability that the walker visits node $i$. The probabilities $p_i$ are high for the most central nodes of group $g_i$ and low for the nodes on the border. It holds
\begin{equation}
	p_i = \sum_{j}\frac{p_j}{k_j^{g_j}}A_{ij}\delta(g_i,g_j),
	\label{eq:prefs}
\end{equation}
where $k_i^{g_i}=\sum_jA_{ij}\delta(g_i,g_j)$ is the number of neighbors of node $i$ in its group $g_i$. Clearly $p_i=k_i^{g_i}$ is the solution of~\eqref{prefs}, but initializing the probabilities as $p_i=1$ and updating their values according to~\eqref{prefs} only when the nodes change their groups $g_i$ gives a different result. This mimics the actual propagation of labels occurring in a random order and keeps the node probabilities $p_i$ synchronized with the node groups $g_i$. \eqref{prefs} can be efficiently implemented in~\algref{lpa} by updating $k_i^{g_i}$.

Label propagation with node preferences defined in~\eqref{prefs} is called defensive propagation~\cite{SB11d} as it restrains the propagation of labels to preserve a larger number of groups by increasing the propagation strength of their central nodes or, equivalently, decreasing the propagation strength of their border nodes. Another strategy is to increase the propagation strength of the border nodes, which results in a more rapid expansion of groups and a smaller number of larger groups. This is called offensive~pro\-pa\-ga\-tion~\cite{SB11d} with the label propagation rule written as
\begin{equation}
	g_i = \argmax_g\sum_{j}(1-p_j)A_{ij}\delta(g_j,g).
	\label{eq:lpo}
\end{equation}

\begin{figure}[t]\centerline{%
	\includegraphics[width=.2375\textwidth]{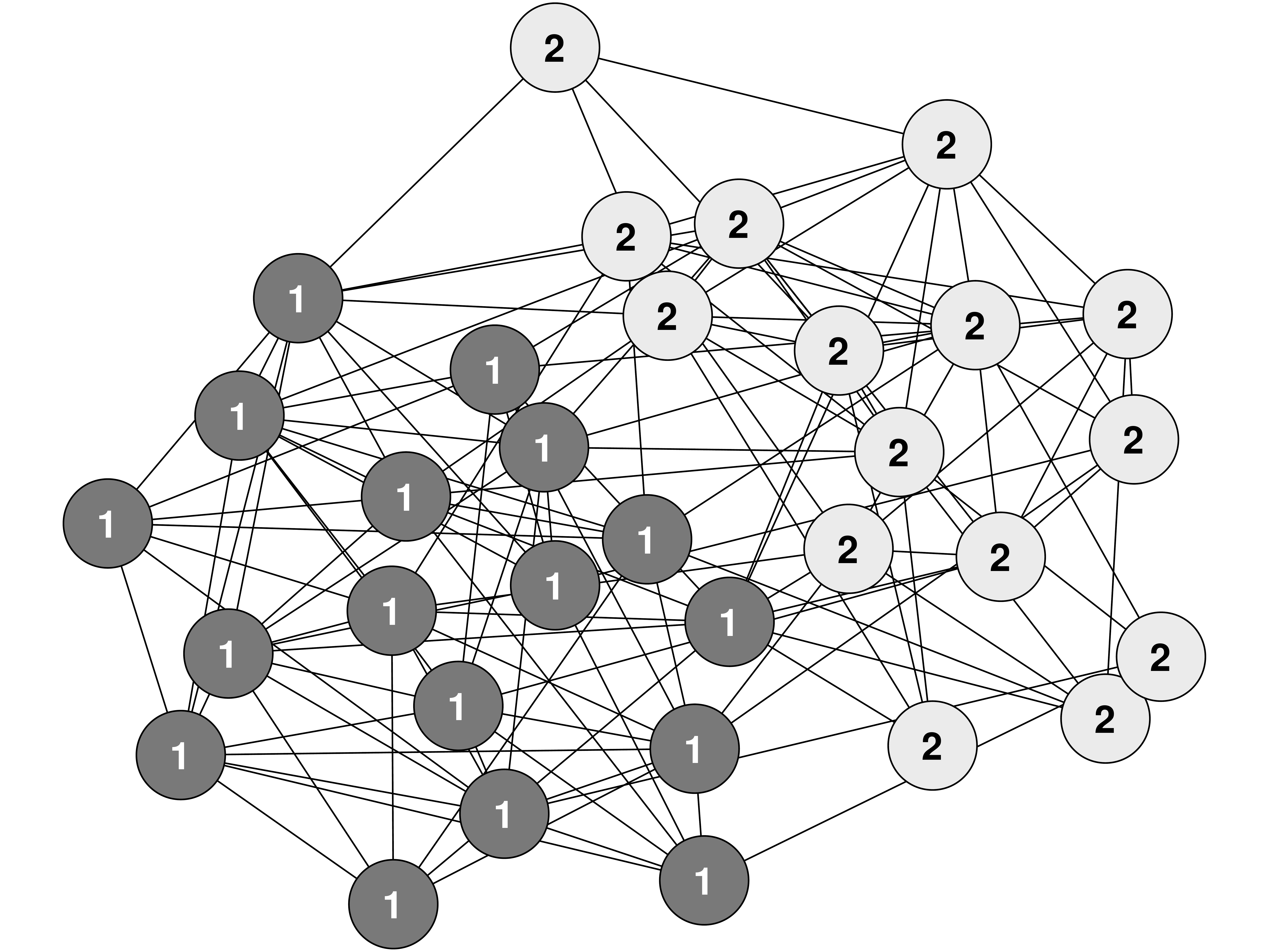}%
	\includegraphics[width=.2375\textwidth]{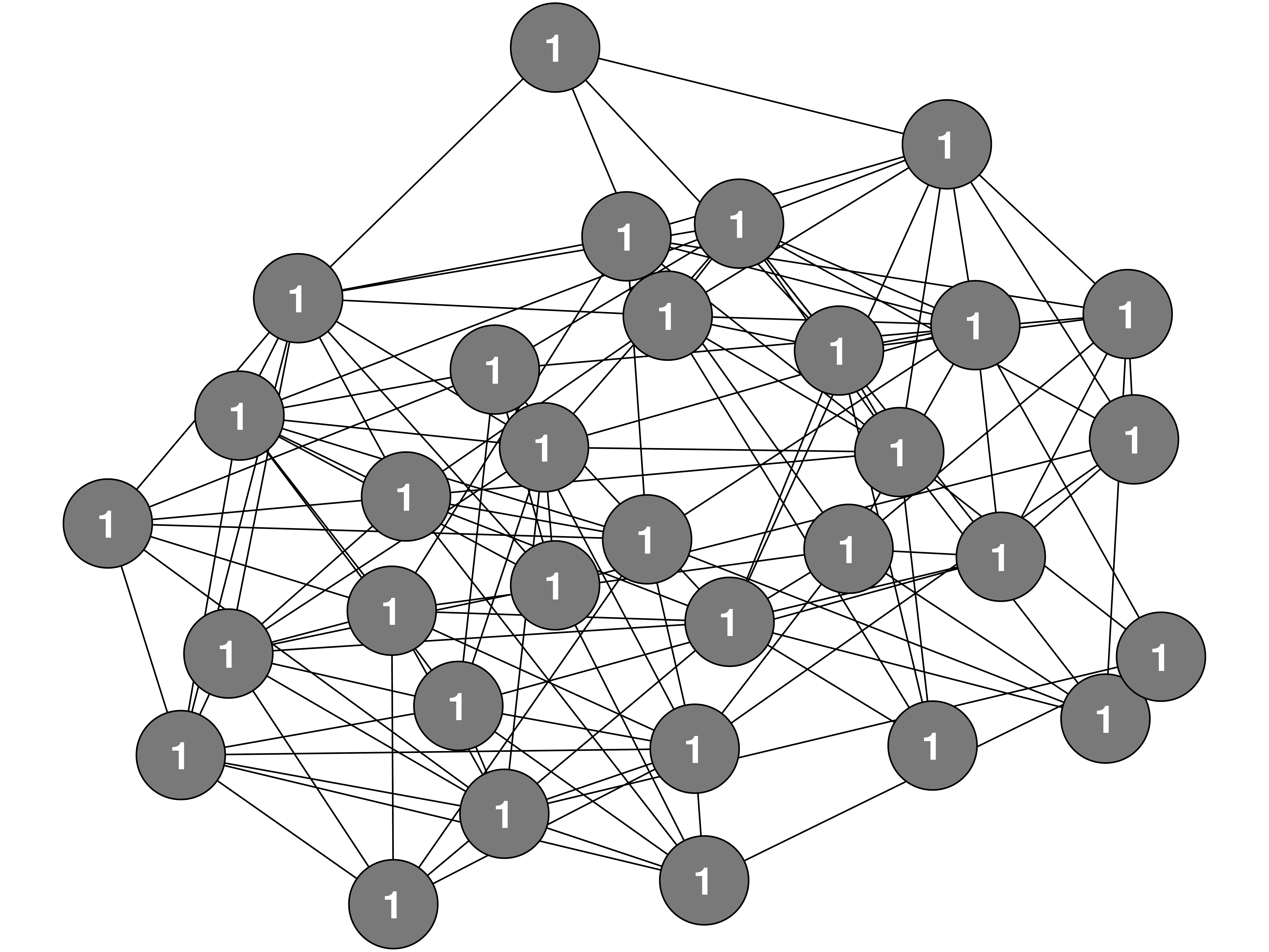}\hskip.05\textwidth
	\includegraphics[width=.2375\textwidth]{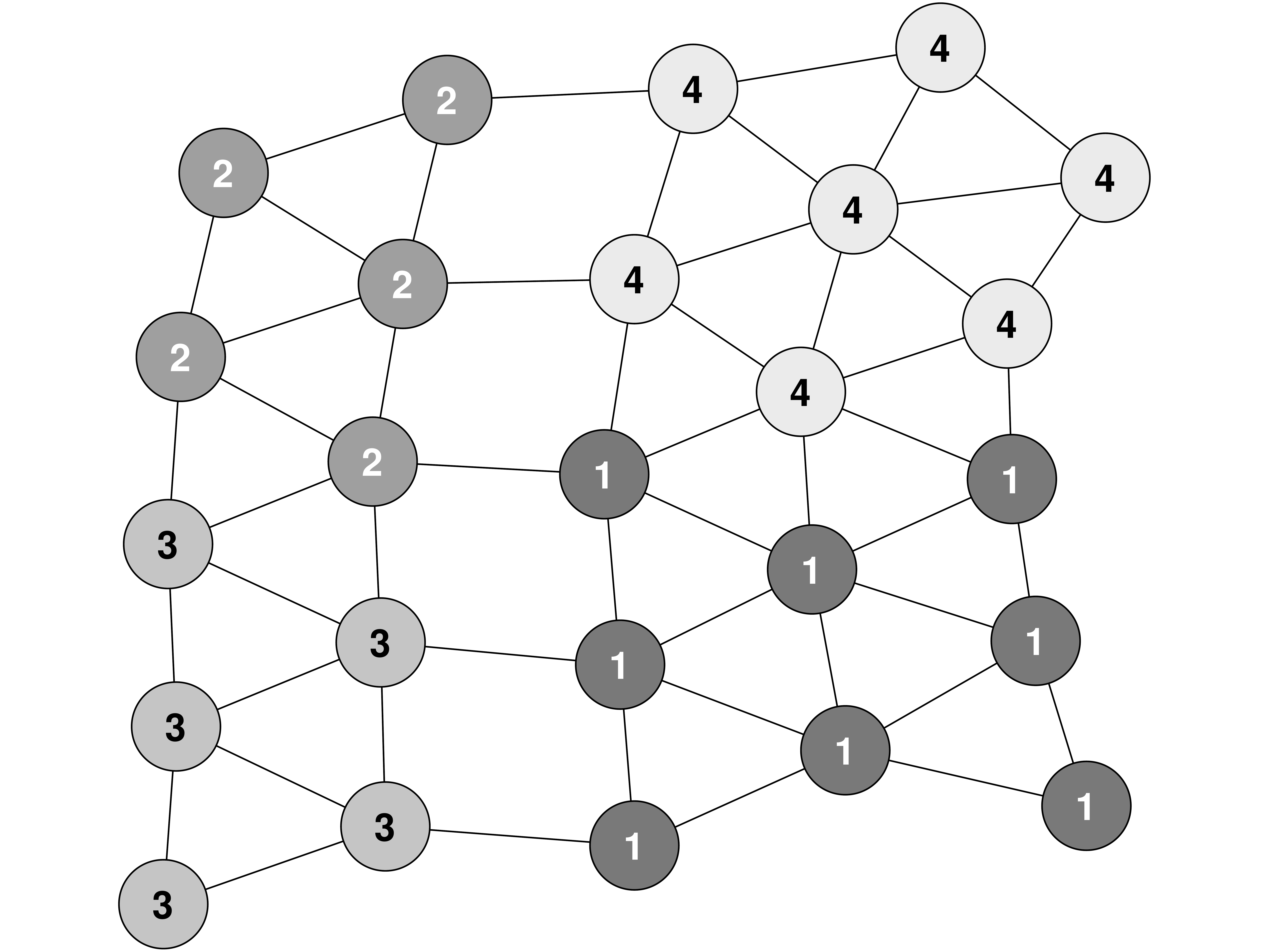}%
	\includegraphics[width=.2375\textwidth]{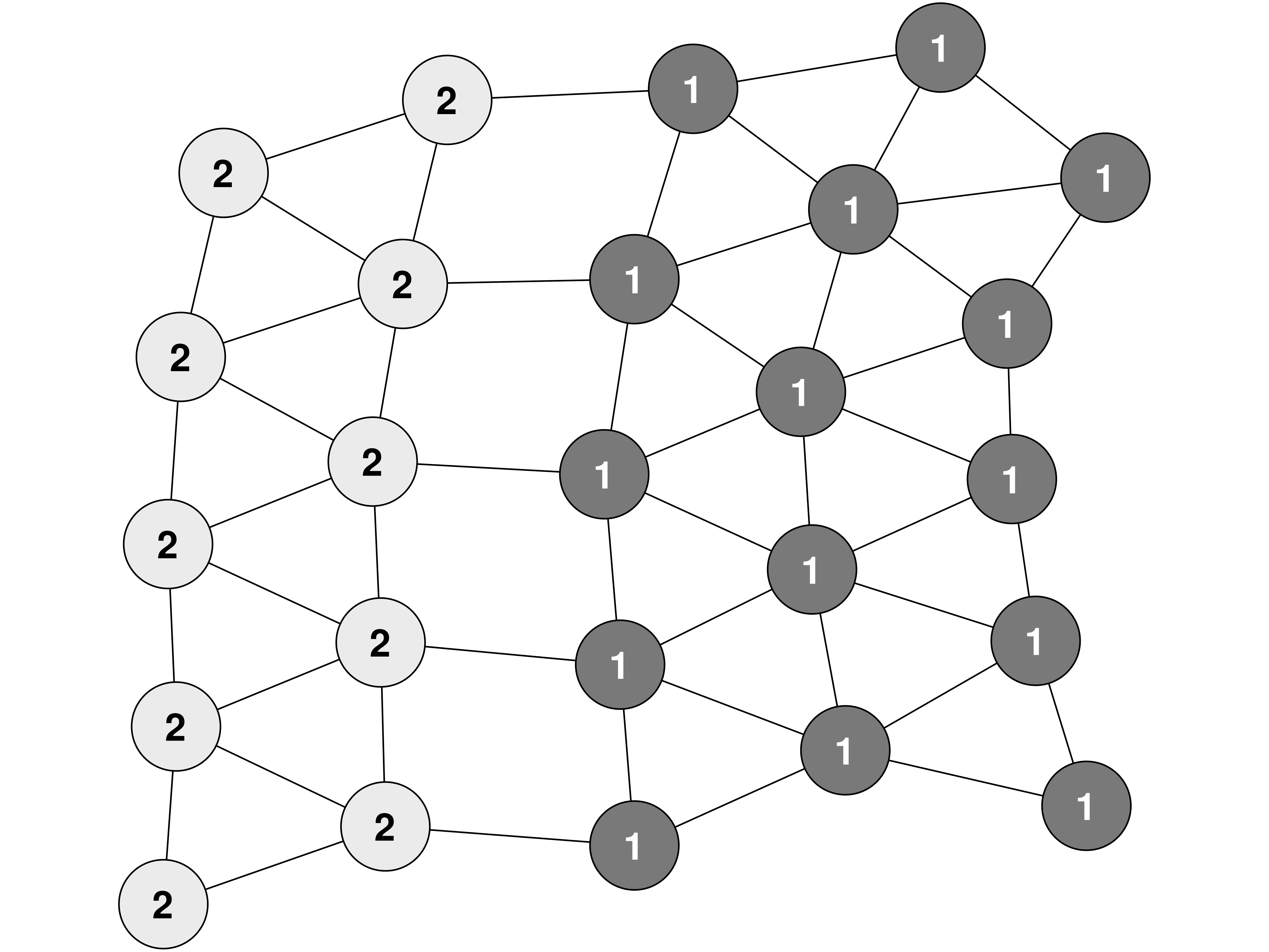}}
	\caption[The defensive and offensive label propagation methods.]{Comparison of defensive and offensive label propagation in artificial networks with a planted community structure and a triangular grid with four missing edges. The labels and shades of the nodes represent communities or groups identified by the two methods.}
	\label{fig:prefs}
\end{figure}

The left side of~\figref{prefs} demonstrates the defensive and offensive label propagation methods in an artificial network with two planted communities that are only loosely separated. While defensive propagation correctly identifies the communities planted in the network, offensive propagation spreads the labels beyond the borders of the communities and reveals no structure in this network. The right side of~\figref{prefs} compares the methods also on a graph partitioning problem. The methods are applied to a triangular grid with four edges removed, which makes a division into two groups the only sensible partition. In contrast, offensive propagation correctly partitions the grid into two groups, whereas defensive propagation overly restrains the spread of labels and recovers four groups. 

\begin{table}[b]
	\caption[Degeneracy diagrams of the label propagation methods.]{Degeneracy diagrams of the label propagation methods displaying the non-degenerate ranges of the revealed groups (thick lines), while the percentages show the fraction of nodes in the tiny groups (left) and in the largest group (right). The values are averages over $25$ runs of the methods.}
	\label{tbl:degs}
	\begin{tabular*}{\textwidth}{@{\extracolsep{\fill}}lcc}\hline
		Method & European Roads & Wikipedia Users \cr\hline
		Standard Propagation & \degs{0.615}{0.991}{61.5\%}{0.9\%}{mid} & \degs{0.058}{0.324}{5.8\%}{67.6\%}{mid} \cr
		Defensive Propagation & \degs{0.536}{0.991}{53.6\%}{0.9\%}{mid} & \degs{0.066}{0.832}{6.6\%}{16.8\%}{dark} \cr
		Offensive Propagation & \degs{0.071}{0.915}{\phantom{0}7.1\%}{8.5\%}{dark} & \degs{0.043}{0.203}{4.3\%}{79.7\%}{mid} \cr\hline
	\end{tabular*}
\end{table}

\tblref{degs} further compares the defensive and offensive label propagation methods on the European road network~\cite{SB11b} with $1,\!174$ nodes and a network of user interactions on Wikipedia~\cite{MAC11} with $126,\!514$ nodes. Both networks are available at KONECT. Degeneracy diagrams in~\tblref{degs} show the non-degenerate or effective ranges of the revealed groups that span the fraction of nodes not covered by the tiny groups with three nodes or less, or the largest group~\cite{SVW16a} (left and right percentages, respectively). Ideally, the thick lines in~\tblref{degs} would span from left to right. Due to the sparse grid-like structure of the road network, defensive propagation partitions $53.6\%$ of the nodes into tiny groups, which is not a useful result. This can be avoided by using offensive propagation, where this percentage equals $7.1\%$. However, in the case of much denser Wikipedia network, offensive propagation returns one giant group occupying $79.7\%$ of the nodes, thus defensive propagation with $16.8\%$ is preferred. Note that the crucial difference between these two networks requiring the use of different methods is their density. A generally applicable approach is first to use defensive propagation and then iteratively refine the revealed groups with offensive propagation~\cite{SB10d,SB11d}, in this order. For example, such approach reveals a partition of the road network with $7.9\%$ of the nodes in the tiny groups and $6.4\%$ of the nodes in the largest group on average.

An alternative definition of defensive and offensive label propagation is to replace the random walk diffusion in~\eqref{prefs} with the eigenvector centrality~\cite{Bon87} defined as
\begin{equation}
	p_i = \kappa_{g_i}^{-1}\sum_{j}p_jA_{ij}\delta(g_i,g_j),
	\label{eq:prefe}
\end{equation}
where $\kappa_{g_i}$ is a normalizing constant equal to the leading eigenvalue of the adjacency matrix $A$ reduced to the nodes in group $g_i$. \zch have shown that defensive label propagation with the eigenvector centrality for the node preferences is equivalent to the maximum likelihood estimation of a stochastic block model with Gaussian weights on the edges. This relates the label propagation method with yet another popular approach in the literature that is more thoroughly described in~\chpref{11}{Stochastic blockmodeling}.


\subsection{\label{sec:perf}Method Stability and Complexity}

Here, we discuss different techniques to improve the performance of the label propagation method by 
either increasing its stability or reducing its complexity.

One of the main sources of instability of the method is the random order of label updates in asynchronous propagation~\cite{RAK07,LHLC09}. Recall that the primary reason for this is to break cyclic oscillations of labels in synchronous propagation as it occurs in~\figref{oscil}. \lljt proposed still to use synchronous propagation that can lead to oscillations of labels, but rather to break the oscillations by making the label propagation rule in~\eqref{lpa} probabilistic. The probability that the node $i$ with group label $g_i$ updates its label to $g$ is defined as
\begin{equation}
	\prbi{i}{g}\propto\delta(g_i,g)+\sum_{j}A_{ij}\delta(g_j,g).
	\label{eq:prob}
\end{equation}
Although this successfully eliminates the oscillations of labels in~\figref{oscil}, probabilistic label propagation can make the method even more unstable. It must be stressed that this instability represents a major issue, especially in very large networks.

\cgg proposed a more elegant solution called semi-synchronous label propagation based on node coloring. A coloring of network nodes is an assignment of colors to nodes such that no two connected nodes share the same color~\cite{New10}. Notice that if two nodes are not connected their labels do not directly depend on one another in~\eqref{lpa} and can therefore be updated simultaneously using synchronous propagation. Given a coloring of the network, semi-synchronous propagation traverses different colors in a random order as in asynchronous propagation. In contrast, the labels of the nodes with the same color are updated simultaneously as in synchronous propagation. For instance, coloring each node with a different color is equivalent to asynchronous propagation, while a simple greedy algorithm can find a coloring with at most $\Delta+1$ colors, where $\Delta$ is the maximum degree in a network. In contrast to synchronous and asynchronous propagation, the convergence of semi-synchronous propagation can be formally proven.

\sbbb observed empirically that updating the labels of the nodes in some fixed order drives the label propagation process towards similar solutions as setting the node preferences in~\eqref{lpp} higher (lower) for the nodes that appear earlier (later) in the order and then updating their labels in a random order as in asynchronous propagation. The node preferences can thus be used as node balancers to counteract the randomness introduced by asynchronous propagation. Let $t_i$ be a normalized position of the node $i$ in some random order, which is set to $1/n$ for the first node, $2/n$ for the second node and so on, where $n$ is the number of nodes in a network. The value $t_i$ represents the time at which the label of node $i$ is updated. Balanced label propagation sets the node preferences using a logistic function as
\begin{equation}
	g_i = \argmax_g\sum_{j}\frac{1}{1+e^{-\gamma(2t_j-1)}}A_{ij}\delta(g_j,g),
	\label{eq:lpb}
\end{equation}
where $\gamma$ is a parameter of the method. For $\gamma=0$, \eqref{lpb} is equivalent to the standard label propagation rule in~\eqref{lpa}, while $\gamma>0$ makes the method more stable, but this increases its time complexity. In practice, one must therefore decide on a compromise between the method stability and its time complexity.

\begin{figure}[t]\centerline{%
	\includegraphics[height=.225\textwidth]{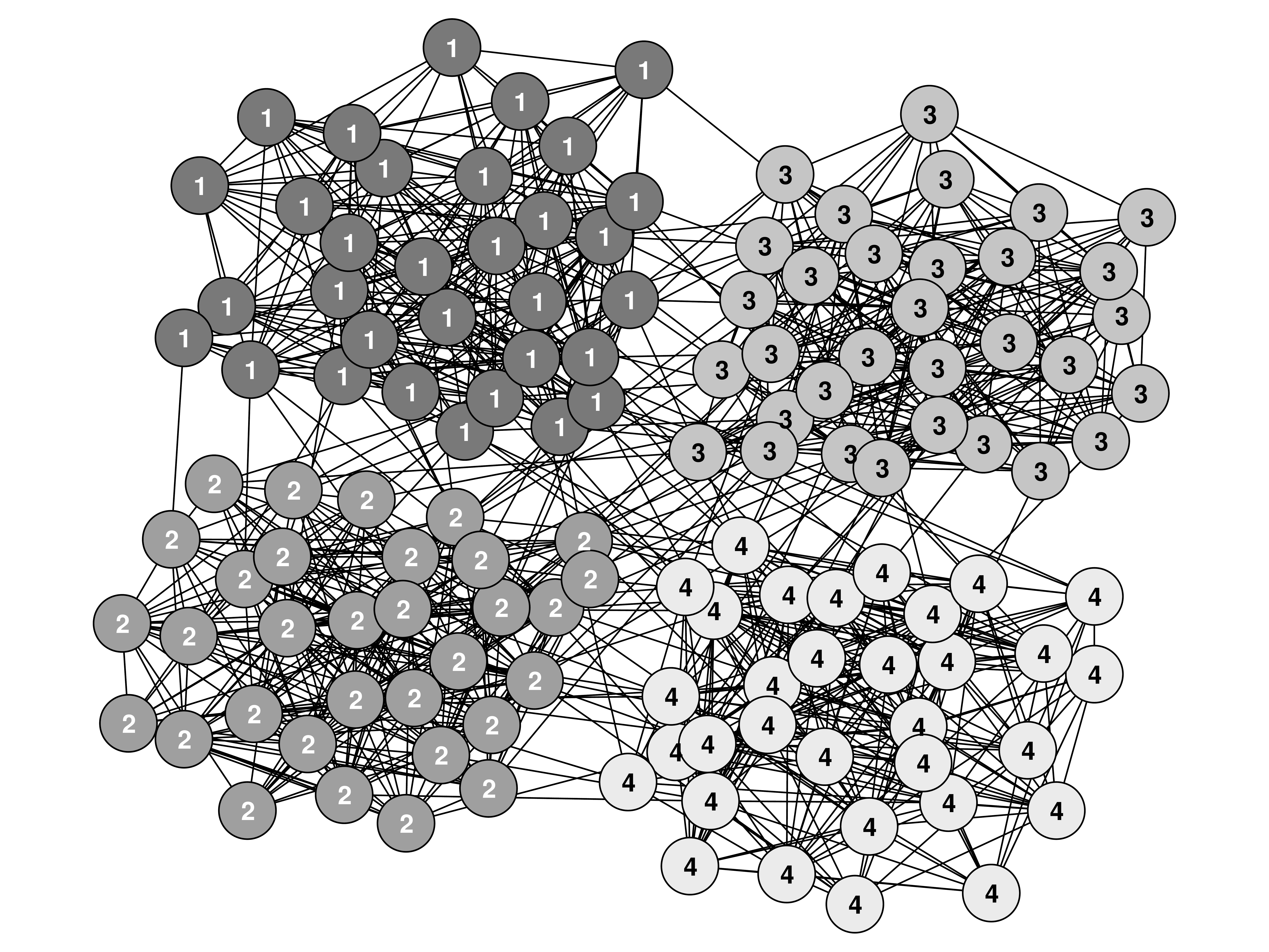}%
	\includegraphics[height=.225\textwidth]{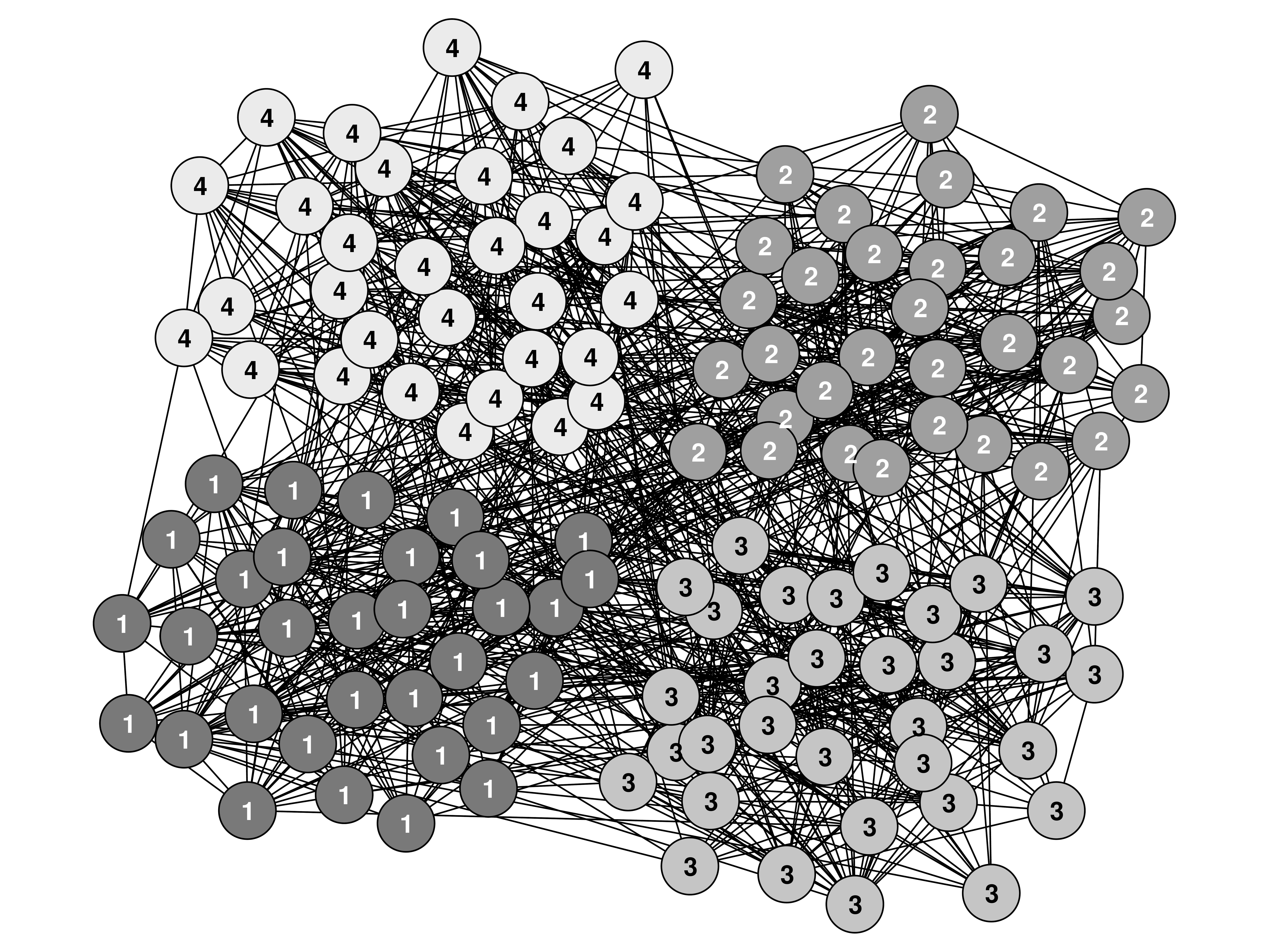}\hskip.05\textwidth
	\includegraphics[height=.25\textwidth]{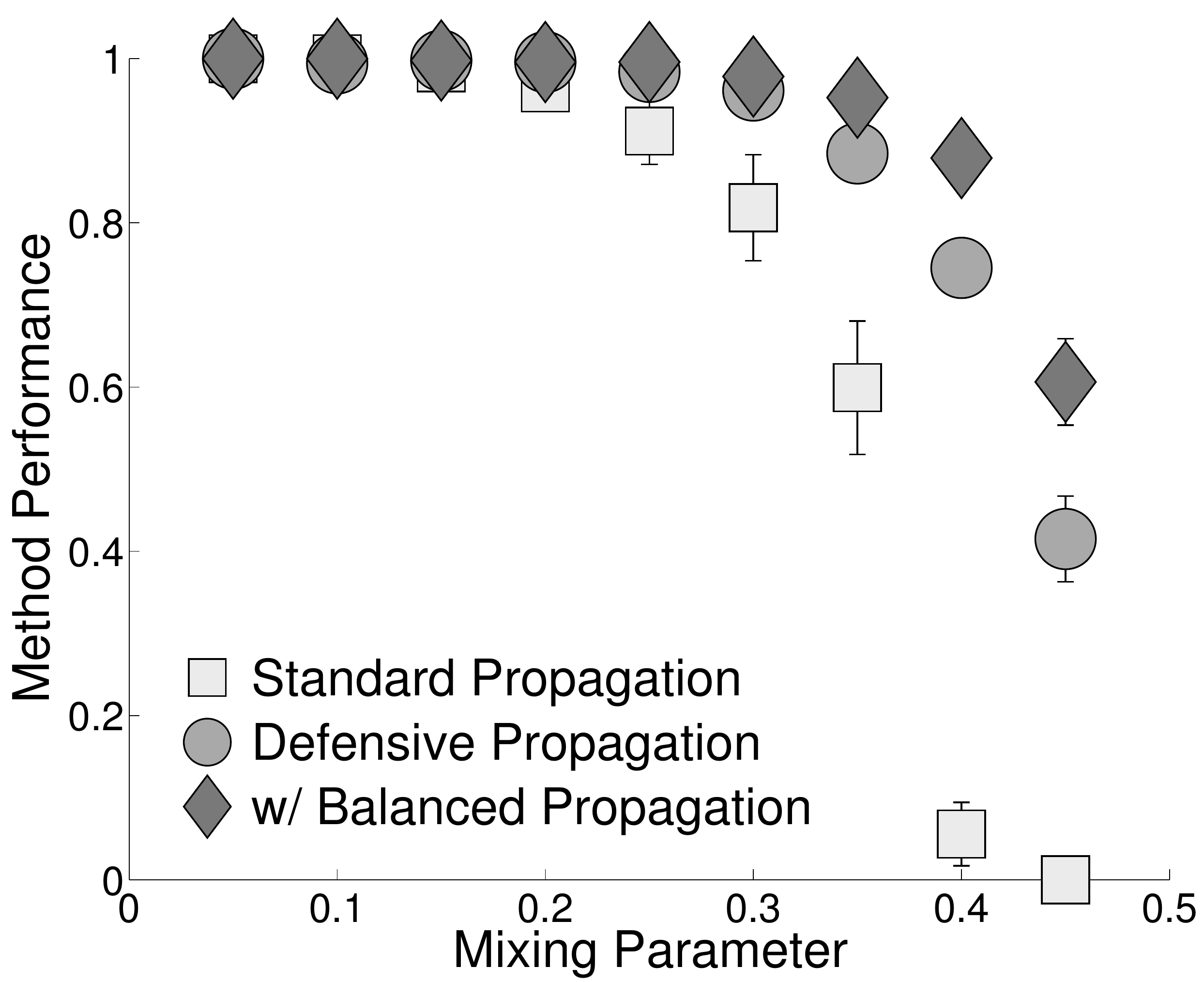}}
	\caption[Community detection with the label propagation methods.]{Performance of the label propagation methods in artificial networks with planted community structure represented by the labels and shades of the nodes. The markers are averages over $25$ runs of the methods, while the error bars show standard errors.}
	\label{fig:perf}
\end{figure}

The method stability is tightly knit with its performance. \figref{perf} compares community detection of the label propagation methods in artificial networks with four planted communities~\cite{GN02}. Community structure is controlled by a mixing parameter $\mu$ that represents the fraction of nodes' neighbors in their own community. For example, the left side of~\figref{perf} shows realizations of networks for $\mu=0.1$ and $0.4$. Performance of the methods is measured with the normalized mutual information~\cite{FH16}, where higher is better (see \cite{FH16} for the exact definition). As seen in the right side of~\figref{perf}, balanced label propagation combined with the defensive node preferences in~\eqref{prefs} performs best in these networks, when $\gamma=1$.

\begin{figure}[b]\centerline{%
	\includegraphics[height=.225\textwidth]{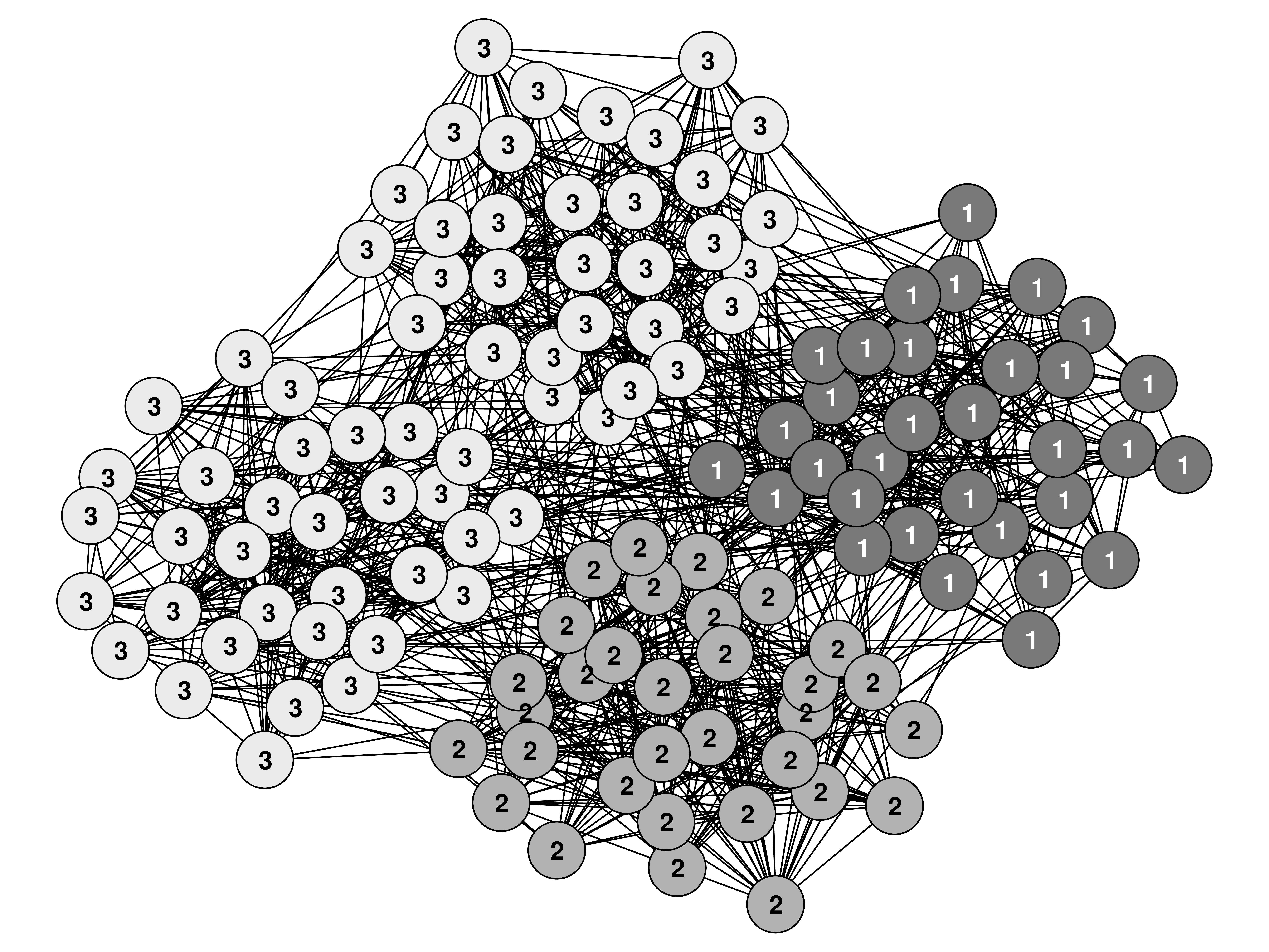}%
	\includegraphics[height=.225\textwidth]{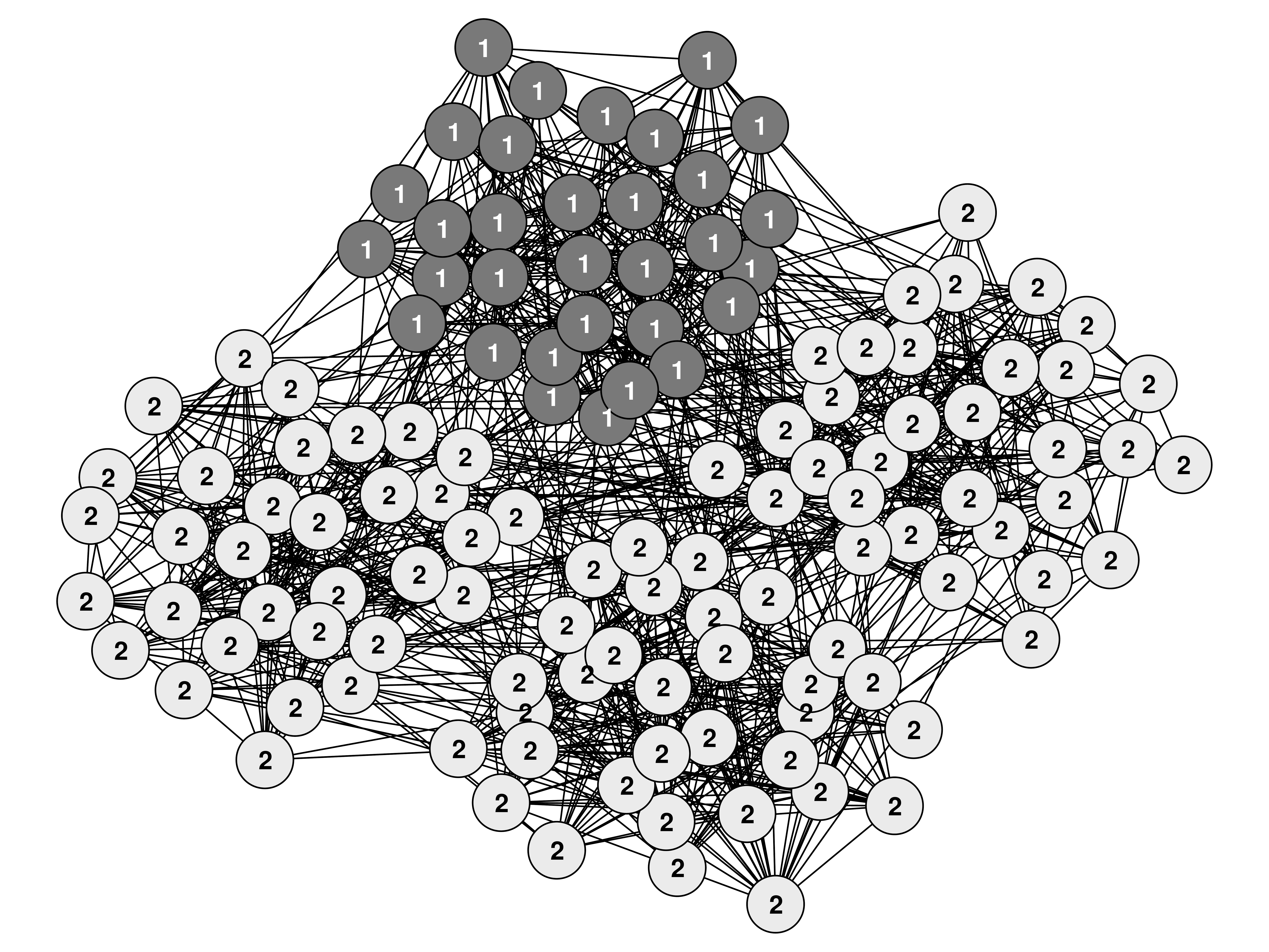}\hskip.025\textwidth
	\includegraphics[height=.225\textwidth]{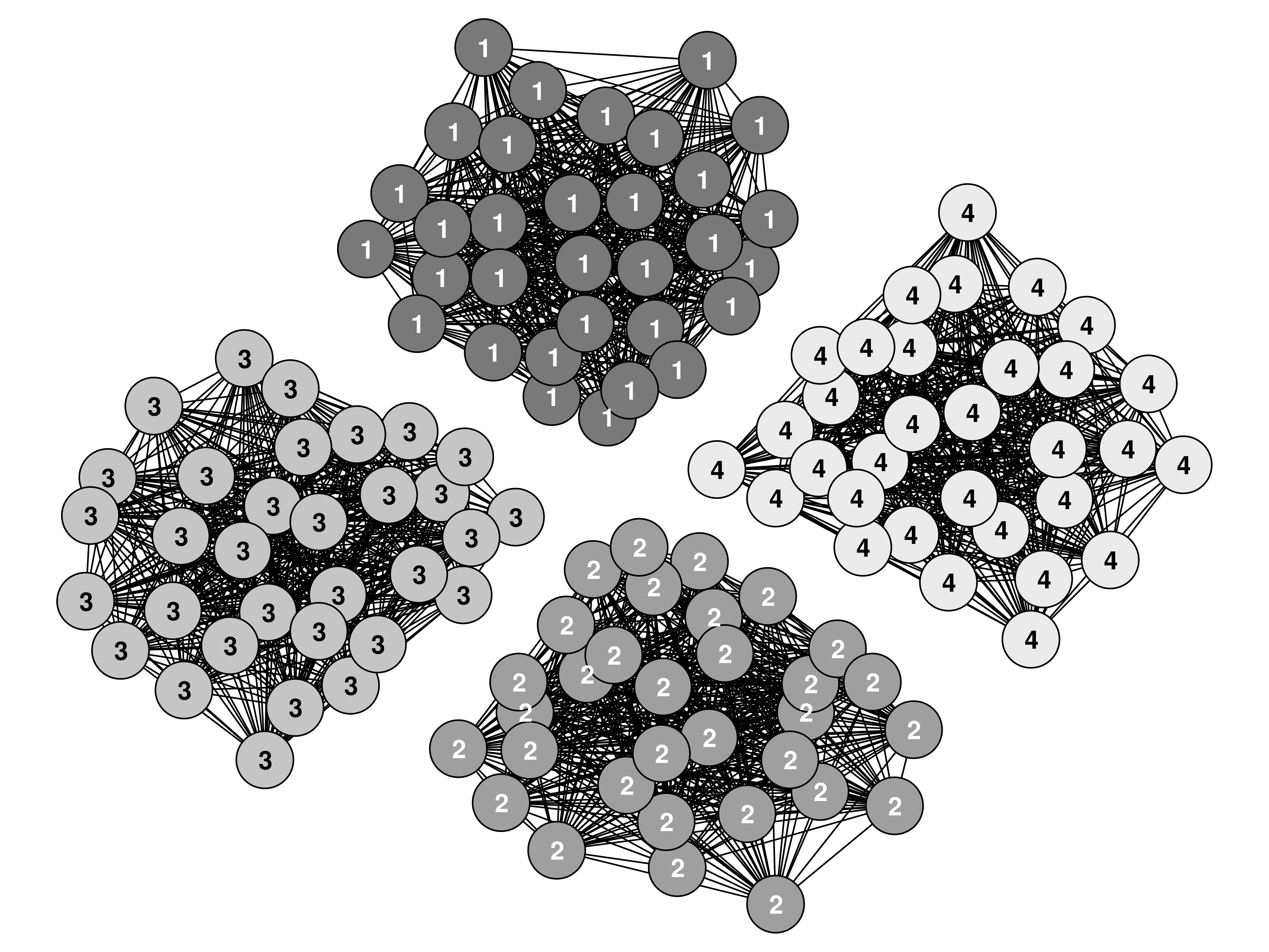}}
	\caption[The label propagation method combined with consensus clustering.]{Label propagation in artificial networks with planted community structure and the corresponding consensus graph. The labels and shades of the nodes represent communities identified by the label propagation method.}
	\label{fig:cons}
\end{figure}

\begin{figure}[t]\centerline{%
	\includegraphics[width=\textwidth]{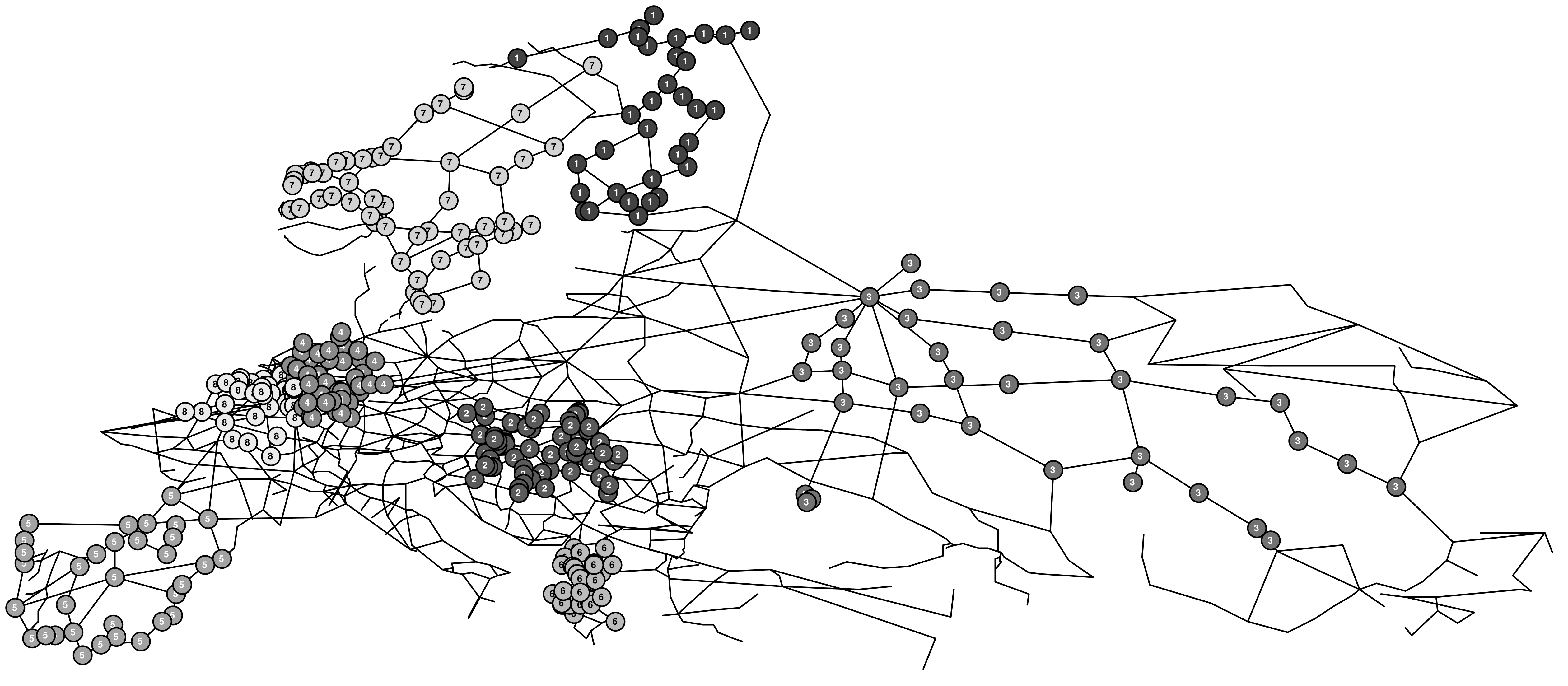}}
	\caption[The offensive label propagation method applied to a road network.]{Offensive label propagation with consensus clustering in the European road network. The labels and shades of the nodes represent the largest eight groups identified by the method.}
	\label{fig:euro}
\end{figure}

Another prominent approach for improving community detection of the label propagation methods is consensus clustering~\cite{LF12,ZJFP14,GCSKXLBNHGMK15}. One first applies the method to a given network multiple times and constructs a weighted consensus graph, where weights represent the number of times two nodes are classified into the same community. Note that only edges with weights above a given threshold are kept. The entire process is then repeated on the consensus graph until the revealed communities no longer change. For example, the left side of~\figref{cons} shows two realizations of groups obtained with the standard label propagation method in~\eqref{lpa} in artificial networks for $\mu=0.33$. Although these do not exactly coincide with the planted communities, label propagation in the corresponding consensus graph recovers the correct community structure as demonstrated in the right side of~\figref{cons}. For another example, \figref{euro} shows the largest connected component of the European road network from~\tblref{degs} and the largest groups revealed by the offensive label propagation method in~\eqref{lpo} with $25$ runs of consensus~clustering.

Note, however, that consensus clustering can substantially increase the method's computational time. Other work has thus considered different hybrid approaches to improve the stability of community detection of the label propagation methods, where community structure revealed by one method is refined by another~\cite{SB10d,SB11d}, possibly proceeding iteratively or incrementally~\cite{LHLC09,CRGP12}. For instance, label propagation under constraints~\cite{LM09b,HLSZD16} has traditionally been combined with a multistep greedy agglomeration~\cite{SC08}.

In the remaining, we also briefly discuss different approaches to reduce the complexity of the label propagation method. Although the time complexity is already nearly linear $\cmp{m^{1.2}}$, where $m$ is the number of edges in a network~\cite{SB11d}, one can still further improve the computational time. As shown in~\figref{comp}, the number of nodes that update their label at a particular iteration of label propagation drops exponentially with the number of iterations. Thus, after a couple of iterations, most nodes already acquire their final label and no longer need to be updated. For instance, one can selectively update only the labels of those nodes for which the fraction of neighbors sharing the same label is below a certain threshold~\cite{LHLC09}, which can make the method truly linear $\cmp{m}$. \xs further formalized this idea using the concept of active and passive nodes. A node is said to be passive if updating would not change its label. Otherwise, the node is active. The labels are therefore propagated only between the active nodes until all nodes become passive.

Due to its algorithmic simplicity, the label propagation method is easily parallelizable, especially with synchronous or semi-synchronous propagation mentioned above. The method is thus suitable for application in distributed computing environments such as Spark\footnote{\url{http://spark.apache.org}}~\cite{BKAFKTK14} or Hadoop\footnote{\url{http://hadoop.apache.org}}~\cite{Ove13} and on parallel architectures~\cite{SN11}. In this way, label propagation has been successfully used on billion-node networks~\cite{BKAFKTK14,WXSW14}.

%
%

\section{\label{sec:nets}Extensions to Other Networks}

Throughout the chapter, we have assumed that the label propagation method is applied to simple undirected networks. Nevertheless, the method can easily be extended to networks with multiple edges between the nodes as in~\eqref{lpa} and networks with weights on the edges as in~\eqref{lpw}. This holds also for the different advances of the propagation methods presented in~\secref{advs}. In contrast, there seem to be no straightforward extension to networks with directed arcs. The reason for this is that propagating the labels exclusively in the direction of arcs enables exchange of labels only between mutually reachable nodes forming a strongly connected component. Since any directed network is a directed acyclic graph on its strongly connected components, the labels can propagate between the nodes of different strongly connected components merely in one direction. Therefore, one usually disregards the directions of arcs when applying the label propagation method to directed networks except in the case when most arcs are reciprocal.

The method can be extended to signed networks with positive and negative edges between the nodes as in the approach of~\dm. In order to partition the network in such a way that positive edges mostly appear within the groups and negative edges between the groups, one assigns some fixed positive (negative) weight to positive (negative) edges and then applies the standard label propagation method for weighted networks in~\eqref{lpw}. According to the objective function in~\eqref{haml}, the method thus simultaneously tries to maximize the number of positive edges within the groups and the number of negative edges between the groups. Still, this does not ensure that the nodes connected by a negative edge are necessarily assigned to different groups, but merely restricts the propagation of labels along the negative edges~\cite{AM15}.

\begin{table}[t]
	\caption[Comparison of the label propagation methods on a signed network.]{Comparison of the label propagation methods on the signed Wikipedia web of trust network. The values are averages over $25$ runs of the methods, while $\hml$ is defined in~\eqref{haml}.}
	\label{tbl:signs}
	\begin{tabular*}{\textwidth}{@{\extracolsep{\fill}}lccc}\hline
		Method & $+$ Edges Within & $-$ Edges Between & Hamiltonian $\hml$ \cr\hline
		Standard Propagation & $96.6\%$ & $6.7\%$ & $-528185.8$ \cr
		Signed Propagation & $90.9\%$ & $56.7\%$ & $-535065.2$ \cr
		w/ Equal Weights & $75.6\%$ & $81.8\%$ & $-460413.1$ \cr\hline
	\end{tabular*}
\end{table}

\tblref{signs} shows the standard and signed label propagation methods applied to the Wikipedia web of trust network~\cite{MAC11} available at KONECT. The network consists of $138,\!587$ nodes connected by $629,\!689$ positive edges and $110,\!417$ negative edges. Standard label propagation ignoring the signs of edges reveals one giant group occupying $89.0\%$ of the nodes on average. Most positive edges are thus obviously within the groups, but the same also holds for negative edges. Signed label propagation with positive and negative weights on the edges reduces the size of the largest group to $60.6\%$ of the nodes on average. Most positive edges remain within the groups, while more than half of negative edges is between the groups. Note that the method assigns weights $1$ and $-1$ to positive and negative edges. Since only $12.0\%$ of the edges in the network are negative, this actually puts more emphasis on the positive edges. To circumvent the latter, one can assign equal total weight to positive and negative edges by using weights $1/m_p$ and $-1/m_n$, where $m_p$ and $m_n$ are the numbers of positive and negative edges. Signed label propagation with equal total weights returns a larger number of groups with $43.2\%$ of the nodes in the largest group, and about the same fraction of positive edges within the groups and negative edges between the groups. For further discussion on partitioning signed networks see~\chpref{8}{Partitioning signed networks}.

\begin{figure}[t]\centerline{%
	\includegraphics[width=.25\textwidth]{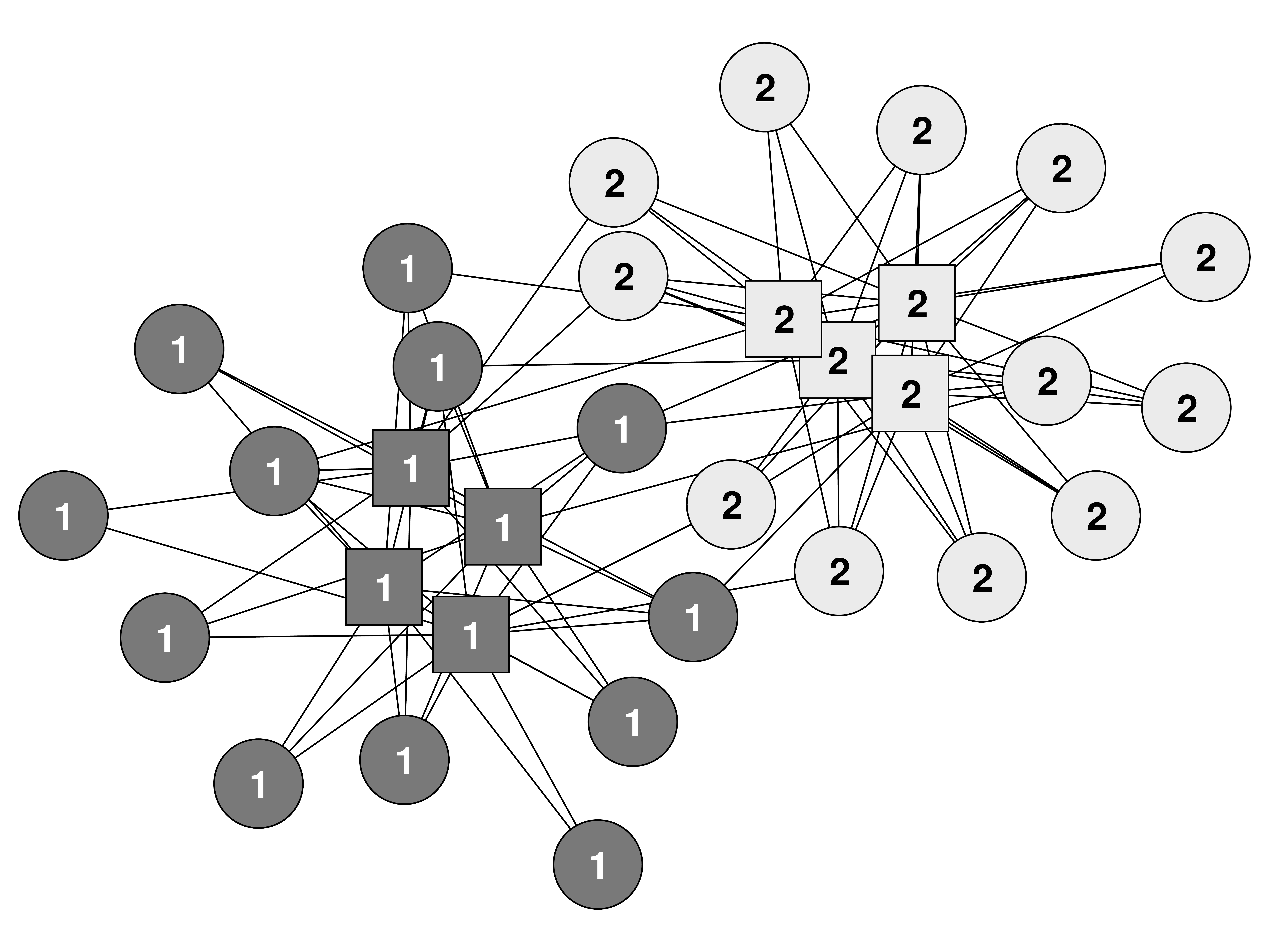}\hskip.05\textwidth
	\includegraphics[width=.25\textwidth]{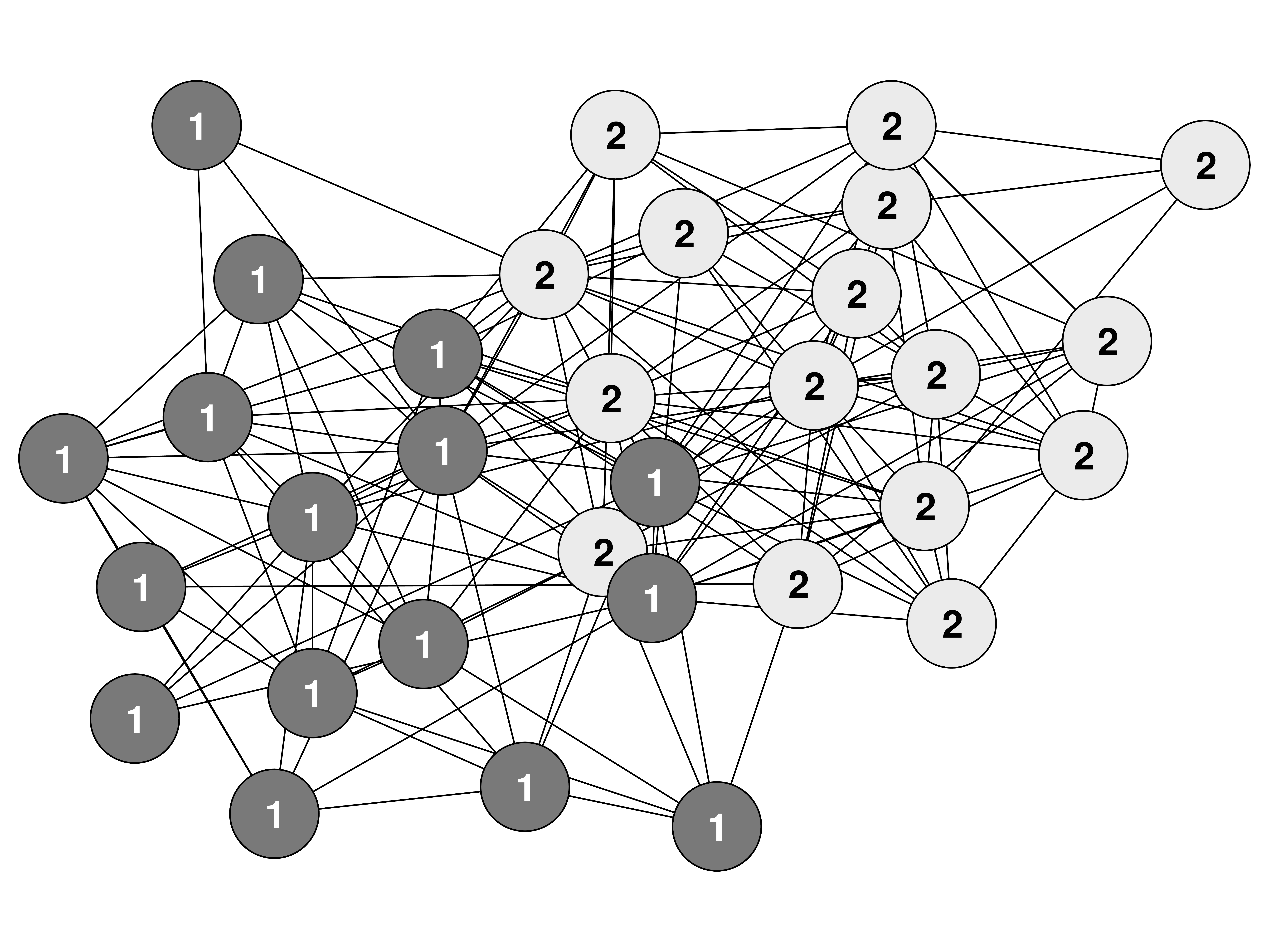}%
	\includegraphics[width=.25\textwidth]{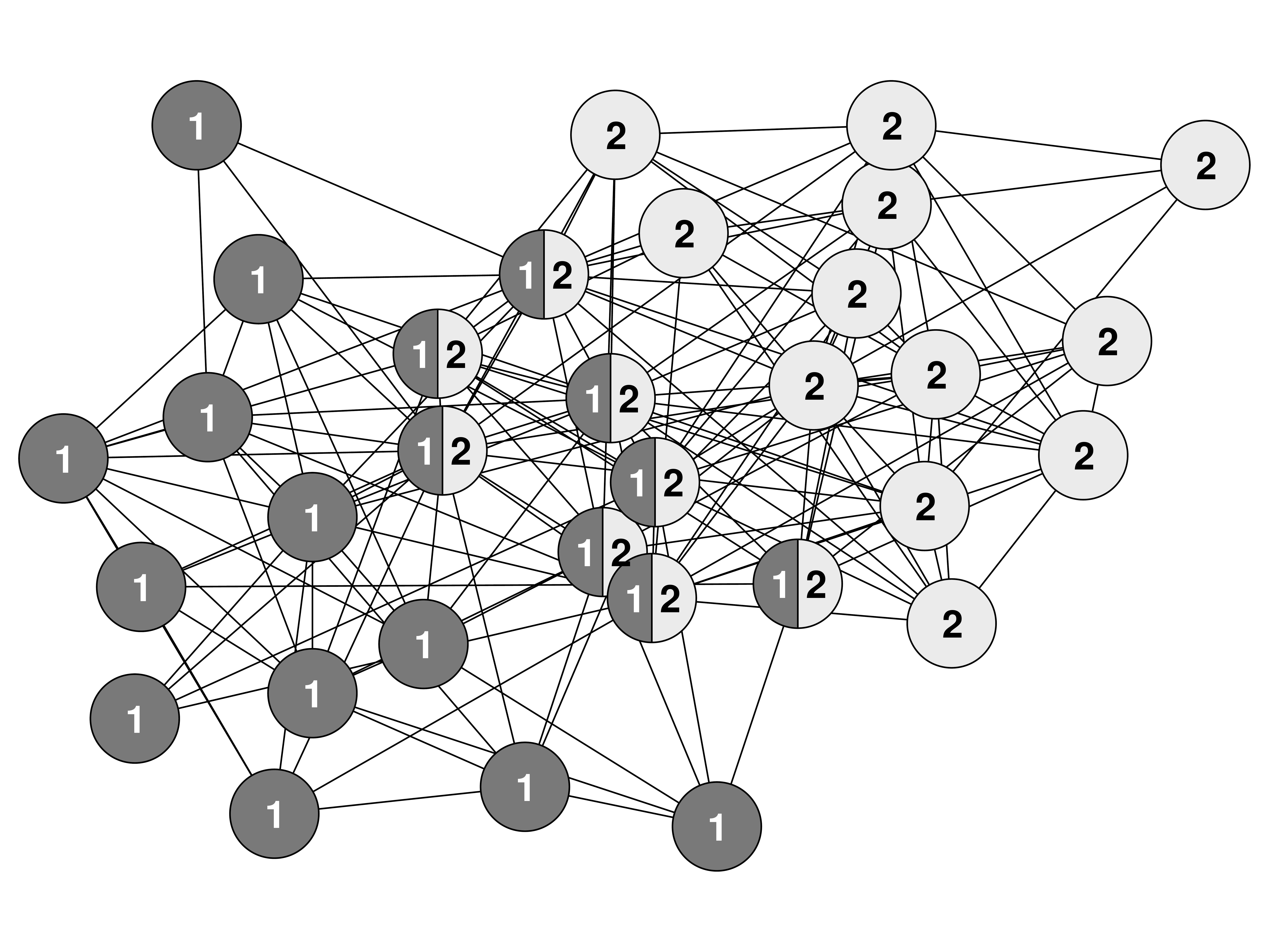}}
	\caption[Overlapping community detection with the label propagation method.]{Non-overlapping and overlapping label propagation in artificial networks with planted community structure. The labels and shades of the nodes represent communities identified by different methods, while the types of nodes of the bipartite network are shown with distinct symbols.}
	\label{fig:alts}
\end{figure}

Any label propagation method can also be used on bipartite networks with two types of nodes and edges only between the nodes of different type as in the left side of~\figref{alts}. For instance, \bc adopted the label propagation methods under constraints to optimize bipartite modularity~\cite{Bar07}. \lmm proposed a proper extension of the label propagation framework to bipartite networks. This is a special case of semi-synchronous propagation with node coloring discussed in~\secref{perf}. Recall that semi-synchronous propagation updates the labels of the nodes with the same color synchronously, while different colors are traversed asynchronously. In bipartite networks, the types of the nodes can be taken for their colors, thus  the method alternates between the nodes of each type, while the propagation of labels always occurs synchronously. The same principle can be extended also to multipartite networks, where again the nodes of the same type are assigned the same color. However, in multirelational or multilayer networks~\cite{BBCGGRSWZ14}, one can separately consider the nodes of different layers, but the propagation of labels within each layer requires asynchronicity for the method to converge.

%
%

\section{\label{sec:grps}Alternative Types of Network Structures}

The label propagation method was originally designed to detect non-overlapping communities in networks~\cite{RAK07,LHLC09}. In the following, we show how the method can be extended also to more diverse network structures. We consider extensions to overlapping groups of nodes, groups of nodes at multiple resolutions that form a nested hierarchy and groups of structurally equivalent nodes. Note that, in contrast to the extensions to other types of networks in~\secref{nets}, this increases the time complexity of the method derived in~\secref{comp}. As shown in the following, the time complexity increases by a factor depending on the type of the groups considered.


\subsection{\label{sec:ovrg}Overlapping Groups of Nodes} 

Extension of the label propagation method to overlapping groups of nodes is relatively straightforward~\cite{Gre10,WLGWT12}. Instead of assigning a single group label $g_i$ to node $i$ as the standard label propagation method in~\eqref{lpa}, multiple labels are assigned to each node. Let $\ovr_i$ be the group function of node $i$ where $\ovrg{i}{g}$ represents how strongly the node is affiliated to group $g$. In particular, the node belongs to groups $g$ for which $\ovrg{i}{g}>0$, while its group affiliations are normalized to one as $\sum_g\ovrg{i}{g}=1$. At the beginning of label propagation, each node is put into its own group by setting $\ovrg{i}{i}=1$. Then, at every iteration, each node adopts the group labels of its neighbors. The affiliation $\ovrg{i}{g}$ of node $i$ to group $g$ is computed as the average affiliation of its neighbors. Hence,
\begin{equation}
	\ovrg{i}{g} = \sum_{j}\frac{\ovrg{j}{g}}{k_i}A_{ij},
	\label{eq:lpv}
\end{equation}
where $A$ is the network adjacency matrix and $k_i$ is the degree of node $i$. \eqref{lpv} can be combined also with an inflation operator raising $\ovrg{i}{g}$ to some exponent~\cite{XS13}. Obviously, the groups can now overlap as the nodes can belong to multiple groups. For example, the right side of~\figref{alts} demonstrates the non-overlapping and overlapping label propagation methods in an artificial network with two planted overlapping communities.

Notice, however, that the label propagation rule in~\eqref{lpv} inevitably leads to every node in a network belonging to all groups. It is therefore necessary to limit the number of groups a single node can belong to. \gre proposed that, after each iteration of label propagation, the group affiliations $\ovrg{i}{g}$ below $1/\nu$ are set to zero and renormalized, where $\nu$ is a method parameter. Since $\sum_g\ovrg{i}{g}=1$ for every node, the nodes can thus belong to at most $\nu$ groups. The parameter $\nu$ can be difficult to determine if a network consists of overlapping and non-overlapping groups. \wlgwt suggested replacing the parameter $\nu$ by a node-dependent threshold $\rho$ to keep node $i$ affiliated to group $g$ as~long~as
\begin{equation}
	\frac{\ovrg{i}{g}}{\max_g\ovrg{i}{g}}\geq\rho.
	\label{eq:ovrg}
\end{equation}

The time complexity of the described overlapping label propagation method is $\cmp{cm\nu}$, where $c$ is the number of iterations of label propagation, $m$ is the number of edges in a network and $\nu$ is the maximum number of groups a single node belongs to. The method is implemented by a popular community detection algorithm COPRA\footnote{\url{http://gregory.org/research/networks/software/copra.html}}~\cite{Gre10}.

It is also possible to detect overlapping groups of nodes by using the standard non-over\-lapp\-ing label propagation method. \xss proposed associating a memory with each node to store group labels from previous iterations. Running the label propagation for $c$ iterations assigns $c$ labels to each node's memory. The probability of observing label $g$ in the memory of node $i$ or, equivalently, the number of occurrences of $g$ in the memory of $i$ can then be interpreted as the group affiliation $\ovrg{i}{g}$ as defined above. Note that label propagation with node memory splits the label propagation rule in~\eqref{lpa} into two steps. Each neighbor $j$ of the considered node $i$ first propagates a random label from its memory, with the label $g$ being selected with probability $\ovrg{j}{g}$, while node $i$ then adds the most frequently propagated label to its memory. The time complexity of the method is $\cmp{cm}$, where $c$ is a small constant set to say $25$. The method is implemented by another popular community detection algorithm SLPA\footnote{\url{http://sites.google.com/site/communitydetectionslpa}}~\cite{XSL11} and its successor SpeakEasy\footnote{\url{http://www.cs.rpi.edu/~szymansk/SpeakEasy}}~\cite{GCSKXLBNHGMK15}.

DEMON\footnote{\url{http://www.michelecoscia.com/?page_id=42}}~\cite{CRGP12} is a well known community detection algorithm that also uses non-overlapping label propagation to detect overlapping groups. Instead of assigning a memory to each node as above, this label propagation method is separately applied to the subnetworks reduced to the neighborhoods of the nodes. All of the resulting groups that are, in general, overlapping are then merged together.


\subsection{\label{sec:hier}Hierarchy of Groups of Nodes} 

\begin{figure}[t]\centerline{%
	\includegraphics[height=.225\textwidth]{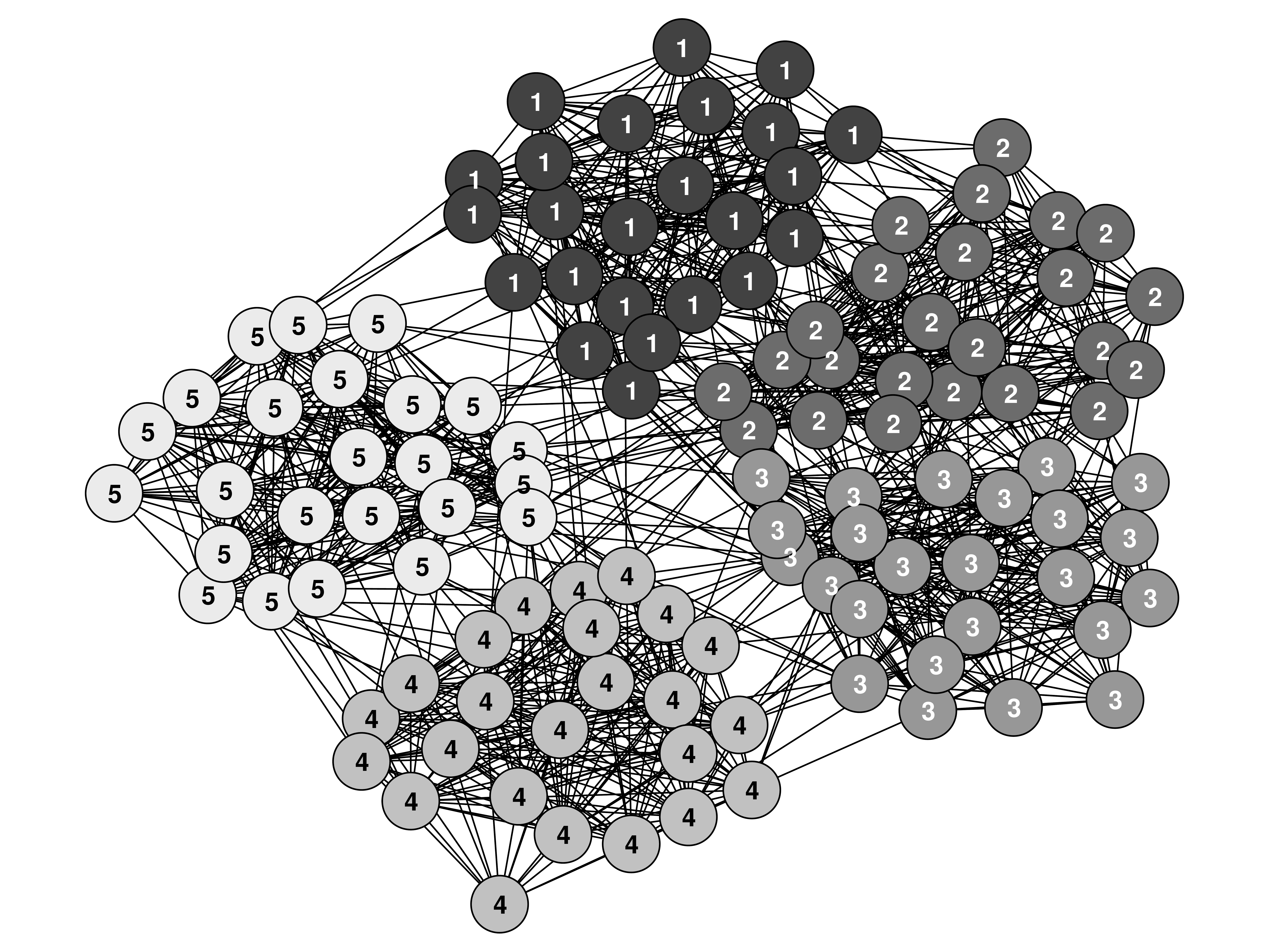}%
	\includegraphics[height=.225\textwidth]{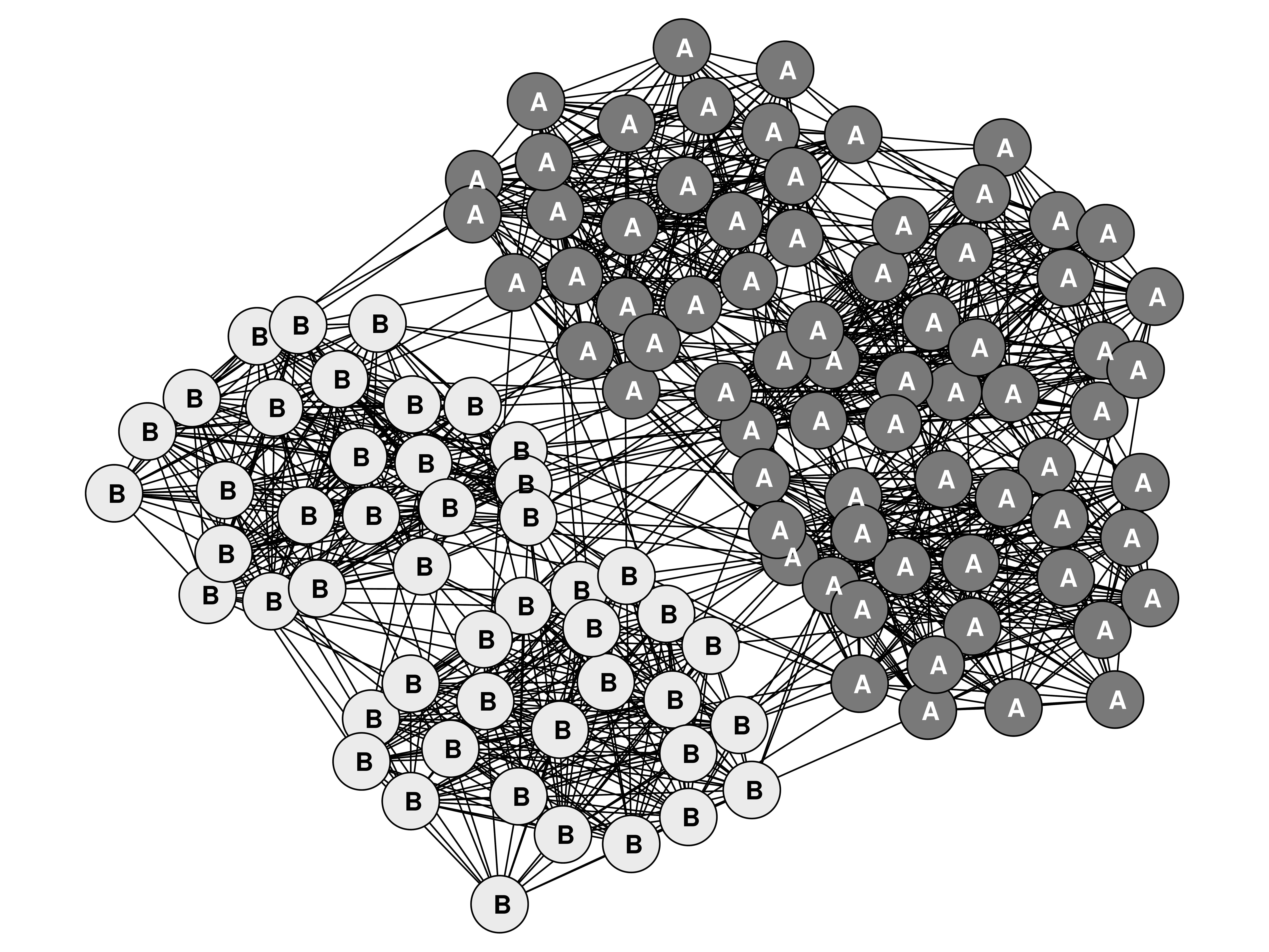}\hskip.05\textwidth
	\includegraphics[width=.225\textwidth,trim=0mm -12mm 0mm 0mm]{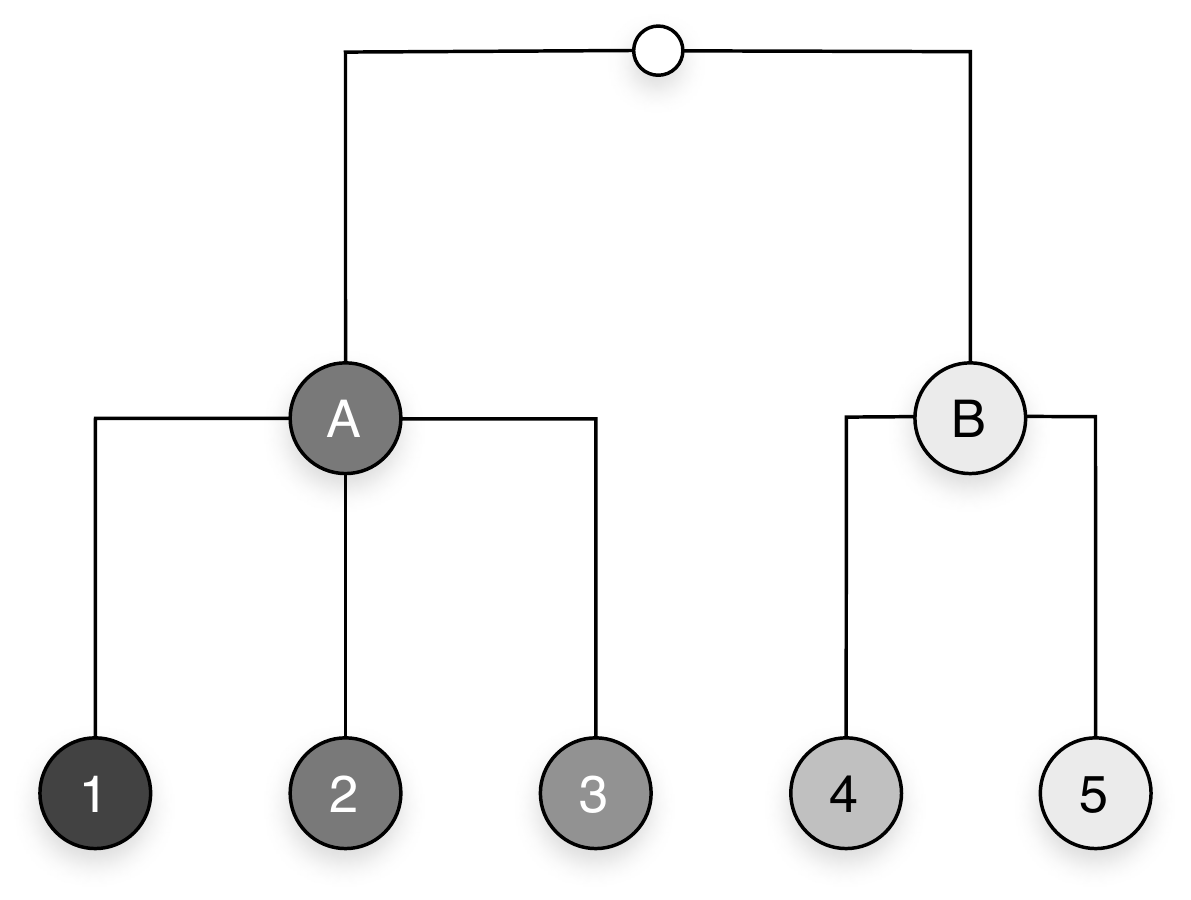}}
	\caption[Hierarchical community detection with the label propagation method.]{Artificial networks with two levels of planted community structure and the corresponding group hierarchy. The labels and shades of the nodes represent communities identified by the label propagation method.}
	\label{fig:hier}
\end{figure}

Label propagation can be applied in a hierarchical manner in order to reveal a nested hierarchy of groups of nodes~\cite{LHLC09,SB11d,SB14g,LHWC17}. The bottom level of such a hierarchy represents groups of nodes. The next level represent groups of groups of nodes and so on. Cutting the hierarchy at different levels results in groups of nodes at multiple resolutions. For example, \figref{hier} demonstrates the hierarchical label propagation method in artificial networks with two levels of planted community structure. Let $G_1,G_2,\ldots$ denote the groups revealed by the basic label propagation method in~\eqref{lpa}, which represent the bottom level of the group hierarchy. One then constructs a meta-network, where nodes correspond to different groups $G_i$ and an edge is put between the groups $G_i$ and $G_j$ if their nodes are connected in the original network. The weight of the edge is set to the number of edges between the groups $G_i$ and $G_j$ in the original network. Similarly, a loop is added to each group $G_i$ with a weight equal to the number of edges within the group $G_i$ in the original network. Finally, one applies the weighted label propagation method in~\eqref{lpw} to the constructed meta-network to reveal groups of groups $G_i$. These constitute the next level of the group hierarchy. The entire process of such bottom-up group agglomeration is repeated iteratively until a single group is recovered, which is the root of the hierarchy. Note that label propagation with group agglomeration is algorithmically equivalent to the famous Louvain modularity optimization method~\cite{BGLL08,Tra15}.

\begin{figure}[t]\centerline{%
	\includegraphics[width=.525\textwidth]{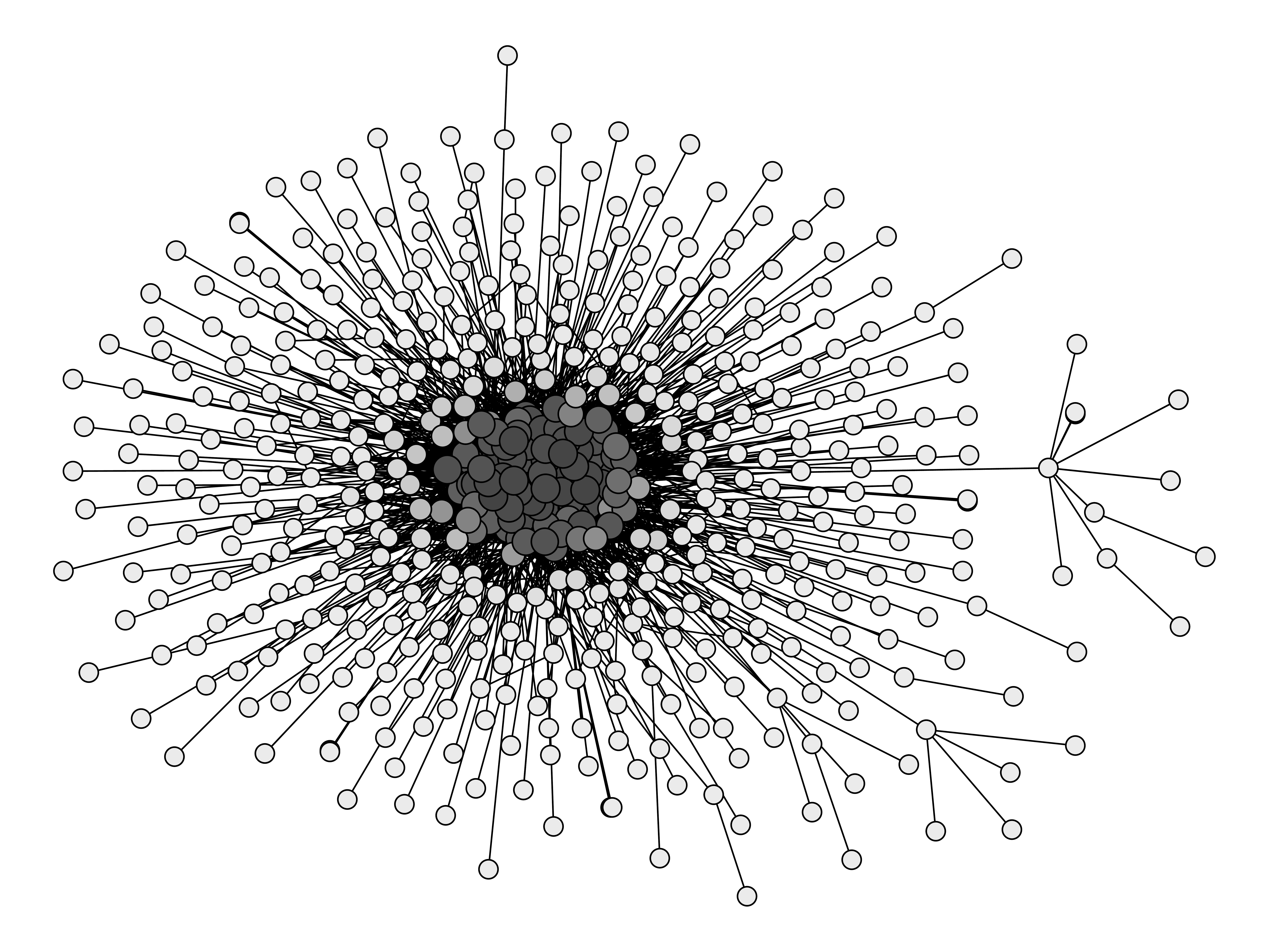}\hskip.025\textwidth
	\includegraphics[width=.475\textwidth]{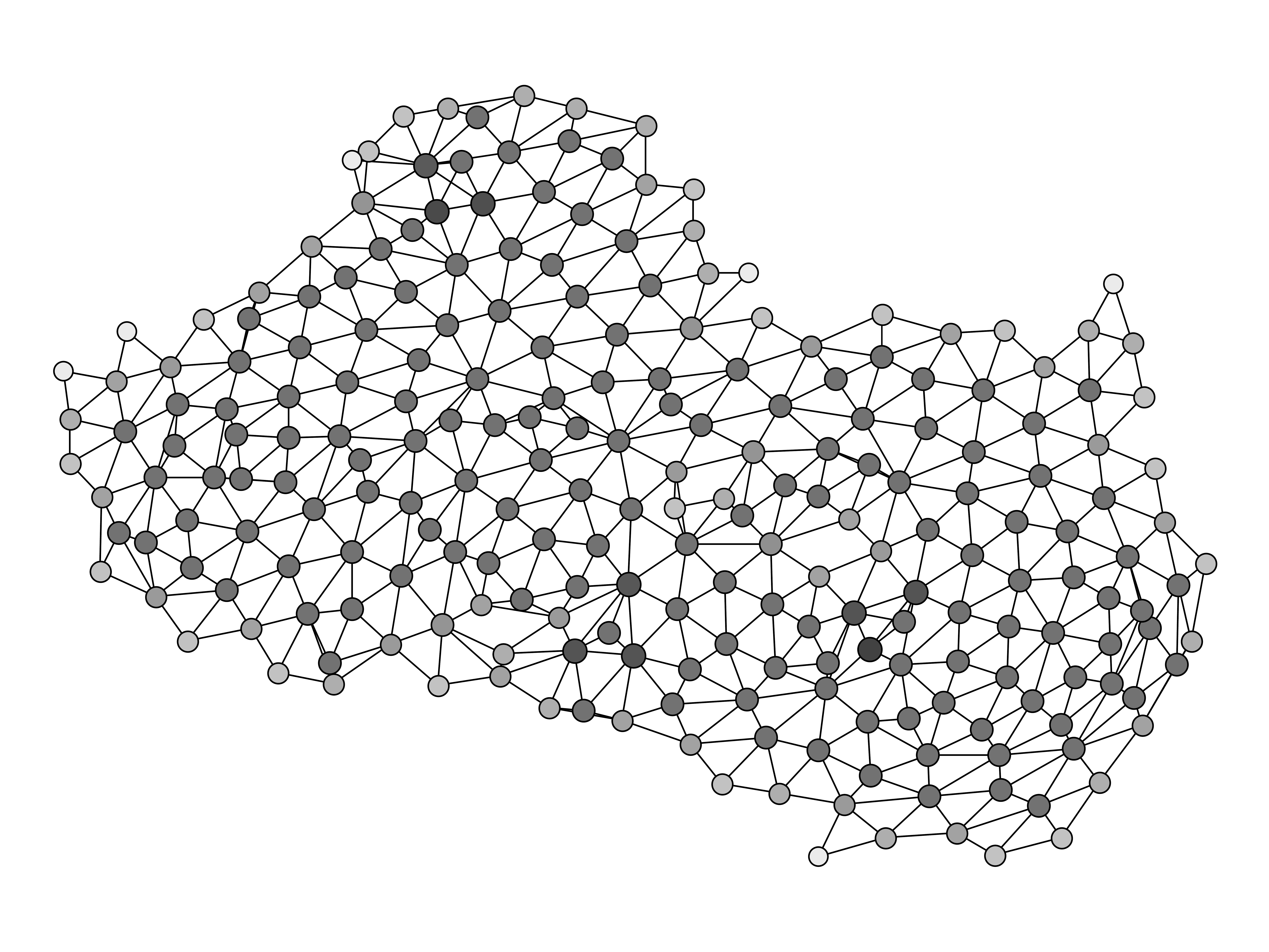}}
	\caption[Meta-networks revealed with the hierarchical label propagation method.]{The meta-networks of the Google web graph and the Pennsylvania road network identified by the hierarchical label propagation method. The shades of the nodes are proportional to their corrected clustering coefficient~\cite{Bat16}, where darker (lighter) means higher (lower).} 
	\label{fig:meta}
\end{figure}

\figref{meta} shows the meta-networks of the largest connected components of the Google web graph from~\figref{comp} with $875,\!713$ nodes and the Pennsylvania road network~\cite{LLDM09} with $1,\!087,\!562$ nodes. Both networks are available at KONECT. The meta-networks were revealed by the hierarchical label propagation method with two and three steps of group agglomeration, and consist of $564$ and $235$ nodes, respectively. Notice that, although the networks are reduced to less than a thousandth of their original size, the group agglomeration process preserves a dense central core of the web graph and a sparse homogeneous topology of the road network~\cite{BSB12}.

Bottom-up group agglomeration can be effectively combined with top-down group refinement~\cite{SB14g,SB14h}. Let $G_1,G_2,\ldots$ be the groups revealed at some step of the group agglomeration. Prior to the construction of the meta-network, one separately applies the label propagation method to the subnetworks of the original network limited to the nodes of groups $G_i$. As this process repeats recursively until a single group is recovered, a sub-hierarchy of groups is revealed for each group $G_i$. Bottom-up agglomeration with top-down refinement enables the identification of a very detailed hierarchy of groups present in a network~\cite{SB14g,GCSKXLBNHGMK15}. One can also further control the resolution of groups by adjusting the weights on the loops in the meta-network~\cite{HLD16}. The time complexity of the described hierarchical label propagation method is $\cmp{cmh}$, where $c$ is the number of iterations and $m$ the number of edges as before, while $h$ is the number of levels of the group hierarchy.


\subsection{\label{sec:equiv}Structural Equivalence Groups} 

\begin{figure}[t]\centerline{%
	\includegraphics[height=.225\textwidth]{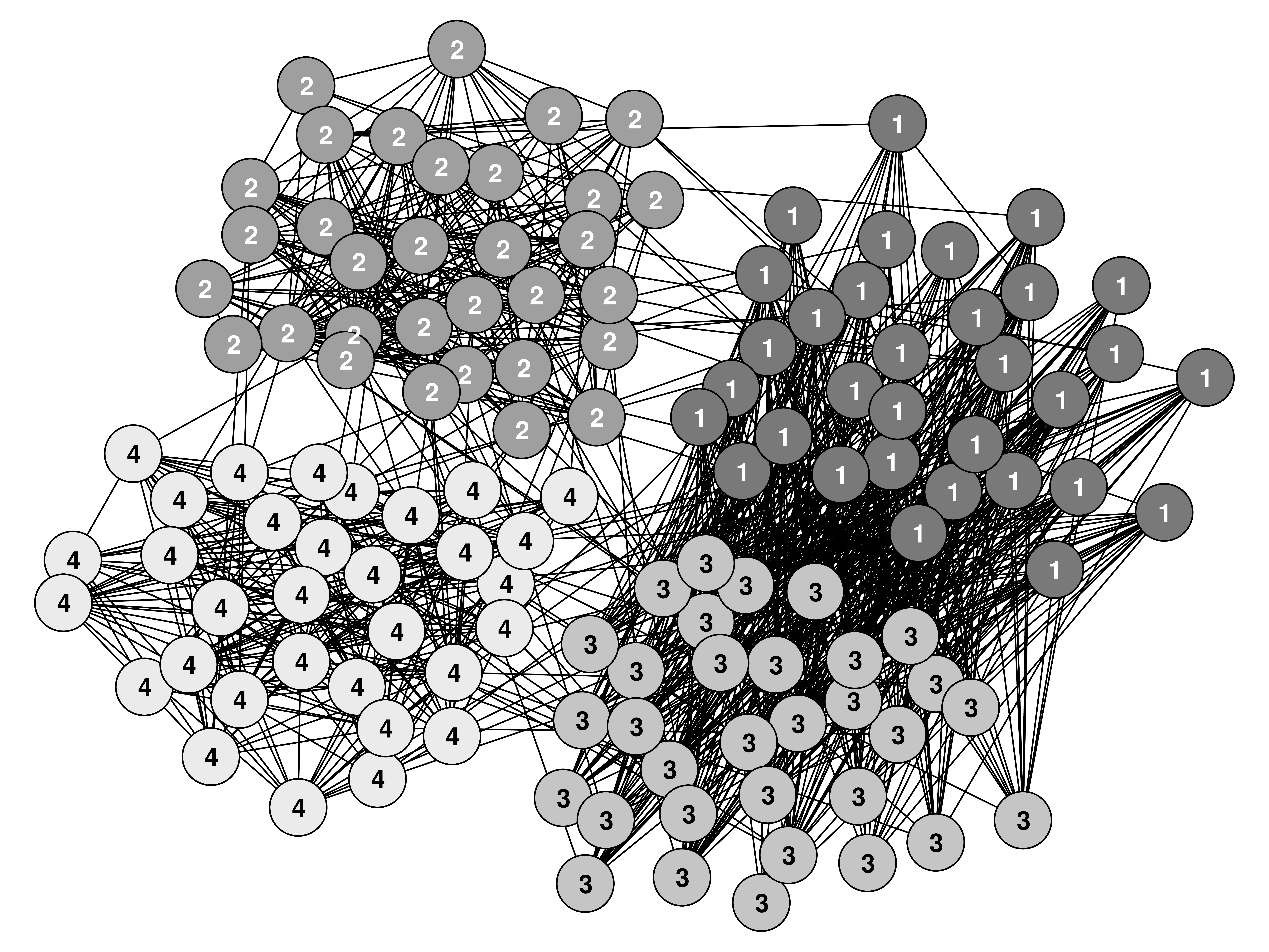}%
	\includegraphics[height=.225\textwidth]{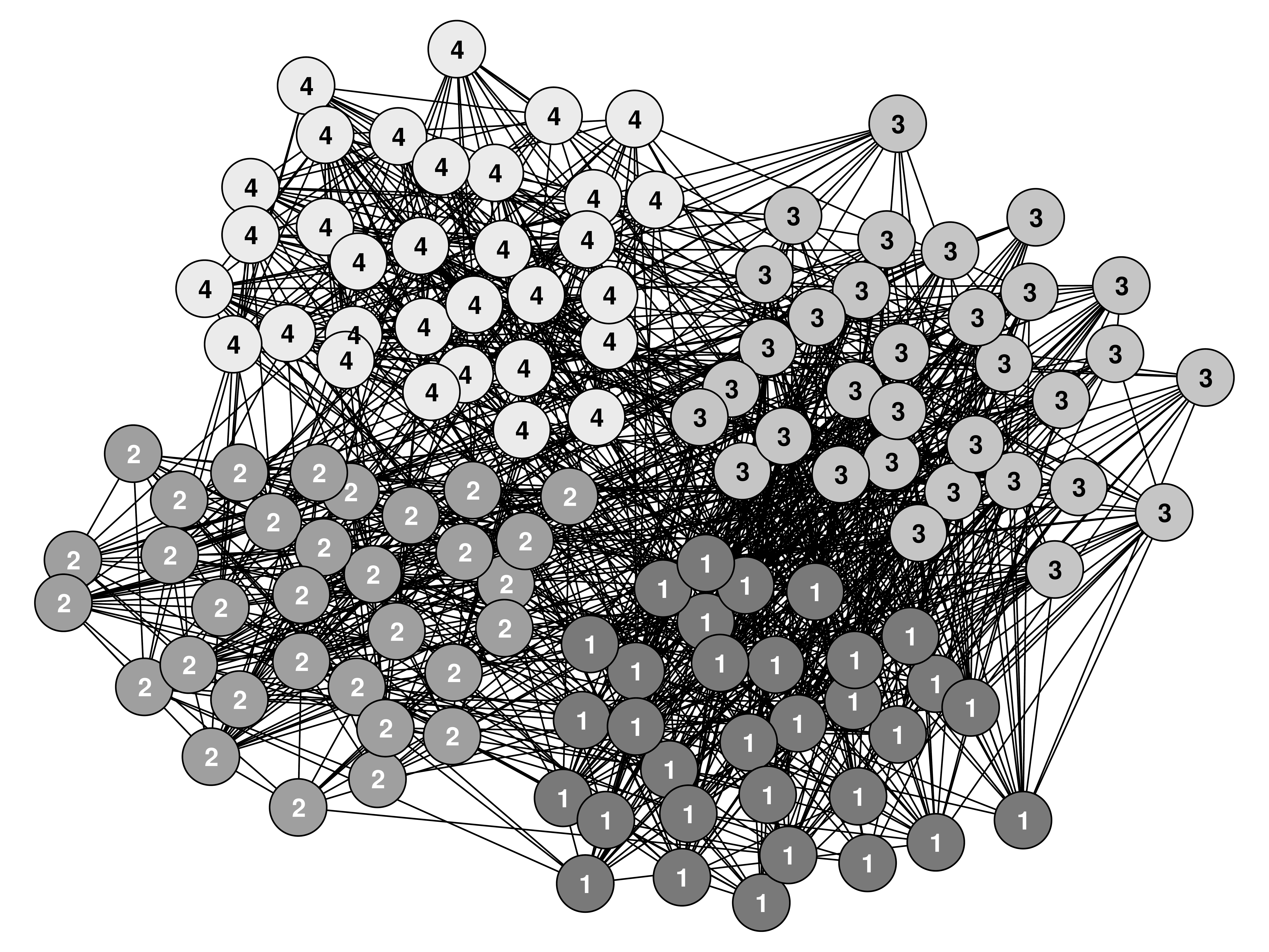}\hskip.05\textwidth
	\includegraphics[height=.25\textwidth]{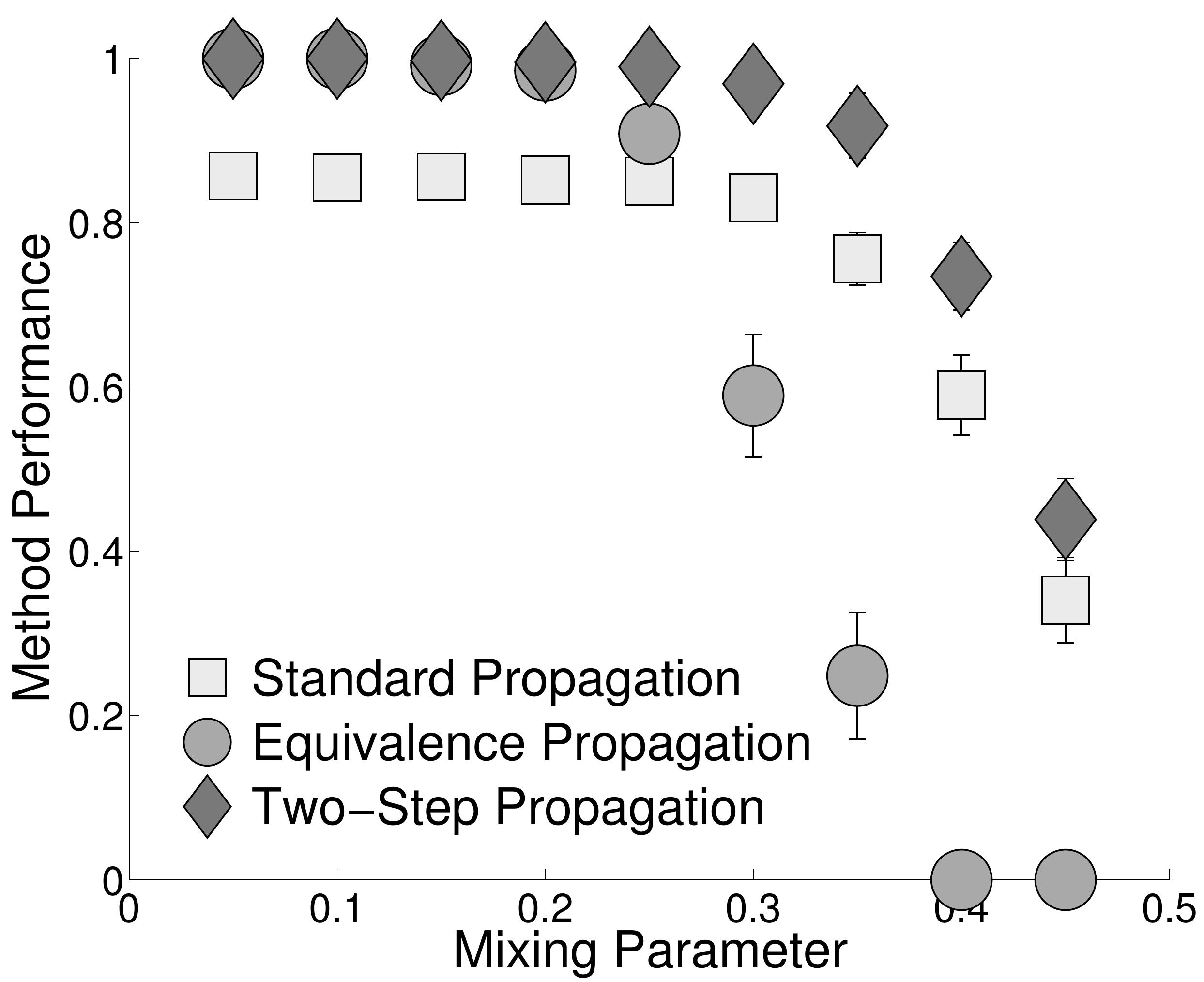}}
	\caption[Structural equivalence detection with the label propagation methods.]{Performance of the label propagation methods in artificial networks with planted communities and structural equivalence groups represented by the labels and shades of the nodes. The markers are averages over $25$ runs of the methods, while the error bars show standard errors.}
	\label{fig:equiv}
\end{figure}

Different label propagation methods presented so far can be used to reveal connected and cohesive groups of nodes in a network. This includes detection of densely connected communities and graph partitioning as demonstrated in~\figref{prefs}. However, the methods cannot be adopted for detection of any kind of disconnected groups of nodes. Therefore, possibly the most interesting extension of the label propagation method is to find groups of structurally equivalent nodes~\cite{SB11g,SB12u,LLZ13,SB14g}. Informally, two nodes are said to be structurally equivalent if they are connected to the same other nodes in the network and thus have the same common neighbors~\cite{LW71,DBF05}, whereas the nodes themselves may be connected or not. We here consider a relaxed definition of structural equivalence in which nodes can have only the majority of their neighbors in common. 
For example, the left side of~\figref{equiv} shows an artificial network with two planted communities of nodes labeled with $2$ and $4$, and two groups of structurally equivalent nodes labeled with $1$ and $3$ that form a bipartite structure. The former are also called assortative groups, while the latter are referred to as disassortative groups~\cite{FH16}.

Let $k_i$ denote the degree of node $i$ and $k_{ij}$ the number of common neighbors of nodes $i$ and $j$. Hence, $k_i=\sum_jA_{ij}$ and $k_{ij}=\sum_kA_{ik}A_{kj}$, where $A$ is the network adjacency matrix. \xs modified the label propagation rule in~\eqref{lpa} as
\begin{equation}
	g_i = \argmax_g\sum_{j}(1+k_{ij})A_{ij}\delta(g_j,g),
	\label{eq:lps}
\end{equation}
which increases the strength of propagation between structurally equivalent nodes. Notice that \eqref{lps} is in fact equivalent to simultaneously propagating the labels between the neighboring nodes as standard and also through their common neighbors represented by the term $k_{ij}$. Yet, the labels are propagated merely between connected nodes, thus the method can still reveal only connected groups of nodes.

\sbg proposed a proper extension of the label propagation method for structural equivalence that separately propagates the labels between the neighboring nodes and through nodes' common neighbors. Let $\tau_g$ be a parameter of group $g$ that is set close to one for connected groups and close to zero for structural equivalence groups. The label propagation rule for general groups of nodes is then written~as
\begin{equation}
	g_i = \argmax_g\left(\tau_g\sum_{j}A_{ij}\delta(g_j,g)+(1-\tau_g)\sum_{kj\neq i}\frac{1}{k_k-1}A_{ik}A_{kj}\delta(g_j,g)\right).
	\label{eq:lpe}
\end{equation}
The lefthand sum propagates the labels between the neighboring nodes $i$ and $j$, while the righthand sum propagates the labels between the nodes $i$ and $j$ through their common neighbors $k$. The degree $k_k$ in the denominator ensures that the number of terms in both sums is proportional to $k_i$. By setting all group parameters in~\eqref{lpe} as $\tau_g=1$, one retrieves the standard label propagation method in~\eqref{lpa} that can detect connected groups of nodes like communities, while setting $\tau_g\approx 0$, the method can detect structural equivalence groups. In the case when a community consists of structurally equivalent nodes as in a clique of nodes, any of the two methods can be used. In practice, the group parameters $\tau_g$ can be inferred from the network structure or estimated during the label propagation process~\cite{SB12u,SB14g}. However, this can make the method very unstable. For this reason, we propose a much simpler approach.

Applying the standard label propagation method to the network in the left side of~\figref{equiv} reveals three groups of nodes, since both structural equivalence groups are detected as a single group of nodes. In general, 
configurations of connected structural equivalence groups are merged together by the method. One can, however, employ this behavior to detect structural equivalence groups using a two-step approach with top-down group refinement introduced before~\cite{SB14g,SB14h}. The first step reveals connected groups of nodes using the standard label propagation method by setting $\tau_g=1$ in~\eqref{lpe}. This includes communities and configurations of connected structural equivalence groups. In the second step, one separately tries to refine each group from the first step using the structural equivalence label propagation method by setting $\tau_g=0$ in~\eqref{lpe}. While communities are still detected as a single group of nodes, configurations of structural equivalence groups are now further partitioned into separate structural equivalence groups.

The right side of~\figref{equiv} compares group detection of the label propagation methods in artificial networks with four groups discussed above~\cite{SB12u}. Network structure is controlled by a mixing parameter $\mu$ that represents the fraction of edges that comply with the group structure, while the examples in the left side of~\figref{equiv} show realizations of networks for $\mu=0.1$ and $0.4$. Performance of the methods is measured with the normalized mutual information~\cite{FH16}, where higher is better. As already mentioned, standard label propagation combines the two structural equivalence groups into a single group. Yet, label propagation for structural equivalence can reveal all four groups, but only when these are clearly defined in the network structure. Finally, the two-step approach performs best in these networks, and can accurately detect communities and structural equivalence groups as long as the latter can first be identified as a single connected group of nodes.

In \secref{nets} we argued that standard label propagation cannot be easily extended to directed networks. In contrast, label propagation for structural equivalence can in fact be adopted for detection of specific groups of nodes in directed networks. For instance, consider a network of citations between scientific papers. Let $A$ be the network adjacency matrix where $A_{ij}$ represents an arc from node $i$ to node $j$ meaning that paper $i$ cites paper $j$. One might be interested in revealing groups of papers that cite the same other papers which is known as cocitation~\cite{BK10,BDFK14}. The label propagation rule for cocitation is
\begin{equation}
	g_i = \argmax_g\sum_{kj\neq i}A_{ik}A_{jk}\delta(g_j,g),
	\label{eq:cocit}
\end{equation}
which propagates the labels between papers $i$ and $j$ through their common citations $k$. An alternative concept is bibliographic coupling~\cite{Jar07}, which refers to groups of papers that are cited by the same other papers. The label propagation rule for bibliographic coupling~is
\begin{equation}
	g_i = \argmax_g\sum_{kj\neq i}A_{ki}A_{kj}\delta(g_j,g).
	\label{eq:bibco}
\end{equation}

\begin{figure}[t]\centerline{%
	\includegraphics[width=.525\textwidth]{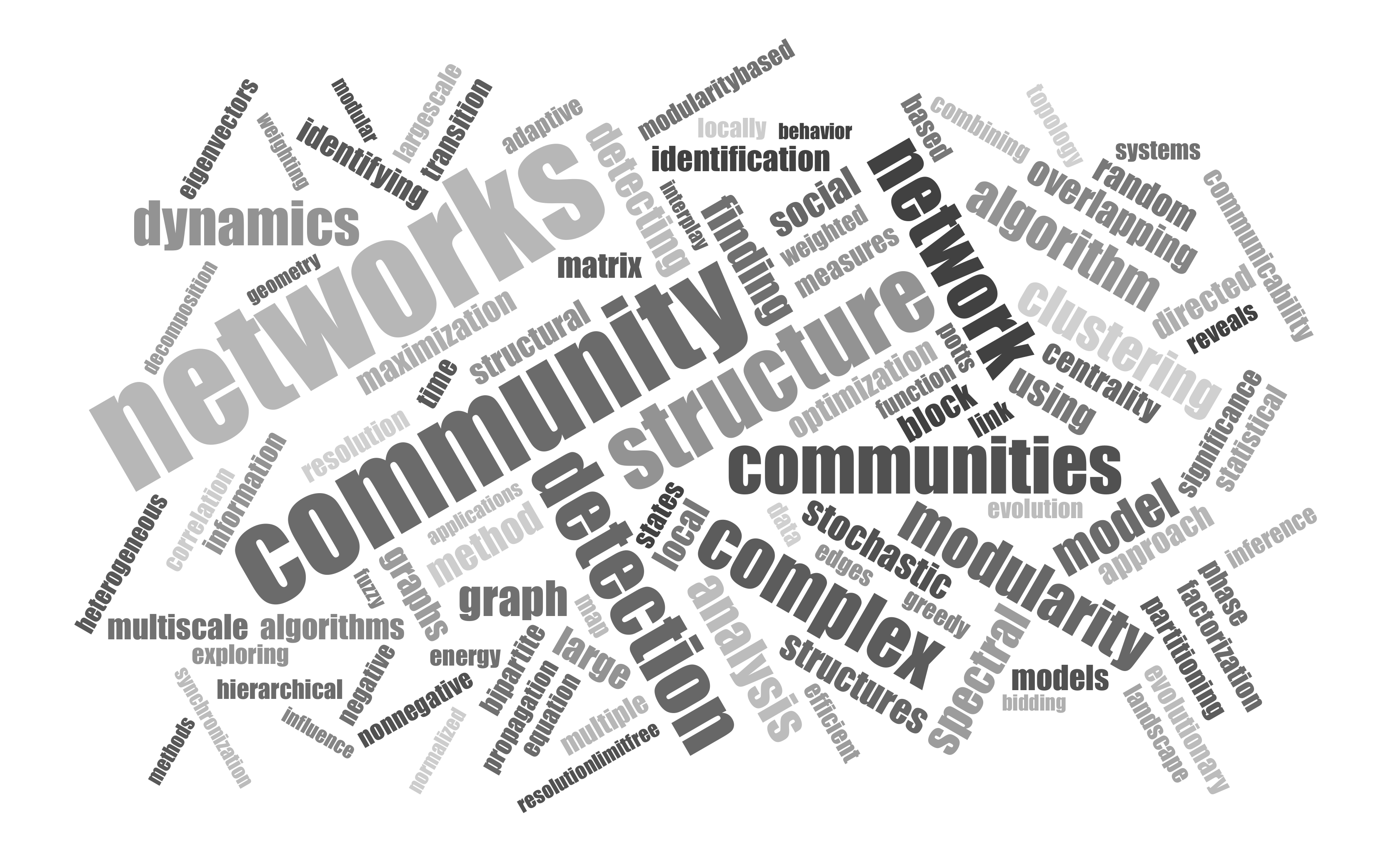}%
	\includegraphics[width=.5\textwidth]{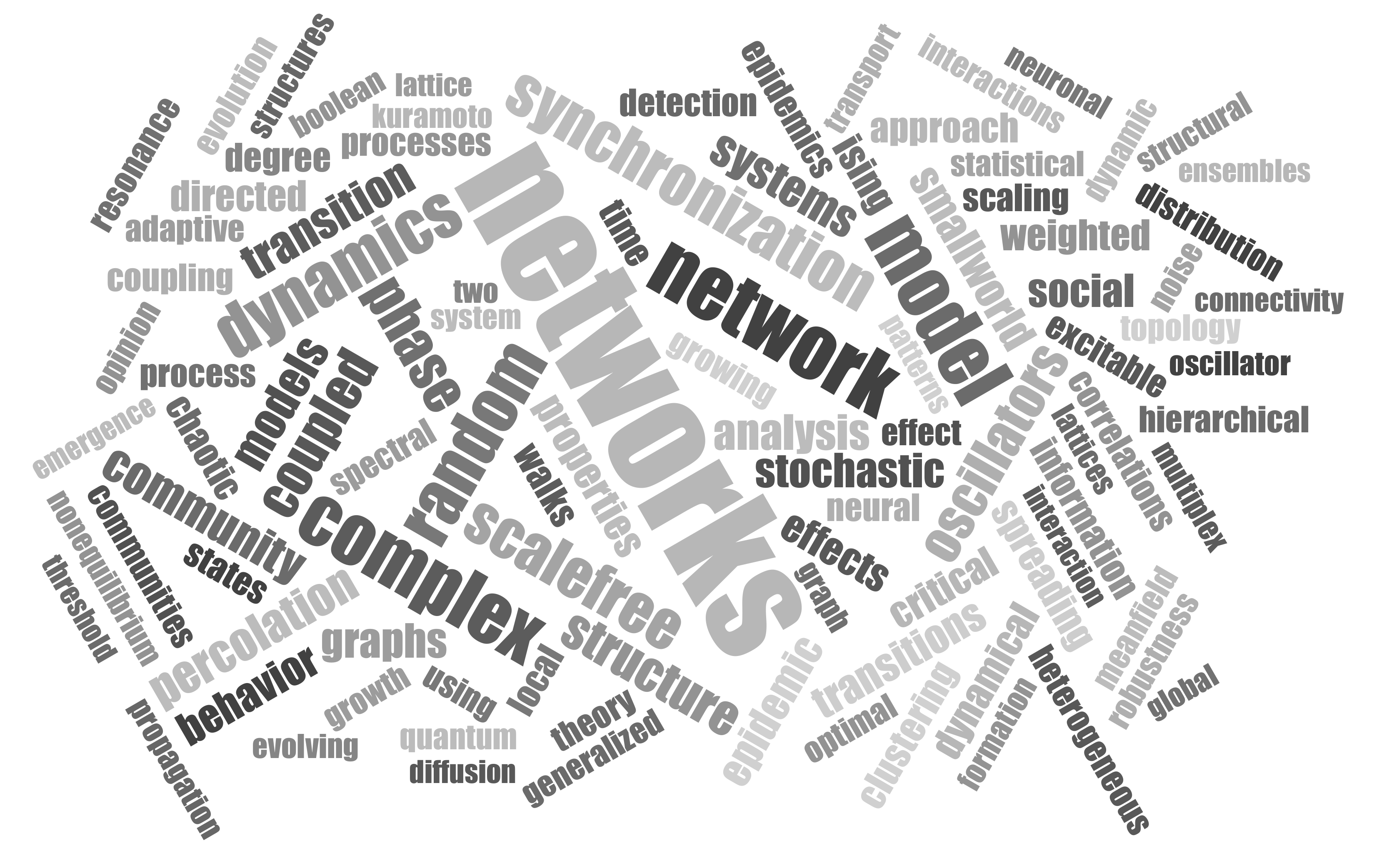}}
	\caption[The label propagation methods applied to a citation network.]{Word clouds demonstrating two of the largest groups of nodes revealed by different label propagation methods in the \emph{Physical Review E} paper citation network. These show the most frequently appearing terms in the titles of the corresponding papers.}
	\label{fig:pre}
\end{figure}

As an example, we constructed a citation network of $26,\!038$ papers published in \emph{Physical Review E}\footnote{\url{http://journals.aps.org/pre}} between the years $2001$ and $2015$. This includes also thirteen references of this chapter namely \crefs{NSW01,NG04,RB06a,RAK07,Bar07,SC08,LFR08,LHLC09,BC09a,RN10,SB11d,TVN11,Tra15}. Twelve of these focus on topics in network community detection and graph partitioning, whereas \cite{NSW01} discusses random graph models. We first ignore the directions of citations and apply the standard label propagation method in~\eqref{lpa} with $25$ runs of consensus clustering introduced in~\secref{perf}. The method reveals $3,\!033$ groups of papers. The largest group consists of $1,\!276$ papers on network structure and dynamics including \cite{NSW01} with the most frequent terms in the titles of the papers being \term{network}, \term{scale-free}, \term{complex}, \term{epidemic}, \term{percolation}, \term{random}, \term{small-world} and \term{social}. The remaining references mentioned above are all included in the fourth largest group with $189$ other papers on network community detection. The left side of~\figref{pre} shows a word cloud generated from the titles of these papers displaying the most frequently appearing terms in an aesthetically pleasing way\footnote{\url{https://www.jasondavies.com/wordcloud}}. These are \term{community}, \term{network}, \term{detection}, \term{modularity}, \term{structure}, \term{complex}, \term{finding} and \term{clustering}.

We next consider also the directions of citations by employing the cocitation label propagation method in~\eqref{cocit} that is again combined with $25$ runs of consensus clustering. The method reveals $1,\!016$ cocitation groups with $2,\!427$ papers in the largest group. The latter consists of papers on various topics in network science including all the thirteen references from above. The right side of~\figref{pre} shows a word cloud generated from the titles of these papers, where the most frequent terms are \term{network}, \term{scale-free}, \term{complex}, \term{synchronization}, \term{community}, \term{random}, \term{small-world} and \term{oscillators}.

As shown in~\secref{comp}, the time complexity of a single iteration of the standard label propagation method is $\cmp{m}=\cmp{\avg{k}n}$, where $n$ and $m$ are the number of nodes and edges in a network, and $\avg{k}=\sum_ik_i/n$ is the average node degree. Since structural equivalence methods presented above propagate the labels also between the nodes two steps apart, the time complexity of a single iteration becomes $\cmp{\avg{k^2}n}$, where $\avg{k^2}=\sum_ik_i^2/n$ is the average node degree squared. The total time complexity of the methods is therefore $\cmp{c\avg{k^2}n}$, where $c$ is the number of iterations of label propagation.

%
%

\section{Applications of Label Propagation}

The label propagation methods are most commonly used for clustering and partitioning large networks with the main goal being network abstraction. In this section, we briefly review also selected other applications of label propagation.

People You May Know is an important feature of the Facebook social service providing recommendations for future friendship ties between its users. Most friendship recommendations are of type \term{friend-of-friend} meaning that the users are suggested other users two hops away in the Facebook social graph~\cite{BL11e}. Due to an immense size of the graph, it is distributed among multiple physical machines thus each machine stores some local part of the graph consisting only of a subset of users. When a friendship recommendation has to be made for a given user, it is desired that the users two steps away in the graph reside at the same machine as the concerned user, in order to minimize the communication between the machines. As reported in $2013$~\cite{UB13}, the users are effectively partitioned among machines using a variant of label propagation under constraints presented in~\secref{const}.

A related application is compression of very large web graphs and online social networks to enable their analysis on a single machine~\cite{BV04b}. Most compression algorithms rely on a given ordering of network nodes such that the edges are mainly between the nodes that are close in the ordering. In the case of web graphs, one can order the nodes representing web pages lexicographically by their URL, whereas no equivalent approach exists for social networks. \brsv adopted the label propagation method in~\eqref{haml3} to compute the ordering of network nodes iteratively starting from a random one. Using such a setting, the authors reported a major improvement in compression with respect to other known techniques. Most social networks and web graphs can be compressed to just a couple of bits per edge, while still allowing for an efficient network traversal. For instance, this compression approach was in fact used to reveal the four degrees of separation between the active users of Facebook in 2011~\cite{BBRUV12}.


%
%

\section{Summary and Outlook}

In this chapter, we have presented the basic label propagation method for network clustering and partitioning, together with its numerous variants and advances, extensions to different types of networks and clusterings, and selected large-scale applications. Due to high popularity of label propagation in the literature, our review here is by no means complete. In particular, we have focused primarily on the results reported in the physics and computer science literature. However, the very same approach is also commonly used in the social networks literature~\cite{BF00,DBF05}, where it is known under the name relocation algorithm or simply as a local greedy optimization. The label propagation method and the relocation algorithm thus provide a sort of common ground between two diverging factions of network science in the natural and social science literature~\cite{Hid16}.

As stated already in the introduction, label propagation is neither the most accurate nor the most robust clustering method. Yet, it is a very fast and versatile method that can readily be applied to largest networks and easily adopted for a particular application. It should be used as the first choice for gaining a preliminary insight into the structure of a network, before trying out more sophisticated and expensive methods. In the case of very large online social networks and web graphs, the label propagation method is in fact often the only choice. Future research should therefore focus more on specific applications of label propagation in large networks, where the use of simple and efficient methods is unavoidable, and less on new \adhoc modifications of the original method, since there are already quite many.

%
%


\todon{Adjust references to Wiley format!}{\begin{chapreferences}{10}
    \bibitem{AM15}
    J.-P. Attal and M.~Malek.
    \newblock A new label propagation with dams.
    \newblock In {\em Proceedings of the {International} {Conference} on {Advances}
      in {Social} {Networks} {Analysis} and {Mining}}, Paris, France, 2015.
    
    \bibitem{BBRUV12}
    L.~Backstrom, P.~Boldi, M.~Rosa, J.~Ugander, and S.~Vigna.
    \newblock Four degrees of separation.
    \newblock In {\em Proceedings of the {ACM} {International} {Conference} on
      {Web} {Science}}, pages 45--54, Evanston, IL, USA, 2012.
    
    \bibitem{BL11e}
    L.~Backstrom and J.~Leskovec.
    \newblock Supervised random walks: {Predicting} and recommending links in
      social networks.
    \newblock In {\em Proceedings of the {ACM} {International} {Conference} on
      {Web} {Search} and {Data} {Mining}}, pages 1--10, Hong Kong, China, 2011.
    
    \bibitem{Bar07}
    M.~J. Barber.
    \newblock Modularity and community detection in bipartite networks.
    \newblock {\em Physical Review E}, 76(6):066102, 2007.
    
    \bibitem{BC09a}
    M.~J. Barber and J.~W. Clark.
    \newblock Detecting network communities by propagating labels under
      constraints.
    \newblock {\em Physical Review E}, 80(2):026129, 2009.
    
    \bibitem{Bat16}
    V.~Batagelj.
    \newblock Corrected overlap weight and clustering coefficient.
    \newblock In {\em Proceedings of the {INSNA} {International} {Social} {Network}
      {Conference}}, pages 16--17, Newport Beach, CA, USA, 2016.
    
    \bibitem{BDFK14}
    V.~Batagelj, P.~Doreian, A.~Ferligoj, and N.~Kej{\v z}ar.
    \newblock {\em Understanding {Large} {Temporal} {Networks} and {Spatial}
      {Networks}}.
    \newblock Wiley, Chichester, 2014.
    
    \bibitem{BF00}
    V.~Batagelj and A.~Ferligoj.
    \newblock Clustering relational data.
    \newblock In {\em Data {Analysis}}, pages 3--15. Springer, Berlin, 2000.
    
    \bibitem{BSB12}
    N.~Blagus, L.~{\v S}ubelj, and M.~Bajec.
    \newblock Self-similar scaling of density in complex real-world networks.
    \newblock {\em Physica A: Statistical Mechanics and its Applications},
      391(8):2794--2802, 2012.
    
    \bibitem{BGLL08}
    V.~D. Blondel, J.-L. Guillaume, R.~Lambiotte, and E.~Lefebvre.
    \newblock Fast unfolding of communities in large networks.
    \newblock {\em Journal of Statistical Mechanics: Theory and Experiment},
      P10008, 2008.
    
    \bibitem{BBCGGRSWZ14}
    S.~Boccaletti, G.~Bianconi, R.~Criado, C.~I. del Genio,
      J.~G{\'o}mez-Garde{\~n}es, M.~Romance, I.~Sendi{\~n}a-Nadal, Z.~Wang, and
      M.~Zanin.
    \newblock The structure and dynamics of multilayer networks.
    \newblock {\em Physics Reports}, 544(1):1--122, 2014.
    
    \bibitem{BRSV11}
    P.~Boldi, M.~Rosa, M.~Santini, and S.~Vigna.
    \newblock Layered label propagation: {A} multiresolution coordinate-free
      ordering for compressing social networks.
    \newblock In {\em Proceedings of the {International} {World} {Wide} {Web}
      {Conference}}, pages 587--596, Hyderabad, India, 2011.
    
    \bibitem{BV04b}
    P.~Boldi and S.~Vigna.
    \newblock The {WebGraph} framework {I}: {Compression} techniques.
    \newblock In {\em Proceedings of the {International} {Conference} on {World}
      {Wide} {Web}}, pages 595--601, New York, NY, USA, 2004.
    
    \bibitem{Bon87}
    P.~Bonacich.
    \newblock Power and centrality: {A} family of measures.
    \newblock {\em American Journal of Sociology}, 92(5):1170--1182, 1987.
    
    \bibitem{BK10}
    K.~W. Boyack and R.~Klavans.
    \newblock Co-citation analysis, bibliographic coupling, and direct citation:
      {Which} citation approach represents the research front most accurately?
    \newblock {\em Journal of the American Society for Information Science and
      Technology}, 61(12):2389--2404, 2010.
    
    \bibitem{BKAFKTK14}
    N.~Buzun, A.~Korshunov, V.~Avanesov, I.~Filonenko, I.~Kozlov, D.~Turdakov, and
      H.~Kim.
    \newblock {EgoLP}: {Fast} and distributed community detection in billion-node
      social networks.
    \newblock In {\em Proceedings of the {IEEE} {International} {Conference} on
      {Data} {Mining} {Workshop}}, pages 533--540, 2014.
    
    \bibitem{CG10}
    G.~Cordasco and L.~Gargano.
    \newblock Community detection via semi-synchronous label propagation
      algorithms.
    \newblock In {\em Proceedings of the {IMSAA} {Workshop} on {Business}
      {Applications} of {Social} {Network} {Analysis}}, pages 1--8, Bangalore,
      India, 2010.
    
    \bibitem{CG11}
    G.~Cordasco and L.~Gargano.
    \newblock Label propagation algorithm: {A} semi{\textendash}synchronous
      approach.
    \newblock {\em International Journal of Social Network Mining}, 1(1):3--26,
      2011.
    
    \bibitem{CRGP12}
    M.~Coscia, G.~Rossetti, F.~Giannotti, and D.~Pedreschi.
    \newblock {DEMON}: {A} local-first discovery method for overlapping
      communities.
    \newblock In {\em Proceedings of the {ACM} {SIGKDD} {International}
      {Conference} on {Knowledge} {Discovery} and {Data} {Mining}}, pages 615--623,
      Beijing, China, 2012.
    
    \bibitem{DBF05}
    P.~Doreian, V.~Batagelj, and A.~Ferligoj.
    \newblock {\em Generalized {Blockmodeling}}.
    \newblock Cambridge University Press, Cambridge, 2005.
    
    \bibitem{DM96}
    P.~Doreian and A.~Mrvar.
    \newblock A partitioning approach to structural balance.
    \newblock {\em Social Networks}, 18(2):149--168, 1996.
    
    \bibitem{ER59}
    P.~Erd{\H o}s and A.~R{\'e}nyi.
    \newblock On random graphs {I}.
    \newblock {\em Publicationes Mathematicae Debrecen}, 6:290--297, 1959.
    
    \bibitem{FH16}
    S.~Fortunato and D.~Hric.
    \newblock Community detection in networks: {A} user guide.
    \newblock {\em Physics Reports}, 659:1--44, 2016.
    
    \bibitem{GCSKXLBNHGMK15}
    C.~Gaiteri, M.~Chen, B.~Szymanski, K.~Kuzmin, J.~Xie, C.~Lee, T.~Blanche, E.~C.
      Neto, S.-C. Huang, T.~Grabowski, T.~Madhyastha, and V.~Komashko.
    \newblock Identifying robust communities and multi-community nodes by combining
      top-down and bottom-up approaches to clustering.
    \newblock {\em Scientific Reports}, 5:16361, 2015.
    
    \bibitem{GN02}
    M.~Girvan and M.~E.~J. Newman.
    \newblock Community structure in social and biological networks.
    \newblock {\em Proceedings of the National Academy of Sciences of United States
      of America}, 99(12):7821--7826, 2002.
    
    \bibitem{Gre10}
    S.~Gregory.
    \newblock Finding overlapping communities in networks by label propagation.
    \newblock {\em New Journal of Physics}, 12(10):103018, 2010.
    
    \bibitem{HLD16}
    J.~Han, W.~Li, and W.~Deng.
    \newblock Multi-resolution community detection in massive networks.
    \newblock {\em Scientific Reports}, 6:38998, 2016.
    
    \bibitem{HLSZD16}
    J.~Han, W.~Li, Z.~Su, L.~Zhao, and W.~Deng.
    \newblock Community detection by label propagation with compression of flow.
    \newblock {\em e-print arXiv:161202463v1}, 2016.
    
    \bibitem{Hid16}
    C.~A. Hidalgo.
    \newblock Disconnected, fragmented, or united? a trans-disciplinary review of
      network science.
    \newblock {\em Applied Network Science}, 1:6, 2016.
    
    \bibitem{Jar07}
    B.~Jarneving.
    \newblock Bibliographic coupling and its application to research-front and
      other core documents.
    \newblock {\em Journal of Infometrics}, 1(4):287--307, 2007.
    
    \bibitem{JMBO01}
    H.~Jeong, S.~P. Mason, A.-L. Barab{\'a}si, and Z.~N. Oltvai.
    \newblock Lethality and centrality of protein networks.
    \newblock {\em Nature}, 411:41--42, 2001.
    
    \bibitem{LF12}
    A.~Lancichinetti and S.~Fortunato.
    \newblock Consensus clustering in complex networks.
    \newblock {\em Scientific Reports}, 2:336, 2012.
    
    \bibitem{LFR08}
    A.~Lancichinetti, S.~Fortunato, and F.~Radicchi.
    \newblock Benchmark graphs for testing community detection algorithms.
    \newblock {\em Physical Review E}, 78(4):046110, 2008.
    
    \bibitem{LLDM09}
    J.~Leskovec, K.~J. Lang, A.~Dasgupta, and M.~W. Mahoney.
    \newblock Community structure in large networks: {Natural} cluster sizes and
      the absence of large well-defined clusters.
    \newblock {\em Internet Mathematics}, 6(1):29--123, 2009.
    
    \bibitem{LHLC09}
    I.~X.~Y. Leung, P.~Hui, P.~Li{\`o}, and J.~Crowcroft.
    \newblock Towards real-time community detection in large networks.
    \newblock {\em Physical Review E}, 79(6):066107, 2009.
    
    \bibitem{LLJT15}
    S.~Li, H.~Lou, W.~Jiang, and J.~Tang.
    \newblock Detecting community structure via synchronous label propagation.
    \newblock {\em Neurocomputing}, 151(3):1063--1075, 2015.
    
    \bibitem{LHWC17}
    W.~Li, C.~Huang, M.~Wang, and X.~Chen.
    \newblock Stepping community detection algorithm based on label propagation and
      similarity.
    \newblock {\em Physica A: Statistical Mechanics and its Applications},
      472:145--155, 2017.
    
    \bibitem{LM09b}
    X.~Liu and T.~Murata.
    \newblock Advanced modularity-specialized label propagation algorithm for
      detecting communities in networks.
    \newblock {\em Physica A: Statistical Mechanics and its Applications},
      389(7):1493, 2009.
    
    \bibitem{LM09c}
    X.~Liu and T.~Murata.
    \newblock Community detection in large-scale bipartite networks.
    \newblock In {\em Proceedings of the {IEEE}/{WIC}/{ACM} {International} {Joint}
      {Conference} on {Web} {Intelligence} and {Intelligent} {Agent} {Technology}},
      pages 50--57, Milano, Italy, 2009.
    
    \bibitem{LM09d}
    X.~Liu and T.~Murata.
    \newblock How does label propagation algorithm work in bipartite networks.
    \newblock In {\em Proceedings of the {IEEE}/{WIC}/{ACM} {International} {Joint}
      {Conference} on {Web} {Intelligence} and {Intelligent} {Agent} {Technology}},
      pages 5--8, Milano, Italy, 2009.
    
    \bibitem{LW71}
    F.~Lorrain and H.~C. White.
    \newblock Structural equivalence of individuals in social networks.
    \newblock {\em Journal of Mathematical Sociology}, 1(1):49--80, 1971.
    
    \bibitem{LLZ13}
    H.~Lou, S.~Li, and Y.~Zhao.
    \newblock Detecting community structure using label propagation with weighted
      coherent neighborhood propinquity.
    \newblock {\em Physica A: Statistical Mechanics and its Applications},
      392(14):3095--3105, 2013.
    
    \bibitem{MAC11}
    S.~Maniu, T.~Abdessalem, and B.~Cautis.
    \newblock Casting a web of trust over {Wikipedia}: {An} {Interaction}-based
      approach.
    \newblock In {\em Proceedings of the {International} {Conference} on {World}
      {Wide} {Web}}, pages 87--88, New York, NY, USA, 2011.
    
    \bibitem{New10}
    M.~E.~J. Newman.
    \newblock {\em Networks: {An} {Introduction}}.
    \newblock Oxford University Press, Oxford, 2010.
    
    \bibitem{NG04}
    M.~E.~J. Newman and M.~Girvan.
    \newblock Finding and evaluating community structure in networks.
    \newblock {\em Physical Review E}, 69(2):026113, 2004.
    
    \bibitem{NSW01}
    M.~E.~J. Newman, S.~H. Strogatz, and D.~J. Watts.
    \newblock Random graphs with arbitrary degree distributions and their
      applications.
    \newblock {\em Physical Review E}, 64(2):026118, 2001.
    
    \bibitem{Ove13}
    M.~Ovelg{\"o}nne.
    \newblock Distributed community detection in web-scale networks.
    \newblock In {\em Proceedings of the {International} {Conference} on {Advances}
      in {Social} {Networks} {Analysis} and {Mining}}, pages 66--73, 2013.
    
    \bibitem{Pot52}
    R.~B. Potts.
    \newblock Some generalized order-disorder transformations.
    \newblock {\em Mathematical Proceedings of the Cambridge Philosophical
      Society}, 48(1):106--109, 1952.
    
    \bibitem{RCCLP04}
    F.~Radicchi, C.~Castellano, F.~Cecconi, V.~Loreto, and D.~Parisi.
    \newblock Defining and identifying communities in networks.
    \newblock {\em Proceedings of the National Academy of Sciences of United States
      of America}, 101(9):2658--2663, 2004.
    
    \bibitem{RAK07}
    U.~N. Raghavan, R.~Albert, and S.~Kumara.
    \newblock Near linear time algorithm to detect community structures in
      large-scale networks.
    \newblock {\em Physical Review E}, 76(3):036106, 2007.
    
    \bibitem{RB06a}
    J.~Reichardt and S.~Bornholdt.
    \newblock Statistical mechanics of community detection.
    \newblock {\em Physical Review E}, 74(1):016110, 2006.
    
    \bibitem{RN10}
    P.~Ronhovde and Z.~Nussinov.
    \newblock Local resolution-limit-free {Potts} model for community detection.
    \newblock {\em Physical Review E}, 81(4):046114, 2010.
    
    \bibitem{RB08}
    M.~Rosvall and C.~T. Bergstrom.
    \newblock Maps of random walks on complex networks reveal community structure.
    \newblock {\em Proceedings of the National Academy of Sciences of United States
      of America}, 105(4):1118--1123, 2008.
    
    \bibitem{SDRL17}
    M.~T. Schaub, J.-C. Delvenne, M.~Rosvall, and R.~Lambiotte.
    \newblock The many facets of community detection in complex networks.
    \newblock {\em Applied Network Science}, 2:4, 2017.
    
    \bibitem{SC08}
    P.~Schuetz and A.~Caflisch.
    \newblock Efficient modularity optimization by multistep greedy algorithm and
      vertex mover refinement.
    \newblock {\em Physical Review E}, 77:046112, 2008.
    
    \bibitem{SN11}
    J.~Soman and A.~Narang.
    \newblock Fast community detection algorithm with {GPUs} and multicore
      architectures.
    \newblock In {\em Proceedings of the {IEEE} {International} {Parallel}
      {Distributed} {Processing} {Symposium}}, pages 568 --579, Anchorage, AK, USA,
      2011.
    
    \bibitem{SB10d}
    L.~{\v S}ubelj and M.~Bajec.
    \newblock Unfolding network communities by combining defensive and offensive
      label propagation.
    \newblock In {\em Proceedings of the {ECML} {PKDD} {Workshop} on the {Analysis}
      of {Complex} {Networks}}, pages 87--104, Barcelona, Spain, 2010.
    
    \bibitem{SB11g}
    L.~{\v S}ubelj and M.~Bajec.
    \newblock Generalized network community detection.
    \newblock In {\em Proceedings of the {ECML} {PKDD} {Workshop} on {Finding}
      {Patterns} of {Human} {Behaviors} in {Network} and {Mobility} {Data}}, pages
      66--84, Athens, Greece, 2011.
    
    \bibitem{SB11b}
    L.~{\v S}ubelj and M.~Bajec.
    \newblock Robust network community detection using balanced propagation.
    \newblock {\em European Physical Journal B}, 81(3):353--362, 2011.
    
    \bibitem{SB11d}
    L.~{\v S}ubelj and M.~Bajec.
    \newblock Unfolding communities in large complex networks: {Combining}
      defensive and offensive label propagation for core extraction.
    \newblock {\em Physical Review E}, 83(3):036103, 2011.
    
    \bibitem{SB12u}
    L.~{\v S}ubelj and M.~Bajec.
    \newblock Ubiquitousness of link-density and link-pattern communities in
      real-world networks.
    \newblock {\em European Physical Journal B}, 85(1):32, 2012.
    
    \bibitem{SB14g}
    L.~{\v S}ubelj and M.~Bajec.
    \newblock Group detection in complex networks: {An} algorithm and comparison of
      the state of the art.
    \newblock {\em Physica A: Statistical Mechanics and its Applications},
      397:144--156, 2014.
    
    \bibitem{SB14h}
    L.~{\v S}ubelj and M.~Bajec.
    \newblock Network group discovery by hierarchical label propagation.
    \newblock In {\em Proceedings of the {European} {Social} {Networks}
      {Conference}}, page 284, Barcelona, Spain, 2014.
    
    \bibitem{SVW16a}
    L.~{\v S}ubelj, N.~J. Van~Eck, and L.~Waltman.
    \newblock Clustering scientific publications based on citation relations: {A}
      systematic comparison of different methods.
    \newblock {\em PLoS ONE}, 11(4):e0154404, 2016.
    
    \bibitem{TK08}
    G.~Tib{\'e}ly and J.~Kert{\'e}sz.
    \newblock On the equivalence of the label propagation method of community
      detection and a {Potts} model approach.
    \newblock {\em Physica A: Statistical Mechanics and its Applications},
      387(19-20):4982--4984, 2008.
    
    \bibitem{Tra15}
    V.~A. Traag.
    \newblock Faster unfolding of communities: {Speeding} up the {Louvain}
      algorithm.
    \newblock {\em Physical Review E}, 92:032801, 2015.
    
    \bibitem{TVN11}
    V.~A. Traag, P.~Van~Dooren, and Y.~Nesterov.
    \newblock Narrow scope for resolution-limit-free community detection.
    \newblock {\em Physical Review E}, 84(1):016114, 2011.
    
    \bibitem{UB13}
    J.~Ugander and L.~Backstrom.
    \newblock Balanced label propagation for partitioning massive graphs.
    \newblock In {\em Proceedings of the {ACM} {International} {Conference} on
      {Web} {Search} and {Data} {Mining}}, pages 507--516, Rome, Italy, 2013.
    
    \bibitem{WXSW14}
    L.~Wang, Y.~Xiao, B.~Shao, and H.~Wang.
    \newblock How to partition a billion-node graph.
    \newblock In {\em Proceedings of the {International} {Conference} on {Data}
      {Engineering}}, pages 568--579, 2014.
    
    \bibitem{WuF82}
    F.~Y. Wu.
    \newblock The {Potts} model.
    \newblock {\em Reviews of Modern Physics}, 54(1):235--268, 1982.
    
    \bibitem{WLGWT12}
    Z.-H. Wu, Y.-F. Lin, S.~Gregory, H.-Y. Wan, and S.-F. Tian.
    \newblock Balanced multi-label propagation for overlapping community detection
      in social networks.
    \newblock {\em Journal of Computer Science and Technology}, 27(3):468--479,
      2012.
    
    \bibitem{XS11}
    J.~Xie and B.~K. Szymanski.
    \newblock Community detection using a neighborhood strength driven label
      propagation algorithm.
    \newblock In {\em Proceedings of the {IEEE} {International} {Workshop} on
      {Network} {Science}}, pages 188--195, West Point, NY, USA, 2011.
    
    \bibitem{XS12}
    J.~Xie and B.~K. Szymanski.
    \newblock Towards linear time overlapping community detection in social
      networks.
    \newblock In {\em Proceedings of the {Pacific}-{Asia} {Conference} on
      {Knowledge} {Discovery} and {Data} {Mining}}, pages 1--13, Kuala Lumpur,
      Malaysia, 2012.
    
    \bibitem{XS13}
    J.~Xie and B.~K. Szymanski.
    \newblock {LabelRank}: {A} stabilized label propagation algorithm for community
      detection in networks.
    \newblock In {\em Proceedings of the {IEEE} {International} {Workshop} on
      {Network} {Science}}, pages 138--143, 2013.
    
    \bibitem{XSL11}
    J.~Xie, B.~K. Szymanski, and X.~Liu.
    \newblock {SLPA}: {Uncovering} overlapping communities in social networks via a
      speaker-listener interaction dynamic process.
    \newblock In {\em Proceedings of the {ICDM} {Workshop} on {Data} {Mining}
      {Technologies} for {Computational} {Collective} {Intelligence}}, 2011.
    
    \bibitem{Zac77}
    W.~W. Zachary.
    \newblock An information flow model for conflict and fission in small groups.
    \newblock {\em Journal of Anthropological Research}, 33(4):452--473, 1977.
    
    \bibitem{ZCH15}
    J.~Zhang, T.~Chen, and J.~Hu.
    \newblock On the relationship between {Gaussian} stochastic blockmodels and
      label propagation algorithms.
    \newblock {\em Journal of Statistical Mechanics: Theory and Experiment},
      2015(3):P03009, 2015.
    
    \bibitem{ZJFP14}
    L.~Zong-Wen, L.~Jian-Ping, Y.~Fan, and A.~Petropulu.
    \newblock Detecting community structure using label propagation with consensus
      weight in complex network.
    \newblock {\em Chinese Physics B}, 23(9):098902, 2014.
\end{chapreferences}}

~\comm{The chapter was compiled on \today\xspace at \currenttime.}


\end{document}